\pdfoutput=1
\documentclass[
twocolumn,
amsmath,amssymb,
aps,
prl,
raggedbottom
]{revtex4-2}

\usepackage[dvipsnames]{xcolor}
\usepackage{tikz}
\usetikzlibrary{arrows.meta, bending, decorations.markings, decorations.pathmorphing}
 
\usepackage{microtype}
\usepackage{mathtools}

\usepackage{bm}
\usepackage[normalem]{ulem}
\usepackage{orcidlink}

\usepackage{hyperref}
\hypersetup{
    pdfencoding=unicode,
	colorlinks=true,
	urlcolor=Maroon,
	linkcolor=RoyalBlue,
	citecolor=Maroon,
	pdftitle={},
	pdfauthor={},
	pdfdisplaydoctitle=true,
	pdfstartview=FitH,
	linktocpage=true
}

\makeatletter
\newcommand\appendixtocname{Supplemental material}
\newcommand\appendixtoc{\@ifstar{\apc@start\@ne}{\apc@start\z@}}
\newcommand\apc@start[1]{\begingroup
    \ifnum#1=\z@ \let\addcontentsline\@gobblethree \fi
    \section*{\appendixtocname}\endgroup
  \begingroup
    \let\appendix\appendix@toc
    \@starttoc{apc}\endgroup
  \ifnum#1=\@ne \addtocontents{toc}{\protect\apc@hide}\fi
  \let\apc@addcontentsline\addcontentsline
  \def\addcontentsline##1##2##3{\apc@addcontentsline{##1}{##2}{##3}\def\apc@tmp{##1}\ifx\apc@tmp\apc@toc \apc@addcontentsline{apc}{##2}{##3}\fi}}
\def\apc@toc{toc}
\newcommand\stopappendixtoc{\@ifundefined{apc@addcontentsline}{}{\let\addcontentsline\apc@addcontentsline
    \addtocontents{toc}{\protect\apc@show}}}
\def\apc@hide{\let\apc@l@section\l@section \let\apc@l@subsection\l@subsection
  \let\apc@l@subsubsection\l@subsubsection
  \let\l@section\@gobbletwo \let\l@subsection\@gobbletwo
  \let\l@subsubsection\@gobbletwo}
\def\apc@show{\@ifundefined{apc@l@section}{}{\let\l@section\apc@l@section \let\l@subsection\apc@l@subsection
    \let\l@subsubsection\apc@l@subsubsection}}
\makeatother

\newcommand{\ba}{\begin{align}}

\newcommand{\be}{\begin{equation}}
\newcommand{\ee}{\end{equation}}

\renewcommand\d{\mathrm{d}}

\newcommand{\nn}{\nonumber}

\newcommand{\e}{\mathrm{e}}
\renewcommand{\O}{\mathcal{O}}

\let\stdle\le \let\stdge\ge
\renewcommand{\le}{\leqslant}
\renewcommand{\ge}{\geqslant}
\renewcommand{\leq}{\leqslant}
\renewcommand{\geq}{\geqslant}
\newcommand{\la}{\langle}
\newcommand{\ra}{\rangle}

\newcommand{\bJ}{\bm{J}}

\newcommand{\bI}{\bm{I}}
\newcommand{\bC}{\bm{C}}
\newcommand{\bG}{\bm{G}}
\newcommand{\bY}{\bm{Y}}
\newcommand{\bR}{\bm{R}}
\newcommand{\bD}{\bm{D}}
\newcommand{\bh}{\bm{h}}

\newcommand{\bkappa}{\bm{\kappa}}
\newcommand{\bK}{\bm{K}}
\newcommand{\bzeta}{\vec{\bm{\zeta}}}

\newcommand{\bx}{\bm{x}}

\newcommand{\bpsi}{\bm{\psi}}
\newcommand{\bxi}{\bm{\xi}}
\newcommand{\bV}{\bm{V}}
\newcommand{\bchi}{\bm{\chi}}
\newcommand{\bet}{\bm{\eta}}
\newcommand{\blambda}{\bm{\lambda}}
\newcommand{\bLambda}{\bm{\Lambda}}
\newcommand{\bSigma}{\bm{\Sigma}}
\newcommand{\bOmega}{\bm{\Omega}}
\newcommand{\bGamma}{\bm{\Gamma}}

\newcommand{\bA}{\bm{A}}
\newcommand{\bW}{\bm{W}}
\newcommand{\bU}{\bm{U}}
\newcommand{\bM}{\bm{M}}
\newcommand{\bc}{\bm{c}}
\newcommand{\bb}{\bm{b}}

\newcommand{\betc}{\bm{\eta}^c}

\newcommand{\bpsic}{\bm{\psi}^c}

\newcommand{\brc}{\bm{R}^c}
\newcommand{\bcc}{\bm{C}^c}
\newcommand{\bv}{\bm{v}}
\newcommand{\bu}{\bm{u}}

\usepackage{amsthm}
\usepackage{enumitem}
\newtheorem{prop}{Proposition}
\newcommand{\ex}[1]{e^{#1}}
\newcommand{\bF}{\bm{F}}
\newcommand{\bQ}{\bm{Q}}
\newcommand{\bsigma}{\bm{\sigma}}
\newcommand{\beps}{\bm{\epsilon}}
\newcommand{\bmm}{\bm{m}}
\newcommand{\by}{\bm{y}}
\newcommand{\bw}{\bm{w}}
\newcommand{\ih}{i\hat\bx}

\newcommand{\obs}{\mathcal O}
\newcommand{\dk}{\delta\bkappa}
\newcommand{\ms}{\bmm^s}
\newcommand{\dkh}{\delta\hat\bkappa}
\newcommand{\sadp}{s}
\newcommand{\csp}{\bC^{s}}
\newcommand{\barkap}{\overline{\bkappa}}
\newcommand{\rsp}{\bR^{s}}
\newcommand{\rkap}{\bR^{(\kappa)}}
\newcommand{\ckap}{\bC^{(\kappa)}}
\newcommand{\Mpop}{\bm{\mathcal M}}
\newcommand{\Rc}{\bR^{(\mathrm{cav})}}
\newcommand{\mc}{\bmm^{(\mathrm{cav})}}
\newcommand{\Cc} {\bC^{(\mathrm{cav})}}
\newcommand{\Sc} {\bSigma^{(\mathrm{cav})}}
\newcommand{\Lc} {\bLambda^{(\mathrm{cav})}}
\newcommand{\Rd}{\bR^\infty}
\newcommand{\Cd}{\bC^\infty}
\newcommand{\md}{\bmm^\infty}
\begin{document}

\title{Connectivity from activity in spiking neural networks}

\author{Albert J. Wakhloo \orcidlink{0009-0003-4669-1572}}
\affiliation{Center for Theoretical Neuroscience, Zuckerman Institute, Columbia University}

\author{Sebastian Mizera \orcidlink{0000-0002-8066-5891}}
\affiliation{Center for Theoretical Physics, Department of Physics, Columbia University}

\begin{abstract}\noindent
We show that a wide range of recurrent neural network models obey a simple set of equations tying synaptic connectivity and single-cell response properties to neural correlations.
Our modeling framework considers networks in a stationary, asynchronous state and contains a broad class of synaptic filters, spiking nonlinearities, and random-plus-structured connectivity.
The resulting equations naturally decompose into contributions from monosynaptic, disynaptic, and higher order paths through the network.
We leverage this decomposition to express the effective monosynaptic connection between two neurons in terms of experimentally measurable statistics of the activity, and in particular, their high-frequency correlations. In this way, we show that it is possible to recover network connectivity from neural activity.
\end{abstract}

\maketitle

\section{Introduction}

\emph{All linear systems are alike, but every nonlinear system is nonlinear in its own way}. In neuroscience, we are faced with nonlinear networks of neurons featuring an immense diversity of synaptic connections and cell-type specific physiologies. Is it possible to derive general relations tying such structure to neural activity in spite of the system's nonlinearity?

In this Letter, we derive a set of equations describing how connectivity gives rise to correlations across a wide range of recurrent neural network models. 
Our results show that the spike-time cross-correlations can be expressed as a function of two quantities: the \emph{auto-correlations} of the cells in the network and the \emph{effective connectivity} \cite{NIPS1993_f7e9050c,PhysRevE.50.3171}, defined as the synaptic filter convolved with the post-synaptic cell's gain. The specific features of a model's nonlinearity or connectivity do not enter beyond their effect on these quantities.

Concretely, we show that at large network size $N$, the Fourier-transformed correlation and linear response functions, $\bC(\omega)$ and $\bR(\omega)$ respectively, are $N \times N$ matrices of the form
\begin{gather} 
    \bC= \bG \bLambda \bG^\dagger\,, \quad \bR = \bG \bSigma\,, 
    \label{eq:introresult}
\end{gather} 
where $\dagger$ denotes the complex-conjugate transpose. In the above, $\bSigma(\omega)$ is a diagonal matrix describing the effective postsynaptic gain, and the resolvent $\bG(\omega)$ is given by
\begin{gather}\label{eq:introG}
    \bG = (\bI - \bV)^{-1},
\end{gather}
where the \emph{effective connectivity} $\bV(\omega)$ is the synaptic weight matrix, multiplied by the postsynaptic gain matrix $\bSigma(\omega)$, and $\bLambda(\omega)$ is a diagonal source matrix that is a function of $\bV$ and the diagonal entries $\mathrm{diag}(\bC)$ and $\mathrm{diag}(\bR)$. Therefore, cross-correlations and responses, $\bC_{ij}$, $\bR_{ij}$ with $i\neq j$ are entirely determined by the effective connectivity and the single unit auto-correlations and responses.

Expanding Eqs.~\eqref{eq:introresult} in powers of $\bV$ and performing an inverse Fourier transform decomposes the correlation and response function into monosynaptic, disynaptic, and higher order paths through the network, as in linear network models \cite{10.1371/journal.pcbi.1002059, 10.1371/journal.pcbi.1002408}. We will see that each of these paths leaves a distinct imprint on $\bC$ and $\bR$. 

In particular, we show that in a certain large-$N$ large-$\omega$ limit, one has
\begin{gather}
    \bm C_{ij} \simeq \bm V_{ij} \bm C_{jj} + \bm C_{ii} \bm V_{ji}^*
\end{gather}
for $i\neq j$. Given a sufficiently precise measurement of $\bm C_{ij}, \bm C_{ii}, \bm C_{jj}$ in high frequency bands, it is therefore possible to infer the effective monosynaptic connections $\bm V_{ij}, \bm V_{ji}$ between a pair of neurons without knowing the underlying model details or observing the entire network.

We prove Eqs.~\eqref{eq:introresult} using a diagrammatic expansion of a path integral formulation of the network dynamics. In this way, we establish the result for a wide range of recurrent network models in an asynchronous, stationary state as is often assumed in models of cortical activity \cite{renart2010asynchronous, ecker2010decorrelated, van1996chaos, brunel2000dynamics}. 
The proof is given in the Supplemental Material (\ref{sm}), and we provide a compressed cavity derivation in the main text. Note that while Eqs.~\eqref{eq:introresult} are given in a self-consistent form---the diagonals of $\bC$ and $\bR$ appear on both sides---it is possible to close the formulae for these matrices through a set of dynamical mean-field theory (DMFT) equations, given in App.~\ref{app:DMFT}. Finally, we show numerical results of our estimator below and in App.~\ref{app:Numerics}. Previous results on reconstructing connectivity from spike data include \cite{perkel1967neuronal, brillinger1976identification, aertsen1985evaluation, csicsvari1998reliability, fujisawa2008behavior, stark2009unbiased, pernice2013reconstruction, english2017pyramidal, kobayashi2019reconstructing, ren2020model, endo2021convolutional, petersen2021cellexplorer, platkiewicz2021monosynaptic, jia2022multiregional, Huszar2022, spivak2022deconvolution, donner2024ensemble, Fink2025.01.16.633450,kobayashi2025inference}. Our work complements these efforts by deriving a theoretically principled approach to connectivity reconstruction.

\section{Model}

We now state our modeling framework in its most general form and show that it contains as special cases various canonical neural network models. 

Our model consists of a recurrent network of $N$ neurons.  We denote the membrane potential of the $i^\text{th}$ neuron by $\bx_i(t)$ and its spike train (or firing rate) by $\bpsi_i(t)$. We will take $\bm\psi_i$ to be an arbitrary, possibly stochastic, causal functional of the membrane potential, $\bm\psi_i(t) = \bm\psi_i[\bx_i](t)$, the precise form of which can vary from cell to cell.  The dynamics of the potential obey the dynamical equations
\begin{gather}
    \mathcal L [\bx_i] =  \bet_i + \bxi_i,
    \label{eq:potential}
\end{gather} 
where $\mathcal L$ is another arbitrary causal operator describing the mapping from synaptic currents to somatic membrane potentials. 

\begin{figure}
\includegraphics[width=0.4\textwidth]{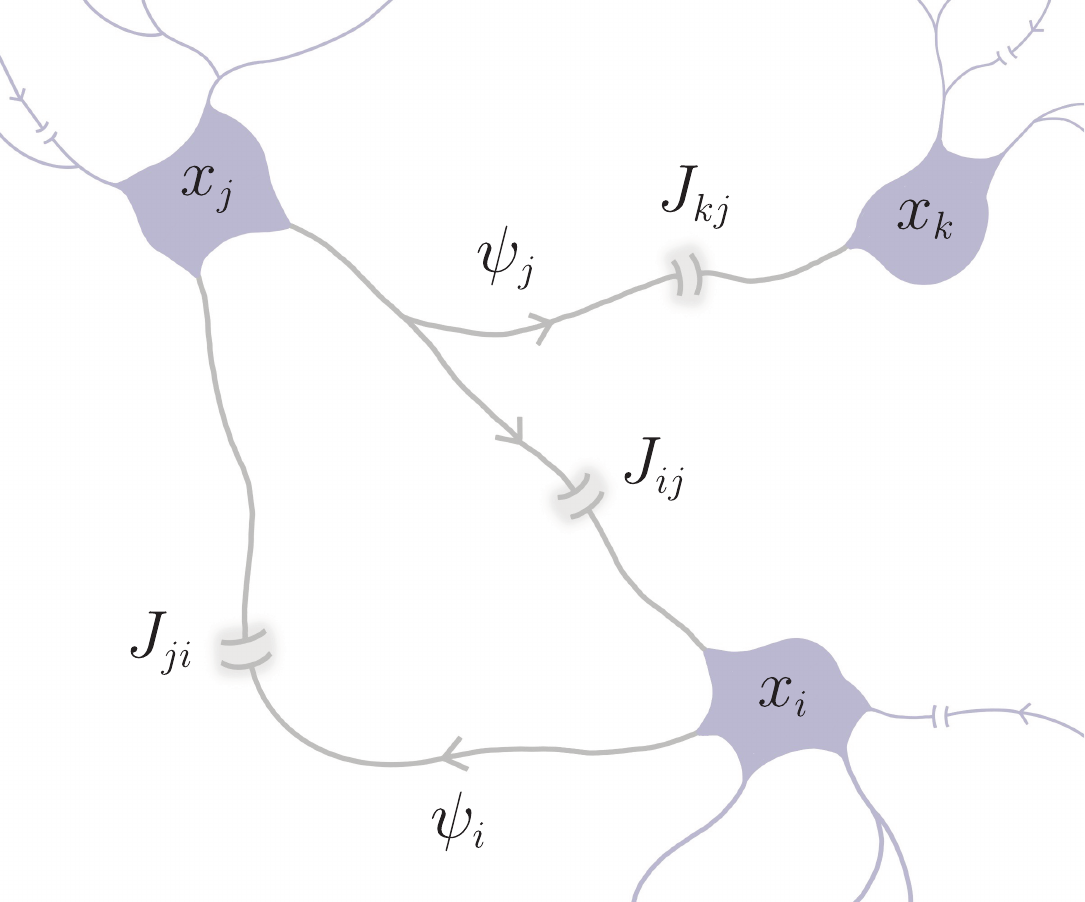}
\caption{\label{fig:neurons}Schematic of our modeling framework (Eq.\ \eqref{eq:potential}): each cell is described by a potential $\bx_i$ and a spike train $\bpsi_i$. The units receive external inputs $\bxi_i$ and communicate through synapses with a filter $\bJ_{ij} = \bW_{ij} \bGamma_{ij}(t-t').$}
\end{figure}

In Eq.~\eqref{eq:potential}, we have split the synaptic current into two components: external inputs $\bxi_i(t)$ into the network and recurrent input $\bet_i(t)$ (Fig.\ \ref{fig:neurons}). The recurrent input evolves according to
\begin{gather}
\bet_i(t) = \sum_{j=1}^N  \int_{-\infty}^t \bJ_{ij}(t-\tau)\, \bpsi_j(\tau) \, \d\tau\, .
\label{eq:rec}
\end{gather} 
In the above, the synapse factorizes as $\bJ_{ij}(t-t') = \bW_{ij} \bGamma_{ij}(t-t')$, where $\bW_{ij}$ are the dimensionless synaptic weights, and $\bGamma_{ij}(\tau)$ describes both synaptic filtering effects as well as time delays in spiking transmission. As we now show, our framework encompasses a number of common models of neural dynamics.  

\textbf{Rate models:} Taking $\bpsi_i(t)$ a smooth function of the potentials, we obtain rate networks. For example, the classic Sompolinsky--Crisanti--Sommers models \cite{SCS1988,CrisantiSompolinsky2018} correspond to taking $\bpsi_i(t) =\phi(\bx_i(t))$ for some nonlinear function $\phi$, the potential operator to be $\mathcal L = 1 +\partial_t$, and $\bGamma_{ij}(\tau) = \delta(\tau)$. The Wilson--Cowan model \cite{WilsonCowan1972,WilsonCowan1973} is similarly a special case. 

\textbf{Phenomenological spiking models:} If we take $\bpsi_i(t)$ to be a sum over delta functions centered at the spike times of the $i$th cell, $\bpsi_i(t) = \sum_s \delta(t- t^i_s)$, one obtains a family of spiking networks. For example choosing $\bpsi_i$ to be a Poisson process whose intensity is a filtered version of the potential gives the Hawkes models \cite{Hawkes1971,Hawkes1971b}. As another example, if we take $\bpsi_i$ to be a spike emission functional with a refractory period and $\mathcal L$ a low-pass filter with a spiking reset term, we obtain networks of leaky-integrate-and-fire (LIF) or adaptive integrate-and-fire neurons.

\textbf{Conditions on disordered connections:} We now describe our assumptions on the connectivity matrix $\bW$, which we split into disordered and structured components: $\bW=\delta\bW + \bM$.  We assume that the disordered components $\delta \bW_{ij}$ are drawn randomly from a zero-mean distribution with zero diagonal $\delta \bW_{jj} = 0$, synapse-specific variances $\langle \delta \bW_{ij}^2\rangle_{\delta\bW} =\sigma^2_{ij}/N$, and partial symmetry $\langle \delta \bW_{ij} \delta\bW_{ji} \rangle_{\delta\bW} = \nu^2_{ij}/N$. The randomness of $\delta\bW$ allows us to control the scaling of various terms in the diagrammatic expansion of the network dynamics (\ref{sm} and Ref.\ \cite{wakhloo2026solutionchaoticneuralnetwork}). We emphasize that while $\delta\bW$ is drawn from a distribution, it is held fixed. Our results \eqref{eq:introresult} do \emph{not} involve an average over $\delta \bW_{ij}$. 

\textbf{Conditions on structured connections:} We assume that the elements $\bM_{ij}$ are at most $\O(N^{-1/2})$ and that $\bM$ is low rank: $\mathrm{rank}(\bM)\ll N$. We similarly assume that the time-varying filter $\bGamma_{ij}(\tau)$ is low rank in the frequency domain and has a finite norm: $\int \d\tau\, \bGamma_{ij}(\tau)^2 < \infty$. 

A key assumption within our framework is that the collective population modes are stationary and stable to external perturbations. More precisely, we define the collective population modes by projecting $\bpsi$ onto the right singular vectors of $\bM\bGamma$. We assume that the dynamics and linear response of the network along these modes are suppressed in powers of $N$---see Sec.\ \ref{sec:setup} for the precise statement. This assumption rules out low-rank nonlinear collective dynamics as in Refs.\ \cite{brunel2003determines, amit1997model, LandauSompolinsky2018}. Our framework includes the low-rank networks of Ref.\ \cite{MastrogiuseppeOstojic2018} in the regime in which the collective modes are at a stable fixed point as well as the balanced excitatory/inhibitory (EI) networks of Ref.\ \cite{KadmonSompolinsky2015} (Sec.\ \ref{sec:Mex}). Intuitively, our results apply for networks in an asynchronous stationary state. 

We emphasize that the above assumptions do not imply that the networks under study are incapable of generating low-rank dynamics. The same network can express nonlinear low-rank dynamics under one set of inputs, while under a different set of inputs, the activity along collective modes remains stationary \cite{MastrogiuseppeOstojic2018}. Our results apply to recordings obtained in the latter state but not the former.

\textbf{Conditions on inputs:} We assume that the external inputs can be written as the sum of a static deterministic component and a fluctuating random component $\bxi_i(t) = \bar\bxi_i + \delta\bxi_i(t)$ with $\delta\bxi_i(t)$ independent across cells. 

We may additionally include quenched correlated inputs by choosing some subset of the $\bpsi_i[\cdot]$ to be stochastic processes that are independent of their inputs $\bet_i, \bxi_i$. In this case, the chosen $\bpsi_i[\cdot]$ act as external inputs to the rest of the network whose correlations are given by the corresponding block of $\bJ \bJ^\dagger$ as described in App.~\ref{SM:corrinputs}.

\section{Spike covariance}

The goal of this letter is to derive a closed-form formula for the spiking covariance matrix $\bC$, also known as the cross-correlogram, with entries
\be
\bC_{ij}(\tau) \equiv \langle \delta\bpsi_i(t+\tau)\, \delta\bpsi_j(t) \rangle\, .
\ee
The notation $\langle \cdot \rangle$ denotes the average over time.
The symbol
\be
\delta\bpsi_i \equiv \bpsi_i - \langle \bpsi_i \rangle
\ee
denotes the centered spike train.
In what follows we work mostly in frequency space, where $\bC$ is the Hermitian matrix: 
\be\label{eq:C}
\bC(\omega) \equiv \int_{-\infty}^{\infty} \e^{i \omega \tau}\, \bC(\tau) \,\d \tau = \langle \delta\bm{\psi}(\omega)\, \delta\bm{\psi}^\dag(\omega) \rangle\, .
\ee
We calculate this matrix in the large-$N$ limit at a fixed instance of the disordered connectivity, $\delta\bW$.

\textbf{Effective connectivity matrix.}
Generalizing \cite{NIPS1993_f7e9050c,PhysRevE.50.3171}, we show that the synaptic connections only appear through the \emph{effective connectivity matrix} $\bV$, which is a product of the synaptic weight $\bW_{ij}$, the filter $\bGamma_{ij}$, and the effective post-synaptic gain, $\bSigma_{ii}$: 
\begin{equation}\label{eq:V}
\bV_{ij}(t-t') \equiv \int_{-\infty}^{t}  \bSigma_{ii}(t-\tau) \, \bJ_{ij} (\tau - t') \, \d\tau
\end{equation} 
with $\bSigma$ given in the frequency domain by 
\begin{equation}\label{eq:Sigma}
\bSigma_{ij}(\omega) = \delta_{ij} \frac{\bR_{ii}}{1 + \bK_i \bR_{ii}}, \ \ \bK_i(\omega) = \sum_{k} \delta \bJ_{ik} \bR_{kk} \delta\bJ_{ki}.
\end{equation}
In writing the above, we have defined $\delta\bJ_{ij}(\omega) = \delta \bW_{ij} \bGamma_{ij}(\omega)$, and $\bR_{ij}(\omega)$ is the response function $\bR_{ij} = \langle \delta \bpsi_i(\omega)/\delta\bxi_j(\omega)\rangle$. The kernel $\bK_i$ represents the first-order feedback a neuron receives through the partial symmetry in the recurrent connections. We can see that the effective gain $\bSigma_{ii}$ is the response function of the $i^\text{th}$ unit, normalized by the feedback kernel.

\section{Derivation}

We now give a derivation of our main results using a two-site cavity method.  The key difference compared to standard mean-field calculations is that we leverage the Gaussianity of the cavity field at a fixed instance of the disorder, $\delta\bW$. Here, we consider the case of a deterministic nonlinearity $\bpsi[\cdot]$ and input $\bxi$, with $\omega = O(1)$ where we work in dimensionless units of frequency, set by the inverse single-neuron timescale. See \ref{sm} for the case of stochastic $\bpsi,\bxi$ with $N,\omega \gg 1$. 

Let us consider the dynamics of the network in which the feedback from two cells $i$ and $j$  is set to zero: $\bJ_{ka} = 0$ for all $k$ and $a \in \{i,j\}$. Denoting the dynamics of this ``cavity network" with $^c$ and expanding $\bpsi_k[\bet]$ for $k\neq i,j$ to first order gives, after Fourier transforming 
\begin{gather}
    \bpsi_k(\omega) = \bpsic_k(\omega) + \sum_{l \neq i,j}\brc_{kl}(\omega)  \sum_{a\in\{i,j\}}\bJ_{la}(\omega) \bpsi_a(\omega),
    \label{eq:offsitecav}
\end{gather}
In effect, the above expansion starts from the cavity network, shifts the external input into $\bpsic_k$ by $\bxi_k(\omega) \mapsto \bxi_k (\omega) +\sum_{a \in \{i,j\}}  \bJ_{ka}(\omega) \bpsi_a(\omega),$ and expands in the shift. To lighten the notation, we suppress the $\omega$ arguments. Unless otherwise specified, every frequency-dependent term is evaluated at the same $\omega$---e.g. $\bJ_{ak} \bpsi_k = \bJ_{ak}(\omega) \bpsi_k(\omega)$.

We may now write the input into the cavity cells $\bpsi_a$ as
\begin{gather}
        \bet_a = \underbrace{\sum_{k \neq i,j} \bJ_{ak} \bpsic_k}_{\betc_a} + \bK_{aa}^c \bpsi_a + (\bK^c_{ab} + \bJ_{ab}) \bpsi_b, 
    \label{eq:cavity}  
\end{gather}
where the feedback kernel is $\bK^c_{ab} = \sum_{k,l \neq i,j} \bJ_{ak} \brc_{kl} \bJ_{lb}$, the index $b\neq a$ is the other cavity cell, and we  set $\bJ_{aa}=0$ (see \footnote{Since we assume $\delta \bW_{ii} = 0$, we may absorb the diagonal $\bM_{ii}$ into the definition of the spiking functional $\bpsi_i[\cdot]$ so that $\bJ_{ii} = 0$.}). We use $\brc, \bcc$ to denote the cavity network's response and correlation functions, which are by construction uncorrelated with $\bJ_{ak}, \bJ_{lb}.$ 

Using our assumption that the response of the network along the structured connections $\bM$ are weak, it follows that the kernel may be written in terms of the centered synapses $\delta\bJ_{ij} = \delta\bW_{ij}\bGamma_{ij}.$ Up to $\O(N^{-1/2})$ errors, one has $\bK^c_{ab} = \sum_{k,l\neq i,j} \delta \bJ_{ak} \brc_{kl} \delta \bJ_{lb}.$ Using the independence of the cavity dynamics from the entries $\delta\bJ_{ak}, \delta \bJ_{lb}$ and assuming that the off-diagonal response function is $\O(N^{-1/2})$ allows us to replace the cavity response with the full network response in the above: $\bK^c_{aa} =\bK_{a}$ up to the same $\O(N^{-1/2})$ errors and $\bK_{a}$ as in \eqref{eq:Sigma}.  The cross-talk between the cavity cells is thus $\O(N^{-1/2})$.

For $k\neq a,b$, the spike trains $\bpsic_k$ do not feel the influence of $\bpsi_a$, and thus, $\bpsic_k$ is uncorrelated with $\bJ_{ak}$. It follows that the field $\betc_a$ in \eqref{eq:cavity} is Gaussian at large $N$. Under a Gaussian measure, the Hermite polynomials are orthogonal and provide a natural basis for a perturbative treatment of $\bpsi_a$. Expanding to first order in the cavity interaction $(\bJ_{ab} + \bK_{ab})\bpsi_b$ in \eqref{eq:cavity} and then performing a Hermite expansion gives 
 \begin{align}\label{eq:decomp}
    \bpsi_a(\omega) = \sum_{n=0}^{\infty} \frac{1}{n!} \int \Big( & H_a^{(n)}(\omega; \omega'_1, \cdots, \omega'_{n})  \\
   &:\! \betc_a(\omega'_1) \cdots \betc_a(\omega'_n)\!: \textstyle\prod_{\alpha=1}^{n}\!\d \omega'_\alpha  \Big)\nn
    \\
    &\!\!\!\!\!\!\!\!\!\!\!\! + \bR_{aa} (\bJ_{ab} + \bK^c_{ab} )\bpsi_b + \mathcal{O}(N^{-1}),
    \nn
\end{align}
where we define the zeroth-order term in the series to be $\langle \bpsi_a\rangle_{\betc}$, and the fluctuations around $\bR_{aa}$ in the first-order cavity interaction expansion additionally shift the Hermite kernels with $n\geq 1$ by $\mathcal{O}(N^{-1/2})$, which will be absorbed into the error terms below (Eq.\ \eqref{eq:Lambda}). We use the notation $:\! \betc_a \cdots \betc_a \! :$ to denote the Hermite polynomials---see App.~\ref{app:Hermite}. We will only need the first Hermite coefficient, which is simply the response $H^{(1)}_a(\omega, \omega') = \bR_{aa}(\omega)\delta(\omega - \omega')$, and the first polynomial $:\betc_a := \delta \betc_a$. 

\textbf{Linear response.}
Let us collect all terms linear in $\delta\bpsi_a$ in Eq.\ \eqref{eq:decomp}.  Using \eqref{eq:offsitecav} to write the $\bpsic_k$ appearing in $\betc_{a}$ in terms of the full dynamics, $\bpsi_k$, gives
\begin{gather}
    [(\bI - \bV) \delta\bpsi]_a = \blambda_a, 
    \label{eq:linear}
\end{gather}
where we have defined $\bV = \bSigma \bJ$, with $\bSigma_{kl}$ as in \eqref{eq:Sigma}. The frequency-dependent vector $\blambda_a$ is simply the sum over the terms $n \geq 2$ in the expansion $\eqref{eq:decomp}$, divided by a factor of $1 + \bK_{a} \bR_{aa}$. The covariance is therefore $\bC = \bG \bLambda \bG^\dag$, as advertised in the introduction (Eq.\ \eqref{eq:introresult}), with  $\bLambda = \langle \bm \blambda \bm \blambda^\dagger\rangle_{\betc}$. 

\textbf{Orthogonality.}
We now close the derivation by solving for $\bLambda$ in terms of $\bV$ and the diagonal of $\bC$. The key is the orthogonality of Wick monomials (equivalently, \emph{Mehler formula}, see, e.g., \cite{Janson1997}), which states that for jointly Gaussian fields $\betc_a(\omega), \betc_b(\omega)$, one has 
\begin{align}
    \langle :\! \betc_a(\omega_1) \cdots \betc_a(\omega_n)\!: \; : \! \betc_b(\omega'_1)\cdots \betc_b(\omega'_m) \! :\rangle_{\bet} \\ = \delta_{nm} \sum_{p \in \mathcal P}\prod_{(\omega,\omega') \in p} \bC_{ab}^{\betc}(\omega)\delta(\omega+\omega')\, , \nonumber
\end{align}
where the sum is over all pairings $p$ of the $\omega_i$ variables with the $\omega'_j$, involving the total of $n!$ permutations. In particular, for $a\neq b$, the source matrix can be written as $\bLambda_{ab} \propto (\bC_{ab}^{\bet})^2$ at the leading order in large $N$. Assuming that cross-correlations are $\bC_{ab}^{\bet} = \mathcal{O}(N^{-1/2})$, it follows that the off-diagonal entries $\bLambda_{ab} = \mathcal{O}(N^{-1})$. The diagonal entries $\bC_{aa}^{\bet}$ and $\bLambda_{aa}$ can be $\mathcal{O}(1)$.

\textbf{Source matrix.}
We now calculate the diagonal of $\bLambda$. Writing the left-hand side of \eqref{eq:linear} in terms of the cavity dynamics $\delta\bpsic_k$, we find $\blambda_a = (\delta\bpsi_a - \bR_{aa}\sum_{k} \bJ_{ak} \delta \bpsic_k)/(1 + \bK_a \bR_{aa}) +\O(N^{-1/2})$. Squaring and averaging both sides, we can use the orthogonality relation $\langle \blambda (\delta\betc)^\dagger\rangle = \mathcal{O}(N^{-1/2})$ together with the independence of the cavity variables  from $\bJ_{ak}, \bJ_{ka}$ to obtain 
\begin{gather}
    \bLambda_{aa} = \frac{\bC_{aa} - |\bR_{aa}|^2 \sum_{k}|\delta  \bJ_{ak}|^2 \bC_{kk}}{|1 + \bK_a \bR_{aa}|^2} + \mathcal{O}(N^{-1/2}),
    \label{eq:Lambda}
\end{gather}
where we have used our assumptions on $\bM$ and the response along collective modes (\ref{sm}) to drop $\bM$-dependent terms in the sums, and we used the independence of $\bC_{kl}^c$ from $\bJ_{ak}, \bJ_{ka}$ together with the assumption that the off-diagonal entries scale as $N^{-1/2}$ to drop cross-terms. The latter scaling is verified self-consistently. Additionally, we have replaced the cavity response and correlations $\brc_{kk}, \bcc_{kk}$ with their full, non-cavity values. This is justified up to $\mathcal{O}(N^{-1/2})$ errors by \eqref{eq:offsitecav}. 

Plugging in the equation for $\bLambda_{ii}$ and following a similar cavity argument for the response we obtain
$\bC = \bG \bLambda \bG^\dag$
and
$\bR = \bG \bSigma$, matching Eq.~\eqref{eq:introresult} from the introduction. 
Note that these equations for $\bC_{ij}, \bR_{ij}$ are given self-consistently in terms of the diagonals $\bC_{ii}, \bR_{ii}$ via \eqref{eq:Sigma} and \eqref{eq:Lambda}. In App.~\ref{app:DMFT}, we derive DMFT equations for the diagonal response and correlation function, closing the loop. In App.~\ref{SM:corrinputs} we extend these results to the case of quenched, correlated external inputs.

\textbf{Higher moments and diagrammatic method.} Note that in deriving \eqref{eq:introresult} we have 
assumed that the errors in our formulae for $\bLambda$ and $\bK$ do not sum coherently in the final formula for $\bC, \bR$. In the SM we give a complete diagrammatic derivation of these results order by order in perturbation theory. The resulting calculation allows one to consider simultaneous large $N,\omega$ limits.

A notable consequence of the calculation is that any cumulant involving $n$ distinct cells scale as $\mathcal{O}(N^{(1-n)/2})$. It follows then that given distinct indices $i_1, \cdots, i_n$ with $n$ even one has the Wick relation 
\begin{gather}
    \langle \delta\bpsi_{i_1} \cdots \delta\bpsi_{i_n} \rangle = \sum_{q \in Q_{n}}\prod_{(i,j)\in q} \bC_{ij} + \O(N^{-(n+2)/4}),
\end{gather}
where $Q_n$ denotes all partitions of $\{i_1, \cdots, i_{n}\}$ into pairs. For $n$ odd the above moment is $\O(N^{-(n+1)/4})$. 

From Eq. \eqref{eq:introresult}, we can see that the cross-correlation and response between two neurons $i,j$ are both fully determined by: (i) the single-unit \emph{auto}-correlations and responses, $\mathrm{diag}(\bC)$ and $\mathrm{diag}(\bR)$, and (ii) the effective connectivity $\bV$. The precise details of the  model and its connectivity  do not enter except through  these quantities.

\section{Multisynaptic decomposition}

The correlation between two neurons depends on their direct monosynaptic coupling as well as di-, tri-, and higher order polysynaptic paths through the network, see, e.g., \cite{10.1371/journal.pcbi.1002059,PhysRevE.85.031916,10.1371/journal.pcbi.1002408,Ostojic10234}. We now show that each of these paths leaves a distinct signature on the cross-correlation and response functions: shorter paths create faster peaks.

The starting point of our analysis is the formal  power series (Born) expansion of the resolvent appearing in \eqref{eq:introG}: 
\begin{gather} 
    \bG = (\bI - \bV)^{-1} = \bI + \bV + \bV^2 + \bV^3 + \cdots
\end{gather} 
This expansion is organized by path length: the matrix element $ \bV_{ij}$ is the monosynaptic coupling $j\to i$, and $(\bV^2)_{ij} = \sum_k \bV_{ik} \bV_{kj}$ sums all disynaptic paths $j \to k \to i$ through which neuron $j$ influences neuron $i$. In general $(\bV^h)_{ij}$ sums over all $h$-hop paths through which $j$ affects $i$. Expanding our formula for the correlation function gives a decomposition in terms of path length: 
\be
    \bm C = \bm \Lambda + \underbrace{ \bV \bLambda + \bLambda \bV^\dagger}_{\mathrm{monosynaptic}} + \underbrace{\bV \bLambda 
    \bV^\dagger + \bV^2 \bLambda + \bLambda (\bV^\dagger)^2}_{\mathrm{disynaptic}} + \cdots
    \label{eq:polysyn}
\ee
The interpretation of this formula is similar: the matrix $\bLambda$ reflects the effective noise driving each cell \footnote{For linear networks, $\bLambda$ is precisely the power spectrum of the noise into each neuron, multiplied by the squared gain $|\bSigma_{ii}|^2$}. This noise hops to the neurons $i$ and $j$ in all possible ways.

How does this expansion help us interpret the cross-correlation function? Recall that each $\bV_{ij}(\omega)$ contains the spiking response function of the post-synaptic cell $i$. Given that synapses filter their inputs causally, we know that for sufficiently large $\omega$: 
\begin{gather} 
    | \bV_{ij}(\omega)| \sim \omega^{-p}
\end{gather}
for some exponent $p > 0$, e.g., LIF networks driven by white noise with linear synaptic filters satisfy $p=3/2$.

\begin{figure}
    \centering    \includegraphics[width=0.45\textwidth]{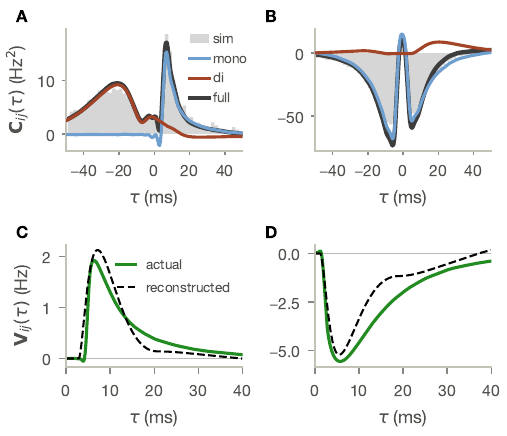}
    \caption{(A--B) Two example cross-correlations from an excitatory-inhibitory LIF network with $N=1000$. Simulated cross-correlations shown in gray, Eq.\ \eqref{eq:introresult} in black, and mono- and disynaptic contributions in blue and red, respectively (Eq.\ \eqref{eq:polysyn}). (A) A pair featuring an $E\to I$ connection. (B) A reciprocal $I\leftrightarrow I$ pair. Monosynaptic connections create fast peaks, disynaptic connections create slower ones, and the full correlation function is a sum over all such contributions. (C--D) Ground-truth and reconstructed $\bV_{ij}(\tau)$ for the pairs in (A--B). We record from a 200 neuron subset of the network for 5.6 hrs and reconstruct $\bV_{ij}(\tau)$ from the cross-correlations.}
    \label{fig:ccgs}
\end{figure}

It follows that monosynaptic connections dominate high-frequency correlations. Fourier transforming \eqref{eq:polysyn} term by term back into the time domain, the result is that monosynaptic connections create fast peaks, while polysynaptic paths create slower ones (Fig.\ \ref{fig:ccgs}).

\textbf{Inverse problem.}  The expansion \eqref{eq:polysyn} gives a systematic way of reconstructing effective connectivity $\bV$ from $\bC$. As an example, consider inversion up to the monosynaptic order.
The diagonal $\bC_{ii}$ already matches $\bLambda_{ii}$ at large frequency due to \eqref{eq:Lambda}.
As shown in the SM, the off-diagonal terms in \eqref{eq:polysyn} give in the joint limit $N,\omega \gg 1$
\be\label{eq:Cij-expanded}
\bC_{ij} \simeq \bV_{ij} \bC_{jj} + \bC_{ii}  \bV_{ji}^\ast\, .
\ee
to leading order in $N,\omega$ and assuming $p>1$. In the special case of tightly balanced E/I networks, one must further impose $\omega^p \gg N^{1/2}$ (\ref{sm}; see also \footnote{A typical cortical neuron receives between 1,000 and 10,000 presynaptic inputs \cite{braitenberg2013cortex}. Using the value of $p=3/2$ for LIF networks with white noise inputs, we therefore expect that taking $\omega \gg N^{1/3}$ with $\omega$ measured in units of the inverse membrane time constant places one in the regime in which \eqref{eq:Cij-expanded} is valid. For the least squares-based reconstruction analysis of Fig.\ \ref{fig:ccgs}, we find that one may use a substantially lower band than predicted by this calculation---see App.\ \ref{app:Numerics}.})

We now use \eqref{eq:Cij-expanded} to reconstruct $\bV$ from $\bC$. The key point is that $\bV_{ij}$ and $\bV_{ji}$ are causal filters without a zero-lag component, owing to the transmission delay between a presynaptic spike and the post-synaptic potential it evokes. We may invert the truncated Eq.\ \eqref{eq:Cij-expanded} using
\be\label{eq:V-ij}
\bV_{ij} = \frac{c_i}{c_j} P_+ \!\left[ \frac{\bC_{ij}}{c_i c_j^\ast} \right] \quad\text{with}\quad c_i \equiv \e^{P_+[\log \bC_{ii}]}\,,
\ee
where $P_+[f] = -\frac{1}{2\pi i }\int f(\omega') (\omega -\omega' + i \varepsilon)^{-1} \d \omega'$ is the causal projector, which amounts to convolving $f$ with the Fourier transform of the Heaviside step function $\theta(\tau)$ (\ref{SM:inverse}). Using analogous techniques, one can find the unique solution of $\bV$ in terms of $\bC$ at any order in the Born expansion (Sec.\ \ref{SM:inverse}). In practice, we estimate $\mathbf V_{ij},\mathbf V_{ji}$ by a least-squares fit of \eqref{eq:Cij-expanded} restricted to a high frequency band (App.~\ref{app:Numerics}).

\section{Discussion}

We have derived equations for the correlation and response functions of general recurrent network models, and we have shown how these formulae can be used to infer connectivity from correlations. The problem of connectivity inference has a long history in both theoretical \cite{PhysRevE.50.3171, NIPS1993_f7e9050c, Hawkes1971} and experimental neuroscience \cite{perkel1967neuronal}. 

Theoretically, exact solutions for correlation functions have been derived for stochastic models \cite{NIPS1993_f7e9050c, PhysRevE.50.3171, Hawkes1971} and more  recently, for chaotic rate networks \cite{ShenHu2025, wakhloo2026solutionchaoticneuralnetwork, clark2026linearequivalencenonlinearrecurrent}. For spiking neural networks, this problem has been studied using linearized and tree-level approximations \cite{10.1371/journal.pcbi.1002059, PhysRevE.85.031916, 10.1371/journal.pcbi.1005583, buice2010systematic, helias2014correlation}.
Interestingly, our formulae mirror those of linear perturbation theory, with the important difference that the gain is replaced by the effective gain $\bSigma$ and the external noise is replaced by effective noise $\bLambda$ generated by the recurrent dynamics.
Our work extends previous literature in the following directions:
(i) we perform a controlled expansion in powers of $1/\sqrt{N}$; (ii) we establish corrections to linear response theory induced by partially symmetric synaptic connectivity $\bK\neq 0$; (iii) we show that an order-by-order large-$\omega$ expansion of the cross-correlations can be leveraged to infer network connectivity without knowledge of the underlying model details (e.g. the filter $\bGamma$). While we have placed conditions on the connectivity, we speculate that our high-frequency reconstruction method works whenever the low-rank collective population modes are sufficiently slow. We are currently extending our results to the case of general nonlinear low-rank  dynamics.

On the experimental side, previous studies have used a variety of statistical methods to infer connectivity from cross-correlations  \cite{Huszar2022, Fink2025.01.16.633450, Kim2026Flows, Palmigiano2025, DasFiete2020, Brinkman2018, english2017pyramidal, medvedeva2026scalable, perkel1967neuronal, fujisawa2008behavior,csicsvari1998reliability, ostojic2009connectivity}. 
Our work complements these lines by introducing a principled, theoretically-grounded approach to connectivity reconstruction. In a forthcoming publication we will show how the multi-synaptic expansion, supplemented by modeling the activity of the unrecorded neurons, can be applied to infer connectivity from extracellular recordings.

\section{Acknowledgments}
We thank L.F.\ Abbott,  Ashok Litwin-Kumar, Roman Husz\'ar, and other members of the Aronov lab for useful discussions and comments on an earlier version of this manuscript. We additionally thank Soomin Kang for designing and illustrating Fig.\ 1.
This material is based upon work supported by the U.S. Department of Energy, Office of Science, under Award Number DE-SC0011941, the Howard Hughes Medical Institute, and the National Institutes of Health grant T32MH126036.

\nocite{braitenberg2013cortex}
\bibliography{references}

@article{SCS1988,
  author  = {Sompolinsky, H. and Crisanti, A. and Sommers, H. J.},
  title   = {{Chaos in random neural networks}},
  journal = {Phys. Rev. Lett.},
  volume  = {61},
  pages   = {259--262},
  year    = {1988},
  doi     = {10.1103/PhysRevLett.61.259}
}

@article{ShenHu2025,
    author        = {Shen, Xuanyu and Hu, Yu},
    title         = {Covariance spectrum in nonlinear recurrent neural networks},
    journal       = {arXiv preprint},
    year          = {2025},
    eprint        = {2508.05288},
    archivePrefix = {arXiv},
    primaryClass  = {q-bio.NC}
}

@article{MastrogiuseppeOstojic2018,
    author        = {Mastrogiuseppe, Francesca and Ostojic, Srdjan},
    title         = {Linking connectivity, dynamics, and computations in low-rank recurrent neural networks},
    journal       = {Neuron},
    volume        = {99},
    pages         = {609--623},
    year          = {2018},
    doi           = {10.1016/j.neuron.2018.07.003},
    eprint        = {1711.09672},
    archivePrefix = {arXiv},
    primaryClass  = {q-bio.NC}
}

@article{KadmonSompolinsky2015,
    author        = {Kadmon, Jonathan and Sompolinsky, Haim},
    title         = {Transition to chaos in random neuronal networks},
    journal       = {Phys. Rev. X},
    volume        = {5},
    pages         = {041030},
    year          = {2015},
    doi           = {10.1103/PhysRevX.5.041030},
    eprint        = {1508.06486},
    archivePrefix = {arXiv},
    primaryClass  = {q-bio.NC}
}

@article{CrisantiSompolinsky2018,
    author        = {Crisanti, Andrea and Sompolinsky, Haim},
    title         = {Path integral approach to random neural networks},
    journal       = {Phys. Rev. E},
    volume        = {98},
    pages         = {062120},
    year          = {2018},
    doi           = {10.1103/PhysRevE.98.062120},
    eprint        = {1809.06042},
    archivePrefix = {arXiv},
    primaryClass  = {cond-mat.dis-nn}
}

@article{LandauSompolinsky2018,
    author  = {Landau, Itamar Daniel and Sompolinsky, Haim},
    title   = {Coherent chaos in a recurrent neural network with structured connectivity},
    journal = {PLoS Comp. Biol.},
    volume  = {14},
    pages   = {e1006309},
    year    = {2018},
    doi     = {10.1371/journal.pcbi.1006309}
}

@article{10.1371/journal.pcbi.1005583,
    doi = {10.1371/journal.pcbi.1005583},
    author = {Ocker, Gabriel Koch AND Josić, Krešimir AND Shea-Brown, Eric AND Buice, Michael A.},
    journal = {PLOS Computational Biology},
    publisher = {Public Library of Science},
    title = {Linking structure and activity in nonlinear spiking networks},
    year = {2017},
    month = {06},
    volume = {13},
    url = {https://doi.org/10.1371/journal.pcbi.1005583},
    pages = {1-47},
    number = {6},

}

@book{Nussenzveig1972,
  author    = {Nussenzveig, H. M.},
  title     = {Causality and Dispersion Relations},
  series    = {Mathematics in Science and Engineering},
  volume    = {95},
  publisher = {Academic Press},
  address   = {New York},
  year      = {1972}
}

@inproceedings{NIPS1993_f7e9050c,
 author = {Ginzburg, Iris and Sompolinsky, Haim},
 booktitle = {Advances in Neural Information Processing Systems},
 editor = {J. Cowan and G. Tesauro and J. Alspector},
 pages = {},
 publisher = {Morgan-Kaufmann},
 title = {Correlation Functions in a Large Stochastic Neural Network},
 volume = {6},
 year = {1993}
}

@article{PhysRevE.50.3171,
  title = {Theory of correlations in stochastic neural networks},
  author = {Ginzburg, Iris and Sompolinsky, Haim},
  journal = {Phys. Rev. E},
  volume = {50},
  issue = {4},
  pages = {3171--3191},
  numpages = {0},
  year = {1994},
  month = {Oct},
  publisher = {American Physical Society},
  doi = {10.1103/PhysRevE.50.3171},
  url = {https://link.aps.org/doi/10.1103/PhysRevE.50.3171}
}

@article{10.1371/journal.pcbi.1002059,
    doi = {10.1371/journal.pcbi.1002059},
    author = {Pernice, Volker AND Staude, Benjamin AND Cardanobile, Stefano AND Rotter, Stefan},
    journal = {PLOS Computational Biology},
    publisher = {Public Library of Science},
    title = {How Structure Determines Correlations in Neuronal Networks},
    year = {2011},
    month = {05},
    volume = {7},
    url = {https://doi.org/10.1371/journal.pcbi.1002059},
    pages = {1-14},
    number = {5},

}

@article{PhysRevE.85.031916,
  title = {Recurrent interactions in spiking networks with arbitrary topology},
  author = {Pernice, Volker and Staude, Benjamin and Cardanobile, Stefano and Rotter, Stefan},
  journal = {Phys. Rev. E},
  volume = {85},
  issue = {3},
  pages = {031916},
  numpages = {7},
  year = {2012},
  month = {Mar},
  publisher = {American Physical Society},
  doi = {10.1103/PhysRevE.85.031916},
  url = {https://link.aps.org/doi/10.1103/PhysRevE.85.031916}
}

@article{10.1371/journal.pcbi.1002408,
    doi = {10.1371/journal.pcbi.1002408},
    author = {Trousdale, James AND Hu, Yu AND Shea-Brown, Eric AND Josić, Krešimir},
    journal = {PLOS Computational Biology},
    publisher = {Public Library of Science},
    title = {Impact of Network Structure and Cellular Response on Spike Time Correlations},
    year = {2012},
    month = {03},
    volume = {8},
    url = {https://doi.org/10.1371/journal.pcbi.1002408},
    pages = {1-15},
    number = {3},

}

@article{Ostojic10234,
	author = {Ostojic, Srdjan and Brunel, Nicolas and Hakim, Vincent},
	title = {How Connectivity, Background Activity, and Synaptic Properties Shape the Cross-Correlation between Spike Trains},
	volume = {29},
	number = {33},
	pages = {10234--10253},
	year = {2009},
	doi = {10.1523/JNEUROSCI.1275-09.2009},
	publisher = {Society for Neuroscience},
	issn = {0270-6474},
	journal = {Journal of Neuroscience}
}

@misc{wakhloo2026solutionchaoticneuralnetwork,
      title={Solution of a chaotic neural network at fixed connectivity}, 
      author={Albert J. Wakhloo},
      year={2026},
      eprint={2604.24141},
      archivePrefix={arXiv},
      primaryClass={cond-mat.dis-nn},
      url={https://arxiv.org/abs/2604.24141}, 
}

@article{Hawkes1971,
  author  = {Hawkes, Alan G.},
  title   = {{Spectra of some self-exciting and mutually exciting point processes}},
  journal = {Biometrika},
  volume  = {58},
  number  = {1},
  pages   = {83--90},
  year    = {1971},
  doi     = {10.1093/biomet/58.1.83}
}

@article{Hawkes1971b,
  author  = {Hawkes, Alan G.},
  title   = {{Point spectra of some mutually exciting point processes}},
  journal = {J. R. Stat. Soc. B},
  volume  = {33},
  number  = {3},
  pages   = {438--443},
  year    = {1971},
  doi     = {10.1111/j.2517-6161.1971.tb01530.x}
}

@article{WilsonCowan1972,
  author  = {Wilson, Hugh R. and Cowan, Jack D.},
  title   = {{Excitatory and inhibitory interactions in localized populations of model neurons}},
  journal = {Biophys. J.},
  volume  = {12},
  number  = {1},
  pages   = {1--24},
  year    = {1972},
  doi     = {10.1016/S0006-3495(72)86068-5}
}

@article{WilsonCowan1973,
  author  = {Wilson, H. R. and Cowan, J. D.},
  title   = {{A mathematical theory of the functional dynamics of cortical and thalamic nervous tissue}},
  journal = {Kybernetik},
  volume  = {13},
  number  = {2},
  pages   = {55--80},
  year    = {1973},
  doi     = {10.1007/BF00288786}
}

@misc{clark2026linearequivalencenonlinearrecurrent,
      title={Linear equivalence of nonlinear recurrent neural networks}, 
      author={David G. Clark},
      year={2026},
      eprint={2604.23489},
      archivePrefix={arXiv},
      primaryClass={cond-mat.dis-nn},
      url={https://arxiv.org/abs/2604.23489}, 
}

@Article{Huszar2022,
author={Husz{\'a}r, Roman
and Zhang, Yunchang
and Blockus, Heike
and Buzs{\'a}ki, Gy{\"o}rgy},
title={Preconfigured dynamics in the hippocampus are guided by embryonic birthdate and rate of neurogenesis},
journal={Nature Neuroscience},
year={2022},
month={Sep},
day={01},
volume={25},
number={9},
pages={1201-1212},
issn={1546-1726},
doi={10.1038/s41593-022-01138-x},
url={https://doi.org/10.1038/s41593-022-01138-x}
}

@article{Fink2025.01.16.633450,
	author = {Fink, Andrew J.P. and Muscinelli, Samuel P. and Wang, Shuqi and Hogan, Marcus I. and English, Daniel F. and Axel, Richard and Litwin-Kumar, Ashok and Schoonover, Carl E.},
	title = {Experience-dependent reorganization of inhibitory neuron synaptic connectivity},
	elocation-id = {2025.01.16.633450},
	year = {2025},
	doi = {10.1101/2025.01.16.633450},
	publisher = {Cold Spring Harbor Laboratory},
	journal = {bioRxiv}
}

@misc{Kim2026Flows,
    author        = "Kim, Timothy Doyeon and Pereira-Obilinovic, Ulises and Wang, Yiliu and Shea-Brown, Eric and S{\"u}mb{\"u}l, Uygar",
    title         = "{Identifying Connectivity Distributions from Neural Dynamics Using Flows}",
    year          = "2026",
    archivePrefix = "arXiv",
    eprint        = "2603.26506",
    primaryClass  = "q-bio.NC"
}

@article{Palmigiano2025,
    author  = "Chau, Ho Yin and Miller, Kenneth D. and Palmigiano, Agostina",
    title   = "{Exact linear theory of perturbation response in a space- and feature-dependent cortical circuit model}",
    journal = "Proc. Nat. Acad. Sci.",
    volume  = "122",
    number  = "31",
    pages   = "e2426758122",
    year    = "2025",
    doi     = "10.1073/pnas.2426758122"
}

@article{DasFiete2020,
    author  = "Das, Abhranil and Fiete, Ila R.",
    title   = "{Systematic errors in connectivity inferred from activity in strongly recurrent networks}",
    journal = "Nature Neurosci.",
    volume  = "23",
    number  = "10",
    pages   = "1286--1296",
    year    = "2020",
    doi     = "10.1038/s41593-020-0699-2"
}

@article{Brinkman2018,
    author        = "Brinkman, Braden A. W. and Rieke, Fred and Shea-Brown, Eric and Buice, Michael A.",
    title         = "{Predicting how and when hidden neurons skew measured synaptic interactions}",
    journal       = "PLoS Comput. Biol.",
    volume        = "14",
    number        = "10",
    pages         = "e1006490",
    year          = "2018",
    doi           = "10.1371/journal.pcbi.1006490",
    archivePrefix = "arXiv",
    eprint        = "1702.00865",
    primaryClass  = "q-bio.NC"
}

@book{Janson1997,
  author    = {Janson, Svante},
  title     = {Gaussian {H}ilbert Spaces},
  series    = {Cambridge Tracts in Mathematics},
  volume    = {129},
  publisher = {Cambridge University Press},
  address   = {Cambridge},
  year      = {1997},
  doi       = {10.1017/CBO9780511526169}
}

@article{english2017pyramidal,
  title={Pyramidal cell-interneuron circuit architecture and dynamics in hippocampal networks},
  author={English, Daniel Fine and McKenzie, Sam and Evans, Talfan and Kim, Kanghwan and Yoon, Euisik and Buzs{\'a}ki, Gy{\"o}rgy},
  journal={Neuron},
  volume={96},
  number={2},
  pages={505--520},
  year={2017},
  publisher={Elsevier}
}

@article{medvedeva2026scalable,
  title={Scalable inference of functional neural connectivity at submillisecond timescales},
  author={Medvedeva, Arina and Balzani, Edoardo and Williams, Alex and Keeley, Stephen},
  journal={Advances in Neural Information Processing Systems},
  volume={38},
  pages={26447--26465},
  year={2026}
}

@article{perkel1967neuronal,
  title={Neuronal spike trains and stochastic point processes: II. Simultaneous spike trains},
  author={Perkel, Donald H and Gerstein, George L and Moore, George P},
  journal={Biophysical journal},
  volume={7},
  number={4},
  pages={419--440},
  year={1967},
  publisher={Elsevier}
}

@article{fujisawa2008behavior,
  title={Behavior-dependent short-term assembly dynamics in the medial prefrontal cortex},
  author={Fujisawa, Shigeyoshi and Amarasingham, Asohan and Harrison, Matthew T and Buzs{\'a}ki, Gy{\"o}rgy},
  journal={Nature neuroscience},
  volume={11},
  number={7},
  pages={823--833},
  year={2008},
  publisher={Nature Publishing Group US New York}
}

@article{csicsvari1998reliability,
  title={Reliability and state dependence of pyramidal cell--interneuron synapses in the hippocampus: an ensemble approach in the behaving rat},
  author={Csicsvari, Jozsef and Hirase, Hajime and Czurko, Andras and Buzs{\'a}ki, Gy{\"o}rgy},
  journal={Neuron},
  volume={21},
  number={1},
  pages={179--189},
  year={1998},
  publisher={Elsevier}
}

@article{ostojic2009connectivity,
  title={How connectivity, background activity, and synaptic properties shape the cross-correlation between spike trains},
  author={Ostojic, Srdjan and Brunel, Nicolas and Hakim, Vincent},
  journal={The Journal of neuroscience},
  volume={29},
  number={33},
  pages={10234--10253},
  year={2009},
  publisher={Society for Neuroscience}
}

@article{buice2010systematic,
  title={Systematic fluctuation expansion for neural network activity equations},
  author={Buice, Michael A and Cowan, Jack D and Chow, Carson C},
  journal={Neural computation},
  volume={22},
  number={2},
  pages={377--426},
  year={2010},
  publisher={MIT Press One Rogers Street, Cambridge, MA 02142-1209, USA journals-info~…}
}

@article{helias2014correlation,
  title={The correlation structure of local neuronal networks intrinsically results from recurrent dynamics},
  author={Helias, Moritz and Tetzlaff, Tom and Diesmann, Markus},
  journal={PLoS computational biology},
  volume={10},
  number={1},
  pages={e1003428},
  year={2014},
  publisher={Public Library of Science San Francisco, USA}
}

@book{altland2010condensed,
  title={Condensed matter field theory},
  author={Altland, Alexander and Simons, Ben D},
  year={2010},
  publisher={Cambridge university press}
}

@book{helias2020statistical,
  title={Statistical field theory for neural networks},
  author={Helias, Moritz and Dahmen, David},
  volume={970},
  year={2020},
  publisher={Springer}
}

@article{hertz2017path,
  title={Path integral methods for the dynamics of stochastic and disordered systems},
  author={Hertz, John A and Roudi, Yasser and Sollich, Peter},
  journal={Journal of Physics A: Mathematical and Theoretical},
  volume={50},
  number={3},
  pages={033001},
  year={2017},
  publisher={IOP Publishing}
}

@article{Wilson1972,
  author  = "Wilson, G. Tunnicliffe",
  title   = "{The Factorization of Matricial Spectral Densities}",
  journal = "SIAM J. Appl. Math.",
  volume  = "23", number = "4", pages = "420--426", year = "1972",
  doi     = "10.1137/0123044"
}

@article{Wilson1978,
  author  = "Wilson, G. Tunnicliffe",
  title   = "{A convergence theorem for spectral factorization}",
  journal = "J. Multivariate Anal.",
  volume  = "8", number = "2", pages = "222--232", year = "1978",
  doi     = "10.1016/0047-259X(78)90073-8"
}

@article{Dhamala2008b,
  author  = "Dhamala, Mukeshwar and Rangarajan, Govindan and Ding, Mingzhou",
  title   = "{Analyzing information flow in brain networks with nonparametric Granger causality}",
  journal = "NeuroImage",
  volume  = "41", number = "2", pages = "354--362", year = "2008",
  doi     = "10.1016/j.neuroimage.2008.02.020"
}

@article{van1996chaos,
  title={Chaos in neuronal networks with balanced excitatory and inhibitory activity},
  author={Van Vreeswijk, Carl and Sompolinsky, Haim},
  journal={Science},
  volume={274},
  number={5293},
  pages={1724--1726},
  year={1996},
  publisher={American Association for the Advancement of Science}
}

@article{stark2009unbiased,
  title={Unbiased estimation of precise temporal correlations between spike trains},
  author={Stark, Eran and Abeles, Moshe},
  journal={Journal of neuroscience methods},
  volume={179},
  number={1},
  pages={90--100},
  year={2009},
  publisher={Elsevier}
}

@article{aertsen1985evaluation,
  author  = {Aertsen, Ad M.H.J. and Gerstein, George L.},
  title   = {Evaluation of neuronal connectivity: Sensitivity of cross-correlation},
  journal = {Brain Research},
  volume  = {340},
  number  = {2},
  pages   = {341--354},
  year    = {1985},
  doi     = {10.1016/0006-8993(85)90931-x}
}

@article{brillinger1976identification,
  author  = {Brillinger, David R. and Bryant, Hugh L. and Segundo, Jos{\'e} P.},
  title   = {Identification of synaptic interactions},
  journal = {Biological Cybernetics},
  volume  = {22},
  number  = {4},
  pages   = {213--228},
  year    = {1976},
  doi     = {10.1007/bf00365087}
}

@article{pernice2013reconstruction,
  author  = {Pernice, Volker and Rotter, Stefan},
  title   = {Reconstruction of sparse connectivity in neural networks from spike train covariances},
  journal = {Journal of Statistical Mechanics: Theory and Experiment},
  volume  = {2013},
  number  = {03},
  pages   = {P03008},
  year    = {2013},
  doi     = {10.1088/1742-5468/2013/03/p03008}
}

@article{kobayashi2019reconstructing,
  author  = {Kobayashi, Ryota and Kurita, Shuhei and Kurth, Anno and Kitano, Katsunori and Mizuseki, Kenji and Diesmann, Markus and Richmond, Barry J. and Shinomoto, Shigeru},
  title   = {Reconstructing neuronal circuitry from parallel spike trains},
  journal = {Nature Communications},
  volume  = {10},
  pages   = {4468},
  year    = {2019},
  doi     = {10.1038/s41467-019-12225-2}
}

@article{ren2020model,
  author  = {Ren, Naixin and Ito, Shinya and Hafizi, Hadi and Beggs, John M. and Stevenson, Ian H.},
  title   = {Model-based detection of putative synaptic connections from spike recordings with latency and type constraints},
  journal = {Journal of Neurophysiology},
  volume  = {124},
  number  = {6},
  pages   = {1588--1604},
  year    = {2020},
  doi     = {10.1152/jn.00066.2020}
}

@article{endo2021convolutional,
  author  = {Endo, Daisuke and Kobayashi, Ryota and Bartolo, Ramon and Averbeck, Bruno B. and Sugase-Miyamoto, Yasuko and Hayashi, Kazuko and Kawano, Kenji and Richmond, Barry J. and Shinomoto, Shigeru},
  title   = {A convolutional neural network for estimating synaptic connectivity from spike trains},
  journal = {Scientific Reports},
  volume  = {11},
  pages   = {12087},
  year    = {2021},
  doi     = {10.1038/s41598-021-91244-w}
}

@article{petersen2021cellexplorer,
  author  = {Petersen, Peter C. and Siegle, Joshua H. and Steinmetz, Nicholas A. and Mahallati, Sara and Buzs{\'a}ki, Gy{\"o}rgy},
  title   = {{CellExplorer}: A framework for visualizing and characterizing single neurons},
  journal = {Neuron},
  volume  = {109},
  number  = {22},
  pages   = {3594--3608.e2},
  year    = {2021},
  doi     = {10.1016/j.neuron.2021.09.002}
}

@article{platkiewicz2021monosynaptic,
  author  = {Platkiewicz, Jonathan and Saccomano, Zachary and McKenzie, Sam and English, Daniel and Amarasingham, Asohan},
  title   = {Monosynaptic inference via finely-timed spikes},
  journal = {Journal of Computational Neuroscience},
  volume  = {49},
  number  = {2},
  pages   = {131--157},
  year    = {2021},
  doi     = {10.1007/s10827-020-00770-5}
}

@article{jia2022multiregional,
  author  = {Jia, Xiaoxuan and Siegle, Joshua H. and Durand, S{\'e}verine and Heller, Greggory and Ramirez, Tamina K. and Koch, Christof and Olsen, Shawn R.},
  title   = {Multi-regional module-based signal transmission in mouse visual cortex},
  journal = {Neuron},
  volume  = {110},
  number  = {9},
  pages   = {1585--1598.e9},
  year    = {2022},
  doi     = {10.1016/j.neuron.2022.01.027}
}

@article{spivak2022deconvolution,
  author  = {Spivak, Lidor and Levi, Amir and Sloin, Hadas E. and Someck, Shirly and Stark, Eran},
  title   = {Deconvolution improves the detection and quantification of spike transmission gain from spike trains},
  journal = {Communications Biology},
  volume  = {5},
  pages   = {520},
  year    = {2022},
  doi     = {10.1038/s42003-022-03450-5}
}

@article{donner2024ensemble,
  author  = {Donner, Christian and Bartram, Julian and Hornauer, Philipp and Kim, Taehoon and Roqueiro, Damian and Hierlemann, Andreas and Obozinski, Guillaume and Schr{\"o}ter, Manuel},
  title   = {Ensemble learning and ground-truth validation of synaptic connectivity inferred from spike trains},
  journal = {PLOS Computational Biology},
  volume  = {20},
  number  = {4},
  pages   = {e1011964},
  year    = {2024},
  doi     = {10.1371/journal.pcbi.1011964}
}

@article{kobayashi2025inference,
  author  = {Kobayashi, Ryota and Shinomoto, Shigeru},
  title   = {Inference of monosynaptic connections from parallel spike trains: A review},
  journal = {Neuroscience Research},
  volume  = {215},
  pages   = {37--46},
  year    = {2025},
  doi     = {10.1016/j.neures.2024.07.006}
}

@article{pillow2005prediction,
  title={Prediction and decoding of retinal ganglion cell responses with a probabilistic spiking model},
  author={Pillow, Jonathan W and Paninski, Liam and Uzzell, Valerie J and Simoncelli, Eero P and Chichilnisky, EJ},
  journal={The Journal of neuroscience},
  volume={25},
  number={47},
  pages={11003--11013},
  year={2005},
  publisher={Society for Neuroscience}
}

@book{coleman1988aspects,
  title={Aspects of symmetry: selected Erice lectures},
  author={Coleman, Sidney},
  year={1988},
  publisher={Cambridge University Press}
}

@article{brunel2003determines,
  title={What determines the frequency of fast network oscillations with irregular neural discharges? I. Synaptic dynamics and excitation-inhibition balance},
  author={Brunel, Nicolas and Wang, Xiao-Jing},
  journal={Journal of neurophysiology},
  volume={90},
  number={1},
  pages={415--430},
  year={2003},
  publisher={American Physiological Society}
}

@article{amit1997model,
  title={Model of global spontaneous activity and local structured activity during delay periods in the cerebral cortex.},
  author={Amit, Daniel J and Brunel, Nicolas},
  journal={Cerebral cortex (New York, NY: 1991)},
  volume={7},
  number={3},
  pages={237--252},
  year={1997}
}

@book{braitenberg2013cortex,
  title={Cortex: statistics and geometry of neuronal connectivity},
  author={Braitenberg, Valentino and Sch{\"u}z, Almut},
  year={2013},
  publisher={Springer Science \& Business Media}
}

@article{renart2010asynchronous,
  title={The asynchronous state in cortical circuits},
  author={Renart, Alfonso and De La Rocha, Jaime and Bartho, Peter and Hollender, Liad and Parga, N{\'e}stor and Reyes, Alex and Harris, Kenneth D},
  journal={science},
  volume={327},
  number={5965},
  pages={587--590},
  year={2010},
  publisher={American Association for the Advancement of Science}
}

@article{ecker2010decorrelated,
  title={Decorrelated neuronal firing in cortical microcircuits},
  author={Ecker, Alexander S and Berens, Philipp and Keliris, Georgios A and Bethge, Matthias and Logothetis, Nikos K and Tolias, Andreas S},
  journal={science},
  volume={327},
  number={5965},
  pages={584--587},
  year={2010},
  publisher={American Association for the Advancement of Science}
}

@article{brunel2000dynamics,
  title={Dynamics of sparsely connected networks of excitatory and inhibitory spiking neurons},
  author={Brunel, Nicolas},
  journal={Journal of computational neuroscience},
  volume={8},
  number={3},
  pages={183--208},
  year={2000},
  publisher={Springer}
}
\bibliographystyle{apsrev4-2}

\clearpage

\appendix

\setcounter{secnumdepth}{2}

\section{\label{app:Hermite}Hermite polynomials}
The Hermite kernels used in \eqref{eq:decomp} are given by the variations
\begin{gather}
    H_i^{(n)}(\omega; \omega'_1,\cdots, \omega'_n) \equiv \bigg\langle \frac{\delta^{n}\bpsi_i (\omega)}{\delta \betc_i(\omega'_1) \cdots \delta \betc_i(\omega'_n)} \bigg\rangle_{\!\bet} . 
    \label{eq:Hker}
\end{gather}
In \eqref{eq:decomp}, the symbol $:\! \bet_i(\omega_1) \cdots \bet_i(\omega_k) \! :$ with $k$ instances of $\bet_i$ denotes the Hermite polynomial of order $k$: 
\begin{align}
    &:\! \bet_{i_1}(\omega_1) \cdots \bet_{i_k}(\omega_k)\!:\, \equiv 
    \mathcal D \, 
    \e^{\int \d\omega [\bh^\top \delta \bet -\frac 1 2 \bh^\top \bC^{\bet}\bh^{*}]  } \big|_{\bh=0}, 
\end{align}
where 
$\mathcal D={\delta^k}/[{\delta \bh_{i_1}(\omega_1) \cdots \delta \bh_{i_k}(\omega_k)}]$ is a differential operator. 
The first couple of Hermite polynomials are given by
\begin{gather} 
: \!\bet_i(\omega)\! : = \delta \bet_i(\omega) , \\
: \!\bet_i(\omega) \bet_j(\omega')\! : = \delta \bet_i(\omega) \delta \bet_j(\omega') - \bC_{ij}^{\bet}(\omega) \delta(\omega+\omega'),
\end{gather}
where $\delta \bet_i = \bet_i - \langle \bet_i\rangle$. Note that in Eq. \eqref{eq:Hker}, the brackets denote an average over Gaussian $\bet_i$ with a stationary covariance $\bC^{\bet}_{ij}(t-t') \equiv \la \delta\bet_i(t) \delta\bet_j(t')\ra$. In the frequency domain: $\bC^{\bet}_{ij}(\omega,\omega') = \bC^{\bet}_{ij}(\omega)\delta(\omega+\omega').$ Note that in the context of this calculation, the covariance $\bC^{\bet}$ is set self-consistently by the DMFT below.

\section{\label{app:DMFT}DMFT for single-site statistics.}

The equations for $\bLambda_{ii},\bSigma_{ii}$ (Eqs.\ (\ref{eq:Sigma},\ref{eq:Lambda})) are given in terms of the diagonals $\bC_{aa}, \bR_{aa}.$ Here we obtain dynamical mean-field equations for these functions, closing the loop. From Eq.\ \eqref{eq:cavity}, it follows that the mean satisfies
\begin{gather}
    \langle \betc_a \rangle_{\betc_{a}} = \sum_{k} \bJ_{ak} \bm m_k - \bK_a \bm m_a  + \O(N^{-1/2}) ,
\end{gather}
where we evaluate at $\omega=0$ since the mean is constant in time.
In the large-$N$ limit, Eq.\ \eqref{eq:offsitecav} implies that the means and correlation matrices for the non-cavity cells converge to their limiting values:  $\bcc_{kk} \to \bC_{kk}$, and $\brc_{kk}\to\bR_{kk}$. Using the fact that $\bcc_{kl}$ is uncorrelated with $\delta \bJ_{ak},\delta \bJ_{ka}$ together with the off-diagonal scaling of $N^{-1/2}$ gives, from the definition $\betc_a = \sum_k \bJ_{ak} \bpsic_k$ the equation for the variance
\begin{gather}
    \langle \delta \betc_a  \delta (\betc_a)^*\rangle_{\betc_{a}} = \sum_{k}| \delta \bJ_{ak}|^2 \bC_{kk} + \O(N^{-1/2}).
\end{gather}
In addition to this Gaussian field, the unit receives the self-feedback $\bK_a\bpsi_a$ of Eq.\ \eqref{eq:cavity}, and the external input $\bxi_a$. Enforcing $\langle \delta\bpsi_a \delta \bpsi_a^*\rangle = \bC_{aa},\, \langle \delta\bpsi_a /\delta \bxi_a\rangle = \bR_{aa},\, \langle \bpsi_a \rangle = \bm m_a $ closes the equations. Note that in writing the above, we have additionally used the fact that the contribution to $\bC_{aa}, \bR_{aa}$ from the $\bM$-dependent terms is subleading. This is shown to follow from our assumptions on $\bM_{ij}$ using a Hessian analysis in Sec.\ \ref{sec:hess}.

\section{\label{SM:corrinputs}Correlated inputs}
Suppose we take $\bpsi_i[\cdot]$ for $1\leq i \leq b$ to be stochastic processes that are independent of their inputs with autocorrelations $\chi_i(\omega)$. It follows that $\bR_{ii} = \bSigma_{ii}= 0$, and hence $\bV_{ij} = 0$ for $i\leq b$. Thus, these cells act as external inputs to the rest of the network that are correlated through the connections  $\bJ_{ik}$ with $i > b$ and $k\leq b$. More precisely, defining the input correlation matrix $\bC^{(\mathrm{ext})}_{ij}(\omega)=\langle \delta\bet^{(\mathrm{ext})}_i\delta \bet^{(\mathrm{ext})*}_j\rangle $ with $i,j>b$ and $\bet^{(\mathrm{ext})}_i = \sum_{k\leq b} \bJ_{ik}\bpsi_k$ the sum over quenched external inputs, we have $ \bC^{(\mathrm{ext})}_{ij} = \sum_{k\leq b} \bJ_{ik}\chi_k(\omega)\bJ_{jk}^*$.

Let us consider the effect on the correlations of the rest of the network. Since the rows of $\bV$ corresponding to the exogenous units vanish, $\bI - \bV$ is block triangular, and \eqref{eq:introresult} gives for the correlations of the recurrent population $\bC^{(>b)} = (\bI - \bV^{(>b)})^{-1}\tilde \bLambda(\bI -\bV^{(>b)})^{-\dagger},$ with $\bV$ restricted to its lower $(N{-}b)\times (N{-}b)$ block, and the source matrix shifted to $\tilde \bLambda_{ij} = \delta_{ij}\bLambda_{ii} +  \bSigma_{ii} \bSigma_{jj}^*\bC^{(\mathrm{ext})}_{ij}$. For spiking networks $\bC_{ii}(\omega)$ and $\bLambda_{ii}$ converge to a constant at large $\omega$, while $\bSigma_{ii} \bSigma_{jj}^*\bC^{(\mathrm{ext})}_{ij} = \O(\omega^{-2p})$. It follows that in the high frequency limit \eqref{eq:Cij-expanded}, the effect of quenched, correlated external inputs is washed out: $\tilde \bLambda_{ij} \to \delta_{ij} \bLambda_{ii}$.

\section{\label{app:Numerics}Simulation protocol}

We simulate an LIF network containing a single excitatory and inhibitory population ($N_E=800, N_I=200$) obeying the following dynamical equations:
\begin{gather} 
(1+\tau_m \partial_t) \bx_i =  \bet_i + \bxi_i  - \tau_m(V_{th} - V_r) \bpsi_i,
\end{gather}
where $\tau_m = 20$ ms, $\bxi_i$ is Gaussian white noise with a cell-type dependent mean ($13.58$ mV for E, $16.44$ mV for I), and covariance $\langle \bxi_i(t) \bxi_i(t')\rangle_\xi = 2\tau_m\sigma^2\delta(t-t')$ with $\sigma= 5.35$ mV, and $V_r=10$ mV is the reset potential.  The spiking functional is the same across all cells $\bpsi_i=\psi[\bx_i]$ and emits a delta spike each time the threshold $V_{th} = 20$ mV is crossed with a 2 ms refractory period while the synapses continue to integrate. See Fig.\ \ref{fig:recon}A-B for summary statistics and a raster of activity.

We record the activity of the network over 5.6 hrs for reconstruction and use a longer 44 hr recording to compute the cross-correlation histograms in Fig.\ \ref{fig:ccgs} (16 independently-seeded runs of length $10^4$ s). The equations of motion were integrated using the exact exponential propagator at a step size of $dt=0.1$ ms. For the reconstruction analyses below (Figs.\ \ref{fig:ccgs}C-D and Fig.\ \ref{fig:recon}C-D), we consider a subsample of the network to $150$ E cells and $50$ I cells and compute their pairwise statistics using the $5.6$ hr subset.

The cross-correlations $\bC_{ij}(\omega)$ were calculated in simulation by mean-centering the spike trains, splitting the recording into subsets of size $T_s=8$ s, and multiplying each segment by $P=7$ Slepian taper functions with $NW=4$. Each taper copy and segment yields a single estimate of the cross-correlation. Averaging over tapers and segments then gives $\bC_{ij}(\omega)$ on a uniform 1 Hz-spaced grid over $\omega/(2\pi) \in [1, 500]$ Hz.  

We now describe the synapses of the network. All E synapses are drawn from a Gaussian with mean $0.319$ mV and standard deviation $0.0957$ mV that we truncate at 0 to enforce Dale's law. The I synapses are drawn from the same distribution with a sign flip and with means and standard deviations multiplied by a factor of 4. All synapses are then randomly set to zero with a probability of 0.9, leading to a typical number of 100 presynaptic partners in the network. We set all autapses to zero. Delays are drawn uniformly from the interval $[0.75, 5]$ ms. Finally, we use exponential synapses $\bGamma_{ij}(\tau) = e^{-(\tau -d)/\tau_c} \theta(\tau-d)/h_c$ with $d$ the randomly chosen synapse-specific delay, $\tau_c$ with $c \in \{E,I\}$ set by the pre-synaptic cell type ($\tau_E = 2$ ms and $\tau_I = 5$ ms), and $h_c=(\tau_c/\tau_m)^{1/(1-\tau_c/\tau_m)}$ a normalization factor chosen so that $\bW_{ij}$ represents the post-synaptic potential of a single spike in mV.

We estimate the effective connectivity using our high frequency expansion technique. We fit $\hat \bV_{ij}(\omega)$ with raised cosine basis functions in log-time \cite{pillow2005prediction}: $\hat \bV_{ij}(\omega) = \sum_{k=1}^K w_k B_k(\omega)e^{i\omega d_{ij}}$ with $w_k$ real coefficients, $d_{ij}$ a fitted onset, and $B_k$ the Fourier transform of $K$ bumps $\tfrac12(1+\cos)$ in $\log(\tau+c)$, $c=2$ ms, centres equally spaced in that variable from the first bump's foot at $\tau=0$ to the last at $t_{max}$, so the kernel $ \sum_k w_k b_k(\tau - d_{ij})$ rises smoothly from zero. For each pair $(i,j)$ we search $d_{ij}, d_{ji}$ jointly on a $0.5$ ms grid from $0.5$ ms to $6$ ms, then on a $0.05$ ms grid within $\pm0.25$ ms of the minimum. At each delay pair, we minimize $\sum_\omega |\bC_{ij}(\omega) - \hat \bV_{ij}(\omega) \bC_{jj}(\omega) - \hat \bV_{ji}^*(\omega) \bC_{ii}(\omega)|^2/(\bC_{ii} \bC_{jj})$ for $\omega$ in a band $[\omega_l, \omega_h]$, and we keep the fit with minimum error.

The hyperparameters are the band, the number of basis functions $K$, and their temporal extent $t_{max}$. We set the band's lower limit to the point at which the LIF response begins to decay in powers of frequency, $\omega_l/(2\pi) = 1/(2\pi\tau_m) \approx 8$ Hz, and the upper limit to half the Nyquist frequency of the 0.5 ms grid of the $\bC_{ij}(\omega)$ estimator, $(\omega_h/2\pi) = 500$ Hz. We leave a more principled band-fitting procedure to future work. We chose $K \in \{3, 4, 6, 8, 12\}$ and $t_{max} \in \{30, 60, 100\}$ ms, shared by all pairs, by 4-fold cross-validation over contiguous blocks, with the held-out $\bC_{ij}$ error summed over all pairs, then refit on all blocks. This gives $K=3, t_{max} = 60$ ms. 

We estimate the ground-truth $\bV_{ij}(\omega)$ using the ground-truth synapses $\bJ_{ij}(\omega)$, together with the DMFT solution of $\bSigma_{ii}$. To do this, we first iterate the self-consistent DMFT equations for single-unit LIF neurons driven by colored Gaussian noise $\bet_i$. After each iteration, the mean and autocovariance of $\bet_i$ are updated using the DMFT equations of App.~\ref{app:DMFT}. We iterate until reaching a relative change across iterations of 1\% for the $\bm{m_i} $ and 2\% for $\bC_{ii}(\omega)$. 

Following convergence, we run the single-site DMFT dynamics for each unit once more for a time of $T_R=10^4$ s. Using $\bSigma_{ii} = \bR_{ii} + O(N^{-1/2})$ for networks without partial symmetry, together with the Furutsu--Novikov theorem gives $\bSigma_{ii}(\omega) \langle |\bet_i(\omega)+\bxi_i|^2 \rangle = \langle\delta \bpsi^{(\mathrm{dmft})}_i(\omega) (\bet_i+\bxi_i)^*(\omega)\rangle$ with $\bpsi_i^{(\mathrm{dmft})}$ a draw from the single-site DMFT problem. We use this identity to estimate $\bSigma_{ii}$ by regressing $\bpsi_i^{(\mathrm{dmft})}$  against the net input $\bet_i+\bxi_i$ in each frequency band. This is done by subsetting the final long DMFT run to segments of length $T_s= 8$ s and forming $P=7$ Slepian-tapered copies of each segment as in the estimation of $\bC_{ij}(\omega).$ Regressing across all segments and taper copies, the regression coefficients yield an estimate of $\bSigma_{ii}(\omega)$ at each frequency $\omega$.  For plotting purposes we additionally used Gaussian smoothing on the final estimate of $\bSigma_{ii}(\omega)$.

\section{\label{app:Recon}Reconstruction error}

\begin{figure}
    \centering
    \includegraphics[width=0.45\textwidth]{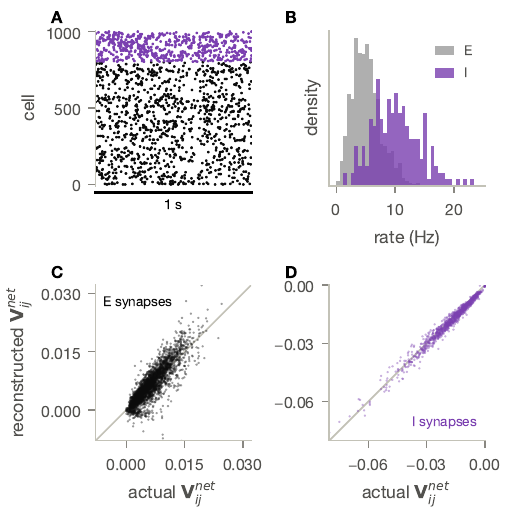}
    \caption{Simulation and reconstruction of an LIF network. (A) Raster plot showing 1 s of simulation time for E cells (black) and I cells (purple). (B) Histogram of baseline firing rates by cell type. (C-D) Actual vs. reconstructed $\bV_{ij}^{(\mathrm{net})}$ for E (C) and I (D) cells.}
    \label{fig:recon}
\end{figure}

To quantify the full connectivity reconstruction error, we define the ``net strength" as $\bV_{ij}^{(\mathrm{net})}= \int_0^{10\, \mathrm{ms}}d\tau\, \bV_{ij}(\tau)$ and compare the actual vs. reconstructed strength in Fig.\ \ref{fig:recon}C-D. In this way we find that we are able to reconstruct the connectivity with high accuracy: the true vs. reconstructed net strengths have a correlation of 0.86 for E synapses and 0.99 for I synapses.

\clearpage
\pagebreak

\makeatletter
\renewcommand{\thesection}{SM.\arabic{section}}
\def\@sectioncntformat#1{\csname the#1\endcsname\quad}
\def\@hangfrom@section#1#2#3{\@hangfrom{#1#2}\MakeTextUppercase{#3}}
\renewcommand{\thesubsection}{\thesection.\arabic{subsection}}
\def\p@subsection{}
\renewcommand{\thesubsubsection}{\thesubsection.\arabic{subsubsection}}
\def\p@subsubsection{}
\makeatother
\setcounter{secnumdepth}{3}

\counterwithout{equation}{section}
\setcounter{equation}{0}
\setcounter{section}{0}
\renewcommand{\theequation}{S\arabic{equation}}

\makeatletter\def\@currentlabel{SM}\makeatother\label{sm}

\appendixtoc

\section{\label{SM:proof}Diagrammatic proof}
\begingroup
\renewcommand{\bet}{\beta}\renewcommand{\bzeta}{\bm{\zeta}}\let\le\stdle \let\leq\stdle \let\ge\stdge \let\geq\stdge
\makeatletter\ltx@footnote@pop\makeatother
\edef\savedfootnotecount{\the\value{footnote}}\setcounter{footnote}{0}

\subsection{Setup}
\label{sec:setup}
We consider the system described by the dynamical equations given in the main text:
\begin{gather}
\mathcal L_i [\bx_i](t) = \sum_j \bW_{ij} \int d\tau\, \bGamma_{ij}(t-\tau) \bpsi_j(\tau)+ \bxi_i(t),  
\end{gather}
where the $\mathcal L_i$ are nonlinear operators, $\bpsi_i = \bpsi_i[\bx_i]$ causal functionals, and $\bxi_i(t) = \bar\bxi_i + \delta \bxi_i(t)$ with $\delta \bxi_i(t)$ independent stationary noise across cells. The MSRDJ functional of such a dynamical system is composed of single-site actions together with a bilinear interaction term: 
\begin{gather}
S[\bx, \hat \bx, \bpsi, \hat \bpsi] = \sum_i S^i_0[\bx_i, \hat \bx_i, \bpsi_i, \hat \bpsi_i] + \sum_{ij} i\hat \bx_i \bW_{ij} \bGamma_{ij} \bpsi_j,
\end{gather}
where we suppress time integrals, and $\hat \bx_i$, $\hat \bpsi_i$ are the response fields conjugate to $\bx_i$ and $\bpsi_i$ in the MSRDJ construction. We work with the It\^o convention, so that equal-time response functions vanish. Note that the nonlinearity and stochasticity of the spike-generation mechanism have been absorbed into the $S^i_0$.

We will assume that the matrix $\bW_{ij} =\bM_{ij} + \delta \bW_{ij}$ is composed of a low-rank deterministic component $\mathrm{rank}(\bM)=r$ with $r\ll N$, together with a zero-mean random matrix $\delta \bW_{ij}$. We will additionally assume that the Fourier transform of the matrix-valued function $\bGamma(\omega)$ is low-rank, $\mathrm{rank}(\bGamma(\omega)) \leq s \ll N$ for all $\omega$, and that each entry $\bGamma_{ij}(\omega)$ is bounded and decays at high frequency, $|\bGamma_{ij}(\omega)|\leq K(1+|\omega|)^{-g}$ for some $g\geq 0$ and a constant $K$ independent of $i$, $j$ and $N$.

To ease the notation, we define $\delta\bJ_{ij}(\tau) = \delta \bW_{ij} \bGamma_{ij}(\tau)$ and $\tilde \bM_{ij}(\tau) = \bM_{ij} \bGamma_{ij}(\tau).$ In this notation the synapse of the main text is $\bJ_{ij}(\tau) = \bW_{ij} \bGamma_{ij}(\tau) = \delta\bJ_{ij}(\tau) + \tilde \bM_{ij}(\tau)$. Note that the rank of $\tilde \bM$ remains small relative to $N$ via the inequality $\mathrm{rank}(\tilde \bM) \leq \mathrm{rank}(\bGamma)\mathrm{rank}(\bM) \ll N$. In the remainder of this section we give a detailed description of our modeling assumptions and state our main results.

\subsubsection{Conditions on  connections and network state}

The disordered component of the connectivity $\delta\bW$ is drawn from an elliptic ensemble 
\begin{gather}
   \langle \delta \bW_{ij}^2\rangle = N^{-1} \sigma_{ij}^2, \quad \langle \delta \bW_{ij} \delta \bW_{ji}\rangle = N^{-1}\nu_{ij}^2, \nonumber
   \\ \langle |\sqrt{N} \delta \bW_{ij}|^k\rangle \sim 1,
 \label{eq:momentassmptn}
\end{gather}
for any integer $k\geq 2$.
We additionally assume no autapses, $\delta\bW_{ii}=0$, and we note that synapse-specific random delays can be absorbed as an additional random term in $\delta\bW_{ij}\bGamma_{ij}$ and as a rank-one rescaling to $\bM\bGamma$. Note that we do not require that the $\bW_{ij}$ are Gaussian.

We now state the assumption on $ \bM$. Writing $\tilde \bM_{ij}(\omega) = \sum_{\rho=1}^{r'} \bu^\rho_i(\omega) \by^\rho_j(\omega)$ with $r' = \mathrm{rank}(\tilde\bM) = O(1)$ we define the collective modes, $\bkappa^\rho(\omega) = \sum_i \by^\rho_i(\omega)  \bpsi_i(\omega)$. We will assume that the dynamics along these modes are, on average, stationary. Thus, averaging over the disorder $\delta\bW$ and any stochasticity in the spiking functional, one has $\langle\bkappa_\rho(\omega)\rangle\propto \delta(\omega)$ in the Fourier domain.

Our second assumption on these modes involves the \emph{collective response function} $\rkap_{\rho\sigma}$, which we define as the response of the mode $\bkappa_\rho$ to an input delivered along $\bu^\sigma$: 
\begin{gather}
    \rkap_{\rho\sigma}(t-t') =  \bigg\langle\frac{\delta \bkappa_\rho(t) }{\delta \bh_\sigma (t')}\bigg\rangle  ,
    \label{eq:rkap}
\end{gather}
That is, $\rkap_{\rho\sigma}$ contains the response of $\bkappa_\rho$ when we deliver a perturbation $\sum_\rho \bu^\rho_i \bh_\rho$ to each neuron in the population.

We assume that the feedback in response to such a perturbation is small:
\begin{gather}
    \tilde u_{\max}\tilde y_{\max} ||\bI + \rkap (\omega)||_{op} = O(N^{-1}\bet). 
    \label{eq:balance}
\end{gather}
where $\tilde u_{\max} = \sup_\omega (1+|\omega|)^{g} \max_{i,\rho} |\bu^\rho_i(\omega)|$ and $\tilde y_{\max} = \sup_\omega \max_{i,\rho} |\by^\rho_i(\omega)|$ are the maximum entries of $\bu^\rho_i, \by^\rho_i$ once the synaptic filter $(1+|\omega|)^{-g}$ has been factored out of $\bu$. We will additionally assume the bound $\beta(\omega,N) \leq \min\{\sqrt{N},(1+|\omega|)^p\}$, so that the collective response cannot be arbitrarily fast, and we assume $\tilde \bM_{ij}$ is at most $O(N^{-1/2})$. As shown in Sec.\ \ref{sec:Mex}, balanced E/I networks saturate this $\beta$ bound.

Intuitively, the above conditions on the network state assert that the dynamics along the collective modes are stationary and resist external perturbations. Indeed, in Sec.\ \ref{sec:hess}, we show that the above conditions amount to the statement that one can replace the recurrent contributions arising from $\bM$ with their time- and disorder-averaged values $\langle \bkappa^\rho \rangle$ up to small corrections that can be treated by a Hessian expansion.  In  Sec.\ \ref{sec:Mex}, we show that popular choices for $\bM_{ij}$ satisfy Eq. \eqref{eq:balance}.

\subsubsection{Conditions on spike-generation nonlinearity}

To consider simultaneous large $N$, large $\omega$ limits, we assume that the  response functions of this theory decay for sufficiently large $\omega$. Recall that in the MSRDJ formalism, correlations involving $i\hat \bx$ correspond to functional derivatives---e.g. $\langle \bpsi_i(t) i\hat \bx_j(t')\rangle = \langle \delta \bpsi_i(t) /\delta \bxi_j(t')\rangle.$ We will assume that a cumulant with $n$-functional derivatives (i.e. $n$ $i\hat \bx$ variables) decays at large frequencies as 
\begin{gather}
    \langle i\hat \bx_i(\omega_1) \cdots i\hat \bx_i(\omega_n) \bpsi_i(\omega_{n+1}) \cdots \bpsi_i(\omega_{n+m}) \rangle_{0,c} \nonumber
    \\ = O\bigg( \prod_{s=1}^n (1+|\omega_s|)^{-p_0}\bigg),
    \label{eq:omassmptn1}
    \end{gather}
where the brackets denote the cumulant with respect to $S_0^i$ above with the mean collective input $\langle \bkappa^\rho \rangle$ retained (i.e. the cumulant $\langle\cdot \rangle_{0,c}$ as defined below). The constant $p_0>0$ is a decay parameter, and the bound in Eq.\ \eqref{eq:omassmptn1} holds at all frequencies independent of the cell index. We make no assumption on the decay of correlation functions in their $\bpsi$ frequencies. We further assume that $\omega$-integrals over such correlation-response functions are convergent. 

Finally, we define the combined decay exponent $p = p_0+g$, which corresponds to the filtered response of a neuron. We will assume that this filtered response decays at sufficiently high frequencies: $p>1$.

\subsubsection{Additional assumptions}

Throughout the calculation we will additionally assume that the correlation and responses of single units do not grow with $N$: $\bC_{ii}(\omega), \bR_{ii}(\omega) = O(1)$, and that the various DMFT quantities are similarly well-conditioned. For a given system, these assumptions must be verified self-consistently.

\subsubsection{Main results}

We compute the leading order behavior of the response and correlation functions in Fourier space. We show that 
\begin{widetext}
\begin{gather}
    \bR_{ij}(\omega) = [(\bI - \bV )^{-1} \bSigma ]_{ij} + O\big([N^{-1/2}\delta_{ij} + N^{-1}\omega^{-p}]\omega^{-p_0}\big),\nonumber \\ \bC_{ij}(\omega) = [(\bI - \bV)^{-1} \bLambda (\bI-\bV)^{-\dagger}]_{ij}  + O\big(N^{-1/2}\delta_{ij} + N^{-1}\omega^{-p}\big),
    \label{eq:twopoint}
\end{gather}
where  $\bV=\bSigma\bJ = \bSigma[\delta\bJ+\tilde \bM]$ and the error estimates are understood to be entrywise, with high probability over the realization of $\delta\bW$. The frequency-dependent diagonal matrices $\bSigma$ and $\bLambda$  are given by
\begin{gather}
    \bSigma_{ij} = \delta_{ij} \frac{ \bR_{ii}(\omega)}{1 + \bR_{ii}\sum_{k} \delta\bJ_{ik}  \bR_{kk} \delta\bJ_{ki}}, \quad \bLambda_{ij} = \delta_{ij}\frac{  \bC_{ii}(\omega) - | \bR_{ii}(\omega)|^2 \sum_{k} |\delta\bJ_{ik}|^2  \bC_{kk}(\omega) }{|1 + \bR_{ii} \sum_k \delta\bJ_{ik}  \bR_{kk} \delta\bJ_{ki}|^2} ,
    \label{eq:fullse}
\end{gather}
where $\bSigma_{ii} = O(\omega^{-p_0})$ and $\bLambda_{ii} = O(1)$ (Prop.\ \ref{prop:sefreq}). Taking the double limit $N, \omega \gg 1$ and expanding Eq.\ \eqref{eq:twopoint} gives the monosynaptic form shown in the main text (Sec.\ \ref{sec:hess}, Eq.\ \eqref{eq:highfreqhess}):
\begin{gather}
    \bR = \bSigma + \bV\bSigma + O([N^{-1/2} \bI + N^{-1}\omega^{-p}\bm 1 ]\omega^{-p_0} + \omega^{-2p-p_0}),\nonumber
    \\
    \bC = \bLambda + \bV\bLambda + \bLambda \bV^\dagger + O(\omega^{-2p} + N^{-1/2} \bI + N^{-1}\omega^{-p}\bm 1), 
    \label{eq:highfreq}
\end{gather}
\end{widetext}
where we additionally impose $\omega^p \gg \sqrt{N}$ in the tightly balanced case $\bu^\rho_i \by^\sigma_j = O(N^{-1/2})$ (see Sec.\ \ref{sec:Mex}). Note that the above formulae depend self-consistently on the diagonal entries of $\bC$ and $\bR$. In Sec.\ \ref{sec:dmft}, we fix these diagonal entries by constructing a dynamical mean-field theory (DMFT) of the system.

The calculation is organized as follows: first we reorganize the expansion around a site-decoupled background $S_0$ in which the $\bkappa^\rho$ are pinned to their mean values (Sec.\ \ref{sec:pert}). In Sec.\ \ref{sec:dyson} we then develop the diagrammatic formalism used in the remainder of the proof, and we prove a set of Dyson equations for the two-point functions. In Sec.\ \ref{sec:largeN}, we systematically drop subleading diagrams in $N$, and we estimate their large $\omega$ behavior. In Sec.\ \ref{sec:se}, we use the preceding large $N$ estimates to obtain closed formulae for the two-point functions for the theory in which $\bkappa$ are pinned to their mean values. This establishes the above results for $\bM=0$, and in Sec.\ \ref{sec:highercumulants} we extend to higher cumulants. Finally, in Sec.\ \ref{sec:hess}, we incorporate fluctuations in the collective modes $\bkappa^\rho$ around their mean values through a Hessian analysis, completing the proof. Sec.\ \ref{sec:dmft} then closes the equations for $\bC_{ii}, \bR_{ii}$ by constructing a DMFT of the system, and Sec.\ \ref{sec:Mex} shows that the low-rank networks of Ref.\ \cite{MastrogiuseppeOstojic2018} in the linearly stable regime and the balanced networks of Ref.\ \cite{KadmonSompolinsky2015} fall within our modeling framework.

\subsection{Perturbative expansion} 
\label{sec:pert}

We begin by constructing a site-decoupled background around which we will expand in $i\hat \bx_i \delta\bJ_{ij} \bpsi_j$. Let us factorize the low-rank deterministic component as $\tilde \bM_{ij} = \sum_{\rho=1}^{r'} \bu^\rho_i  \by^\rho_j$ and introduce the collective fields $\bkappa_\rho = \sum_j \by^\rho_j \bpsi_j$ (Eq.\ \eqref{eq:rkap} and above) using delta functions. Given an observable $\obs$, we write its average as
\begin{gather}
    \langle \obs \rangle = \int D(\bkappa, \hat \bkappa) \ex{-i\bkappa \hat \bkappa} \int D(\bx,\hat \bx, \bpsi,\hat \bpsi) \ex{S_0 + S_{int}} \obs,
    \label{eq:fixedkapact}
    \\
    S_0 =  \sum_i S^i_0[\bx_i, \hat \bx_i, \bpsi_i, \hat \bpsi_i] + \sum_{\rho i} i\hat \bx_i\bu^\rho_i \bkappa^\rho + \sum_{\rho i}i\hat \bkappa^\rho \by^\rho_i \bpsi_i,\nonumber
    \\
    S_{int} = \sum_{ij} i\hat \bx_i \delta\bJ_{ij} \bpsi_j,\nonumber
\end{gather}
Assuming that the matrix $\bM_{ij}$ follows the above assumptions, we will show that the fields $\bkappa, \hat \bkappa$ concentrate around their $\delta\bJ$-averaged saddle-point values. To make this precise, let us define $\bmm_j = \langle \bpsi_j \rangle$ with the average calculated with respect to the action $S_0 + S_{int}$ at a fixed value of $\bkappa, \hat \bkappa.$ Thus, $\bmm_j$ is a functional of $\bkappa, \hat \bkappa$.

In what follows, we will fix $\bkappa, \hat \bkappa$ to their $\delta\bJ$-averaged saddle point values:
\begin{gather}
    \bkappa^\rho = \sum_j \by^\rho_j \langle \bmm_j\rangle_{\delta\bJ} , \quad \hat \bkappa^\rho = 0.
    \label{eq:kapsaddle}
\end{gather}
Note that by averaging over $\delta\bm J$, the variables $\bkappa_\rho$ are \emph{not a function of the connections} $\delta\bJ_{ij}$. They are purely deterministic. As such $S_0$ has no $\delta\bJ_{ij}$ dependence. In Sec. \ref{sec:hess}, we systematically expand around this saddle point to incorporate residual fluctuations in $\bkappa,\hat\bkappa$. In what follows, we will use $\csp, \rsp$ to denote the full two-point functions evaluated at the $\bkappa,\hat\bkappa$ saddle point.

The result of replacing the $\bkappa$ with their saddle-point values is that the average over an arbitrary observable $O[\bx,\hat \bx, \bpsi, \hat \bpsi]$ may be written in terms of the decoupled background $S_0$: 
\begin{align}
    \langle \obs \rangle &=  \int D(\bkappa,\hat\bkappa)e^{-i\bkappa\hat\bkappa} \nonumber
    \\ &\times \langle \obs  \sum_k \frac 1 {k!} \bigg(\sum_{ij} i\hat \bx_i \delta\bJ_{ij} \bpsi_j \bigg)^k \rangle_0,
    \label{eq:obs}
\end{align}
where $\langle \cdot \rangle_0$ denotes the average with respect to $S_0$ at fixed $\bkappa,\hat\bkappa$.

\subsection{Diagrammatic notation and Dyson equations}
\label{sec:dyson}
We now introduce the diagrammatic notation. Each directed edge in a diagram corresponds to a factor of $\delta\bJ_{ij}$ and the vertices to cumulants. A vertex with $a$ incoming half-edges and $b$ outgoing half-edges corresponds to $\langle i \hat \bx_i(t_1) \cdots i \hat \bx_i ( t_a) \bpsi_i(t_{a+1}) \cdots \bpsi_i(t_{a+b})\rangle_{c,0}.$

\begin{center}
    \includegraphics{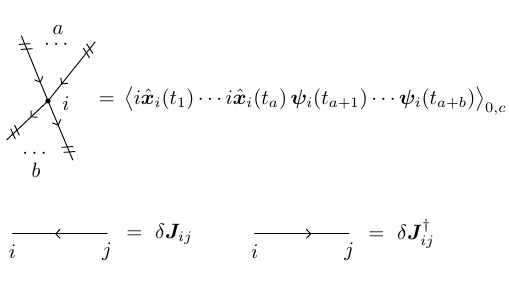}
\end{center}

\noindent where we use directed, hatched legs to denote fixed $i\hat \bx_i, \bpsi_i$ variables whose time variables are not integrated out and whose site indices are not summed.

A diagram's value then follows by setting connected indices equal and multiplying by the appropriate combinatorial factor. To make this precise, consider a diagram $D$ with $E$-many edges and a set of vertices $V$. Labeling the half-edges with indices $i_1, \cdots, i_E, j_1, \cdots, j_E$ the value of $D$ is 
\begin{gather}
    D = \sum_{(\mathbf{i} , \mathbf{j}) \in \lambda} \frac 1 {s_\lambda} \delta\bJ_{i_1 j_1} \cdots \delta\bJ_{i_E j_E} \nonumber
    \\ \times \prod_{v\in V} \langle i\hat \bx_{i_v}, \cdots, i\hat \bx_{i_v}, \bpsi_{i_v}, \cdots, \bpsi_{i_v}\rangle_{0,c}
    \label{eq:diagdefn}
\end{gather}
where $i_v$ is the index corresponding to vertex $v$, $\lambda$ is the set of $\mathbf i, \mathbf j \in \{1,\dots,N\}^{E}$ such that indices corresponding to half-edges incident to the same vertex are set equal, $s_\lambda$ is the number of automorphisms of the diagram that fix the rooted legs, and the frequency variables associated with non-rooted vertices are integrated out. 

\begin{figure*}
    \centering
    \includegraphics{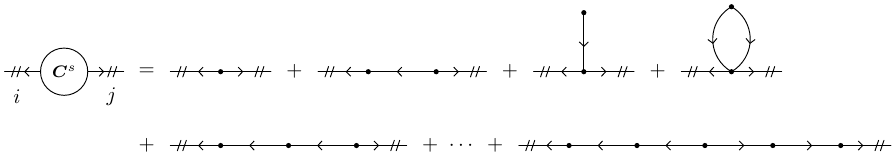}
    \caption{First few terms in the diagrammatic expansion of the connected correlation function $\csp_{ij}$; the corresponding terms are written out in Eq.\ \eqref{eq:example}.}
    \label{fig:Cexpansion}
\end{figure*}

The first few terms in the expansion of the connected correlation function $\csp_{ij}$ may therefore be represented by the diagrams of Fig.\ \ref{fig:Cexpansion}, which correspond to the terms 
\begin{widetext}
\begin{gather}
    \csp_{ij} =  \langle \bpsi_i \bpsi_j\rangle_{0,c} + \langle \bpsi_i i\hat \bx_i\rangle_{0,c} \delta\bJ_{ij} \langle \bpsi_j \bpsi_j\rangle_{0,c} + \langle \bpsi_i \bpsi_j i\hat \bx_i\rangle_{0,c} \sum_k \delta\bJ_{ik} \langle  \bpsi_k \rangle_0 \label{eq:example}
    \\ + \frac 1 2 \langle \bpsi_i i\hat \bx_i i\hat \bx_i \bpsi_j \rangle_{0,c} \sum_k \delta\bJ_{ik}^2\langle \bpsi_k \bpsi_k\rangle_{0,c}
    + \langle \bpsi_i i\hat \bx_i \rangle_0 \sum_k  \delta\bJ_{ik} \langle \bpsi_k i\hat \bx_k \rangle_{0} \delta\bJ_{kj} \langle \bpsi_j \bpsi_j \rangle_{0,c} + \cdots
    \nonumber
   \\ + \sum_{klm} \langle \bpsi_i i\hat \bx_i\rangle_{0,c} \delta\bJ_{ik} \langle \bpsi_k i\hat \bx_k\rangle_{0,c} \delta\bJ_{kl} \langle \bpsi_l \bpsi_l \rangle_{0,c} \delta\bJ_{ml} \langle i\hat \bx_m \bpsi_m \rangle_{0,c}  \delta\bJ_{jm} \langle i\hat \bx_j \bpsi_j\rangle_{0,c}
    . \nonumber
\end{gather}
\end{widetext}
Note the factor of $1/2$ coming from the symmetry between the legs. We will refer to the hatched legs corresponding to the fixed $\bpsi_i, i\hat \bx_i$ as ``rooted legs," and the vertices that they are incident to ``rooted vertices." 

Note that in the last sum of Eq. \eqref{eq:example}, the terms with $k=m, i=j$ scale as $\sum_{k} \delta\bJ_{ik}^2  \sum_{l} \delta\bJ_{kl}^2=O(1)$ contribution, while the terms with $k\neq m$ and $l\neq i$ scale as $\sum_{k\neq m\neq l,\, l\neq i} \delta\bJ_{ik}\delta\bJ_{im}\delta\bJ_{kl} \delta\bJ_{ml}  = O(N^{-1/2})$, as all variables are independent with elements scaling as $O(N^{-1/2})$ (see Sec.\ \ref{sec:largeN} for a rigorous argument).

For the large $N$ arguments to come, it is therefore useful to separate out the various terms within a single diagram based on the pattern of equal indices.  This is done using a dashed line, so that the $k=m$ term in the expansion of $\langle \bpsi_i \bpsi_j\rangle_c$ in the example from the preceding paragraph is drawn as: 
\begin{center}
    \includegraphics{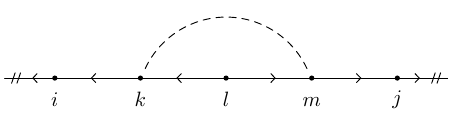}
\end{center}
We call a diagram featuring dashed lines a \emph{refinement} of the original diagram $D.$  Formally, a refined diagram corresponds to a subset of the sum in Eq. \eqref{eq:diagdefn} in which indices connected by a dashed line are forced equal and are otherwise unequal.

We now reorganize the expansion using a Dyson equation. Let us define $\bSigma^{s}_{ij}(t-t')$ as the sum over diagrams with a fixed $\bpsi_i$ on the left and $i\hat \bx_j$ on the right that cannot  be split into two pieces that are each connected to a fixed ``root" leg by cutting a single edge (1PI). We similarly define $\bLambda^{s}_{ij}(t-t')$ as the sum over 1PI diagrams with fixed legs $\bpsi_i, \bpsi_j$, and $\tilde \bLambda^{s}_{ij}(t-t')$ as the sum with $i\hat \bx, i\hat \bx$ fixed. As with $\csp, \rsp$, the superscript $s$ indicates that these self-energies are evaluated at the $\bkappa,\hat\bkappa$ saddle point. We reserve $\bSigma, \bLambda$ for the full self-energies of Eq.\ \eqref{eq:fullse}. The first few terms are shown in Fig.\ \ref{fig:seexpansion}.
Note that our definition includes diagrams that can be split into two pieces by a single edge cut, so long as one piece contains both rooted legs. 
\begin{figure*}
    \centering
    \includegraphics{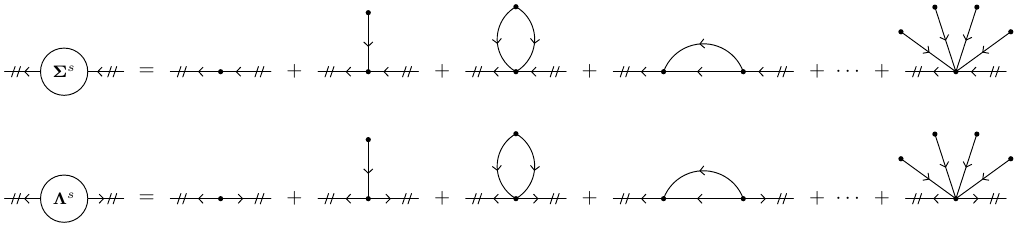}
    \caption{First few terms in the expansions of the self-energies $\bSigma^{s}_{ij}$ (top) and $\bLambda^{s}_{ij}$ (bottom).}
    \label{fig:seexpansion}
\end{figure*}

We now show that knowledge of $\bSigma^{s}, \bLambda^{s}$ is enough to reconstruct the two-point functions of interest (Fig.\ \ref{fig:dyson}).
\begin{figure*}
    \centering
    \includegraphics{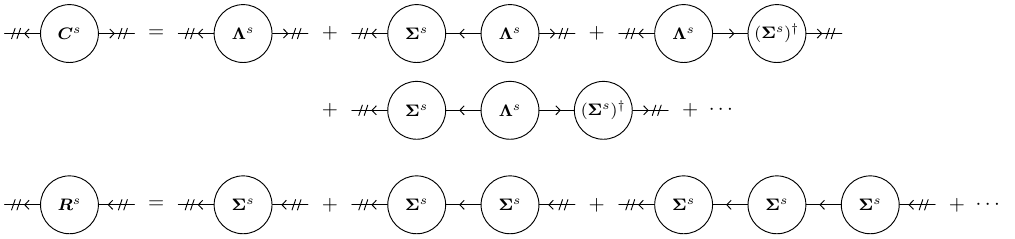}
    \caption{Graphical expansion of the two-point functions $\csp$ and $\rsp$ in terms of the self-energies $\bLambda^{s}$ and $\bSigma^{s}$; the resummed form is Eq.\ \eqref{eq:dysonresum}.}
    \label{fig:dyson}
\end{figure*}

\begin{prop}
Any diagram whose fixed legs are all $i\hat \bx$ vanishes. In particular, $\tilde \bLambda^{s}_{ij} = 0.$
\label{prop:causality}
\end{prop}
\begin{proof}
    Recall that in the MSRDJ formalism, an $i\hat \bx_i(t)$ is equivalent to a functional derivative $\delta/\delta\bxi_i(t)$. Hence if the largest time variable at a vertex comes from an $i\hat \bx$ leg, the resulting cumulant is zero by causality.
    
    Consider a diagram with $k$ $i\hat \bx$ rooted legs evaluated at $t_1, \cdots, t_k.$ In order for the vertex (or vertices) attaching to the fixed $i\hat \bx$ variables to be non-zero, there must be a $\bpsi$ leg evaluated at a later time $\bpsi(\tau)$ with $\tau > t_i$ attaching. This implies that somewhere else in the diagram, there exists a vertex with an $i\hat \bx(\tau)$ leg inserting. For this vertex to be non-zero, there must be a $\bpsi$ leg attaching evaluated at $\tau ' > \tau.$ Proceeding by induction, we find that there is always at least one vertex featuring an $i\hat \bx$ variable evaluated at $\bar \tau$, where $\bar \tau$ is the latest time. Causality then forces this vertex to be zero.
\end{proof}

\begin{prop}
    At the $\bkappa$ saddle point, the correlation and response obey the exact equations:
    \begin{gather}
        \rsp(\omega) = \big(\bI -\bSigma^{s} \delta\bJ\big )^{-1} \bSigma^{s},
        \\
        \csp(\omega) = \big (\bI -\bSigma^{s} \delta\bJ\big )^{-1}\bLambda^{s} \big(\bI -\bSigma^{s} \delta\bJ\big)^{-\dagger}.
        \label{eq:dysonresum}
    \end{gather}
\end{prop}
\begin{proof}
    We begin with the response equation. Starting from the fixed $i\hat \bx$ leg and moving left, cut the diagram at the first edge that splits it into two pieces, each connected to a rooted leg. If there is none, the diagram is a member of $\bSigma^{s}$. If there is, the outgoing leg must be an $i\hat \bx$ (otherwise an $i\hat \bx, i\hat \bx$ 1PI block forms). It follows that any diagram in the expansion of $\rsp$ can be broken into chains of $\bSigma^{s}$ diagrams with $\delta\bJ$ edges connecting them. Since there are no symmetries between 1PI blocks it follows that we can factorize the factor $s_\lambda$ in Eq.\ \eqref{eq:diagdefn} to obtain
    \begin{gather}
        \rsp(\omega) = \sum_k [\bSigma^{s}(\omega) \delta\bJ(\omega) ]^k \bSigma^{s}(\omega) .
    \end{gather}
    Resumming gives the response equation. 

    We now consider the $\bpsi$ correlator. As above, start from the right end of the diagram and walk left, cutting at the first edge whose removal separates the two rooted legs. If no such edge exists, the diagram is a member of $\bLambda^{s}$. If it does, the resulting 1PI after cutting may be either a member of $\bLambda^{s}$ if an $i\hat \bx$ leg exits or of $\bSigma^{s}$ if a $\bpsi$ leg exits. In the former case, the only additional 1PI diagrams that can be attached must be members of $\bSigma^{s}$ with $\delta\bJ_{ij}$ edges in between. In the latter case, one may continue attaching $(\bSigma^{s})^\dagger$ diagrams until a $\bLambda^{s}$ diagram is inevitably reached. The result is
    \begin{gather}
        \csp(\omega) = \sum_{kl} [\bSigma^{s}(\omega) \delta\bJ(\omega)]^k \bLambda^{s}(\omega) \nonumber
        \\ \times [(\bSigma^{s}(\omega)\delta\bJ(\omega) )^\dagger]^l
    \end{gather}
    Resumming gives the second equation. 
\end{proof}
As a reminder, the above equations for $\csp,\rsp$ are written at the saddle point of $\bkappa,\hat\bkappa$. This is why the $\bM$ terms are absent.

\subsection{Large-$N$ analysis} 
\label{sec:largeN}
In this section we prove simple estimates for the self-energy matrices. In this way, we will pass from asymptotic formulae for the entries of the self-energies $\bLambda^{s}, \bSigma^{s}$ to formulae for the entries of the two-point functions $\bC, \bR$. In principle, this step requires a stronger form of control---e.g., control over operator norms rather than entries. Given that the self-energies are defined as the sum over 1PI diagrams, this step amounts to assuming that one may freely add and subtract subleading diagrams to the perturbative series, Eq.\ \eqref{eq:obs}. In other words, we work order by order in perturbation theory as is common in diagrammatic calculations. Note that this step neglects the possibility of non-perturbative phenomena such as instantons.

We estimate the magnitude of a diagram via its root mean square $D \sim \sqrt{\langle |D|^2\rangle_{\delta\bJ}}$ \cite{wakhloo2026solutionchaoticneuralnetwork}. In this way we consider the large $N$ scaling of both the mean and the fluctuations of $D$. This is done by associating a set of multigraphs, $M_D$, to the diagram and using their properties to bound the scaling of $D$.

Throughout this section we will treat $\delta\bJ_{ij}(\omega)$ as scalars. This is justified by the bound $|\bGamma_{ij}(\omega)|\leq K(1+|\omega|)^{-g}$ of Sec.\ \ref{sec:setup}. In the frequency counting of this and the following sections we fold the $\bGamma_{ij}(\omega)$ factor into the $i\hat \bx$ leg. That is, non-rooted $i\hat \bx$ legs bring a factor of $(1+|\omega|)^{-p}$, and rooted $i\hat \bx$ legs a factor of $(1+|\omega|)^{-p_0}$.

To begin, consider the sum: 
\begin{gather}
    \langle |D|^2\rangle_{\delta\bJ} = \sum_{(\mathbf i, \mathbf j), (\mathbf i', \mathbf j') \in \lambda}
    \frac{1}{s_\lambda^2} \nonumber
    \\ \times \langle \delta\bJ_{i_1 j_1}  \cdots \delta\bJ_{i_E j_E} \delta\bJ^*_{i'_1 j'_1} \cdots \delta\bJ^*_{i'_E j'_E}  \rangle \prod_{v \in V} k_ v \prod_{v' \in V} k_{v'}^*,
\end{gather}
where we use the shorthand $k_v$ to stand for the corresponding cumulants. Note that $|D|^2$ itself is simply a disconnected diagram featuring two copies of $D$ with one conjugated. Schematically 
\begin{center}
    \includegraphics{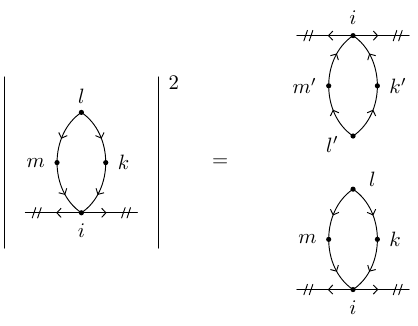}
\end{center}
We now partition the various terms in this sum based on the pattern of equal indices among the $i,j,i',j'$. That is, each partition runs over choices of $i,j,i',j'$ such that a given pair of indices are either always equal or always unequal. It follows that the $N$-scaling of $\langle |D|^2\rangle_{\delta\bJ}$ is simply the maximum scaling among all partitions. (See below for an example.)

To each one of these partitions, draw a multigraph $M_{|D|^2}$ in which the vertices are the unique $i,j$ indices after performing the index identifications implied by the partition, and an undirected edge from $i\to j$ corresponds to a factor of $\delta\bJ_{ij}$. The scaling of the partition is now controlled by the topology of $M_{|D|^2}$.

In the example above, the following multigraphs correspond to the terms in the sum where $k=m, k' =m',$ with all other indices distinct, and $k=k', m=m', l=l'$, with the rest distinct:
\begin{center}
    \includegraphics{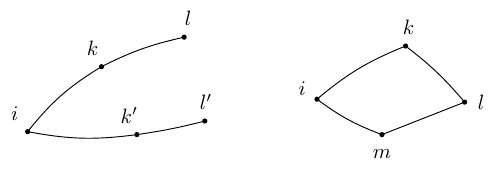}
\end{center}
Algebraically the two graphs correspond to the $\delta\bJ$ contractions
\begin{gather}
    \sum_{k,k',l,l'} \delta\bJ_{ik}^2 \delta\bJ_{kl}^2 (\delta\bJ_{ik'}^*)^2 (\delta\bJ_{k'l'}^*)^2 , \nonumber
    \\
    \sum_{k, m,l} |\delta\bJ_{ik}|^2 |\delta\bJ_{im}|^2 |\delta\bJ_{kl}|^2 |\delta\bJ_{ml}|^2 ,
\end{gather}
where the indices are constrained to be all distinct, and we have not written the cumulants. 

The next proposition will be our main tool for estimating the large-$N$ behavior of the various self-energy diagrams.

\begin{prop}
   Let $\langle |D|^2\rangle_{\delta\bJ}$ be the $\delta\bJ$-averaged mean square of a diagram with $r\geq 1$ distinct rooted indices and consider a partition of its index contractions. Associate a multigraph $M_{|D|^2}$ to this partition in which the vertices are the unique indices and the undirected edges are the $\delta\bJ_{ij}$ factors. Let $|E|$ be the number of unique edges in the corresponding multigraph, let $m_E$ be the multiplicity of edge $E$, and let $c$ be the cycle rank of the graph formed by its unique edges. Then:
    \begin{enumerate}
        \item Each partition scales as $O(N^{-\sum_E m_E /2 + |E| + 1 -r -c})$ if $m_E\geq 2$ for all $E$, otherwise the partition vanishes in the $\delta\bJ$-average.
        \item The leading order diagrams for any number of fixed indices $r$ are precisely those capable of generating multigraphs $M$ that are trees with $m_E=2$ for all $E$. They scale as $D =O(N^{(1-r)/2}).$ The next order diagrams are at most $O(N^{-r/2}).$
    \end{enumerate}
    \label{prop:count}
\end{prop}
\begin{proof}
   If $m_E=1$ for any edge, then a $\langle \delta\bJ_{ij} \rangle =0$ factor forms, since we assume that indices corresponding to distinct vertices are distinct within the sum corresponding to each partition. So we consider the case $m_E\geq 2$ for all $E.$
   
   Each $\delta\bJ_{ij}$ factor brings a power of $N^{-1/2}$ to the mean square (Eq. \eqref{eq:momentassmptn}), accounting for the $-\sum_E m_E/2$ factor in the exponent.  Conversely, the sum over every non-rooted index brings a factor of $N$. Thus the partition scales as $O(N^{-\sum_E m_E/2 + L -r})$ where $L$ is the number of vertices in the multigraph. Since $D$ is connected and the two copies of $D$ in the mean square share the rooted indices ($r\geq 1$), it follows that so too is $M_{|D|^2}$, and thus the relation $c = |E| - L +1$ establishes (1).

   Since the scaling of a diagram is governed by its root mean square, $D= O(\langle |D|^2\rangle^{1/2})$, it follows that the leading order diagrams for any choice of $r$ are simply those with $m_E = 2$ for all $E$ and $c=0$--i.e. trees such that each edge has multiplicity 2. Such diagrams scale as $D=O(N^{(1-r)/2}).$ 

   Let us then consider the next order terms---i.e. the graphs that do not form trees with $m_E=2$. Any graph containing a cycle with $c>0$ and $m_E\geq 2$ is at most $O(N^{-r/2})$. If $m_E > 3$ for any edge, the $O(N^{-r/2})$ bound is immediate, so let us consider those graphs with at least one edge $E$ with $m_E=3$. Since each edge in either copy of $D$ corresponds to an edge in $M_{|D|^2}$, we can see that $\sum_E m_E$ is even. Thus if $m_E=3$ there must be at least one more edge, $E'$, of odd multiplicity. Since any multigraph with $m_{E'}=1$  vanishes, it follows that $m_{E'} \geq 3$, establishing that the entire multigraph is at most $O(N^{-r/2})$. 
\end{proof}
It follows that the leading order diagrams are $O(1)$ on the diagonal and $O(N^{-1/2})$  off it.

Let us now relate the properties of $M_{|D|^2}$ back to the scalings of the various self-energy diagrams $D$. To do this let us break up the diagram $D$ into its refinements as defined in the previous section. We then associate a multigraph $G_D$ in exactly the same way, with a vertex for each unique index in $D$ (i.e. for each set of vertices connected by dashed lines), and a single edge for every unique $\delta\bJ$ factor, keeping separate track of the multiplicities of each edge.

Any multigraph $M_{|D|^2}$ appearing in the mean-square can be generated by drawing two copies of $G_D$ and identifying vertices that are set equal to each other in the corresponding partition. Using a refined version of the example diagram above, we have the following graph $G$ and leading order partition $M$ as: 
\begin{center}
    \includegraphics{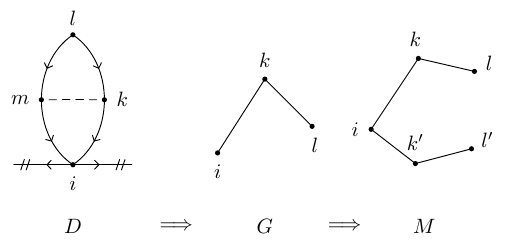}
\end{center}
We now use this to show that all off-diagonal self-energy diagrams are subleading.

\begin{prop}
    All off-diagonal self-energy diagrams are at most $O(N^{-1})$. 
    \label{prop:offdiagse}
\end{prop}
\begin{proof}
    We argue by contradiction. Let $D$ be an off-diagonal self energy diagram appearing in either $\bSigma^{s}_{ij}$ or $\bLambda^{s}_{ij}$ with $i\neq j$ . Consider the corresponding multigraph $G_D$. From the connectedness of the diagram it follows that $G_D$ contains a path from $i$ to $j.$
    
    Assume $D$ is leading order, $D=O(N^{-1/2})$. It follows that there exists a corresponding mean square tree $M_{|D|^2}$ constructed by identifying the vertices of two copies of $G_D$ that are joined at the root. From the tree condition, it follows that there is a single path in $M_{|D|^2}$ from $i$ to $j$ with each edge in the path having multiplicity exactly 2. Since there is only one such path, each edge must belong to one copy of $G_D$. This implies that $G_D$ contains at least one edge with multiplicity 1 connecting $i$ to $j$. This is a contradiction, as we have assumed that $D$ is a self-energy diagram and thus has no such edges.  
    
\end{proof}

We now estimate the scaling of the self-energy diagrams as $\omega \to \infty$. To begin, recall that we call a leg rooted if it carries one of the fixed $\bpsi_i(\omega), i\hat \bx_i (\omega)$ variables, and we call the vertices incident to such legs rooted vertices. The  neuronal index of a rooted leg is not summed and its frequency is not integrated. 

Owing to the stationarity of the background action $S_0$, the sum over all $\omega$ variables at an arbitrary vertex must be zero (see e.g. Ref.\ \cite{altland2010condensed} Ch. 5 and Ref.\ \cite{helias2020statistical}). As such, if a vertex carries only a single rooted leg, then there must be at least one other non-rooted leg carrying a fast $\omega$. Schematically:
\begin{center}
    \includegraphics{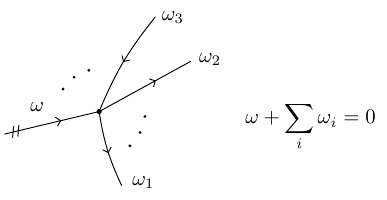}
\end{center}
Walking along the non-rooted ``large $\omega$ leg" to the other vertex it is incident to, it follows that there will be at least one other vertex receiving the large $\omega$ associated to this leg. As we now show, this mechanism shows that the off-diagonal self-energy diagrams are suppressed in powers of $\omega$ as well as powers of $N$. 

\begin{prop}
    The off-diagonal self-energy diagrams scale as
    \begin{align}
        \bSigma^{s}_{ij} = O(N^{-1} \omega^{-p-p_0}),
        \quad \bLambda^{s}_{ij} = O(N^{-1} \omega^{-p}),
    \end{align}
    where $i\neq j$. On the diagonal 
    \begin{align}
        \bSigma^{s}_{ii} = O(\omega^{-p_0}), \quad \bLambda^{s}_{ii} = O(1).
    \end{align}
    \label{prop:sefreq}
\end{prop}
\begin{proof}
    Let us consider the off-diagonal response self-energy $\bSigma^{s}_{ij}$ first. Off the diagonal, there must be two distinct rooted vertices associated to the rooted $\bpsi_i(\omega)$ and $i\hat \bx_j(-\omega)$ legs.  The rooted leg $i\hat \bx_j(-\omega)$ brings $|\omega|^{-p_0}$.
    
    Given that all $\omega$ variables incident to a single vertex sum to zero, each of these vertices must have at least one additional leg associated to a large $\omega$ variable. For the $j$ vertex, if the other ``large $\omega$ leg" is an $i\hat \bx$ leg, it brings an additional $|\omega|^{-p}$, so the factor of $|\omega|^{-p-p_0}$ follows immediately. If on the other hand the large $\omega$ leg is a $\bpsi$ leg, let us walk along the direction of this non-rooted large $\omega$ leg. We arrive at a non-rooted $i\hat \bx$ leg carrying the same large value of $\omega$, which brings $|\omega|^{-p}$ (note that the frequency scaling survives any internal integration since $\int da (1+|a|)^{-p}(1+|\omega-a|)^{-p} \leq C_p (1+|\omega|)^{-p}$ for $p>1$). The total scaling is therefore at least $\omega^{-p-p_0}$. In other words, the rooted legs create a large-$\omega$ $i\hat x$ leg either on the rooted vertex or elsewhere in the diagram. Schematically: 
    \begin{center}
    \includegraphics{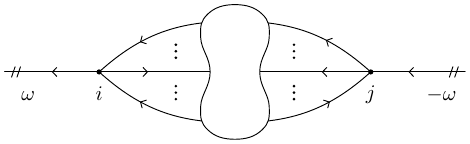}
\end{center}
    with the blob standing for arbitrary structures between the two rooted vertices.
    
    Let us now consider the off-diagonal $\bLambda^{s}_{ij}$ self-energy. In this case, one has two rooted $\bpsi$-legs, $\bpsi_i(\omega), \bpsi_j(-\omega)$. Again, each of these must have one associated leg attached to the same vertex with a large $\omega$ variable. Following the same line of reasoning as above, if both large $\omega$ legs are $i\hat \bx$ legs, one has a factor of $|\omega|^{-2p}$. Otherwise, if one of these legs is a $\bpsi$ leg, then walking along the corresponding edge to the $i\hat \bx$ leg, we find that there is at least a factor of $|\omega|^{-p}$ associated to the diagram. The stated scaling follows. Combining the frequency counting with Prop.\ \ref{prop:offdiagse} gives the stated result.

    Finally, for $\bSigma^{s}_{ii}$ the rooted $i\hat \bx_i(-\omega)$ leg gives $\omega^{-p_0}$, and the bound $\bLambda^{s}_{ii}=O(1)$ follows from the boundedness of the all-$\bpsi$ cumulants.
\end{proof}

In what follows, we will see that the cavity construction requires that we consider diagrams in which at least one of the non-rooted vertices is pinned to a specific value, corresponding to the ``cavity" neuron. The following proposition shows that all such diagrams are subleading.

\begin{prop}
    Consider a diagram $D$ with fixed indices $k,l$ and at least one non-rooted vertex. (The case $k=l$ corresponds to a single fixed index.) Restrict its sum (Eq.\ \eqref{eq:diagdefn}) to be over index vectors $\bm i, \bm j$ such that at least one of the indices is set equal to a ``cavity index," $c\neq k,l$ that is not summed. The resulting ``cavity diagram" $D_c$ is suppressed by a factor of $N^{-1/2}$ relative to the original one: $D_c = O(D N^{-1/2})$ for leading $D$ and $D_c = O(N^{-(r+1)/2})$ otherwise, where $r$ is the number of distinct fixed indices of $D$.
    \label{prop:restrictse}
\end{prop}
\begin{proof}
    Split $D$ into a sum over maximally refined diagrams. Restricting the diagram $D$ to include at least one index set equal to $c$ is equivalent to considering the sum over all maximally refined diagrams in which at least one set of vertices connected by dashed lines is set equal to $c$. The multigraphs $M_{|D|^2}$ corresponding to each of these maximally refined diagrams will all have an additional rooted index. Prop.\ \ref{prop:count} then gives the stated bounds for $D_c$.
\end{proof}

\subsection{Self-energy formulae}
\label{sec:se}
In this section we show that the saddle point self-energies can be written entirely in terms of the diagonal two-point functions $\rsp_{ii}, \csp_{ii}$. (Recall we use the superscript $^s$ to denote the correlation functions at the $\bkappa$ saddle, Eq.\ \eqref{eq:kapsaddle}.) We then use these equations with a cavity argument to obtain solutions of $\rsp, \csp.$

To begin, let us define the resolvent matrix $\bA = (\bI -\bSigma^{s} \delta\bJ)^{-1}$. We will further define $\delta\bJ^{(i)}, \bSigma^{(i)}, \bLambda^{(i)}$ as the $N-1 \times N-1$ dimensional matrices obtained by deleting the $i$th row and column from $\delta\bJ, \bSigma^{s},$ and $\bLambda^{s}$ respectively. Defining $\bA^{(i)} = (\bI -\bSigma^{(i)}\delta\bJ^{(i)})^{-1}$ and neglecting the off-diagonal self-energies $\bSigma^{s}_{ij}$, $i\neq j$ (Prop.\ \ref{prop:offdiagse}), so that $\bSigma^{s}$ is diagonal, the Schur complement formula gives
\begin{gather}
    \bA_{ii} = \frac 1 {1 - \sum_{kl \neq i} [\bSigma^{s} \delta\bJ]_{ik} (\bA^{(i)})_{kl} [\bSigma^{s} \delta\bJ]_{li}},
    \nonumber
    \\
    \bA_{ij} = \bA_{ii} \sum_k [\bSigma^{s} \delta\bJ]_{ik} \bA^{(i)}_{kj}
    \label{eq:schur}
\end{gather}
where $\sum_{kl\neq i}$ denotes the sum over all $k,l$ $\neq i$, and $j\neq i$. 

We will similarly define $\bR^{(i)}, \bC^{(i)}$ through the Dyson identities, \eqref{eq:dysonresum}, with the resolvents $\bA$ replaced by $\bA^{(i)}$ and with $\bSigma^{s}, \bLambda^{s}$ replaced by $\bSigma^{(i)}, \bLambda^{(i)}.$ The above formula immediately gives the following identities after neglecting subleading off-diagonal self-energy terms, $\bSigma^{s}_{ij}, \bLambda^{s}_{ij}$:

\begin{prop}
    The diagonal two-point functions have the Schur representations:
    \begin{gather}
        \rsp_{ii} = \frac{\bSigma^{s}_{ii}}{1- \bSigma^{s}_{ii} K_i} + O(N^{-1/2}\omega^{-p-p_0}), \\
        \csp_{ii}  = |1+K_i \bR^{s}_{ii}|^2 \bLambda^{s}_{ii} + |\bR^{s}_{ii}|^2 G_i +O(N^{-1/2}\omega^{-p}).
        \label{eq:diagresponse}
    \end{gather}
    where the feedback kernels $K_i, G_i$ are given by: 
    \begin{gather}
        K_i = \sum_{kl \neq i } \delta\bJ_{ik} \bR^{(i)}_{kl} \delta\bJ_{li}, \quad G_i = \sum_{kl\neq i} \delta\bJ_{ik} \bC_{kl}^{(i)} \delta\bJ_{il}^* .
        \label{eq:sekernels}
    \end{gather}
\end{prop}

Our goal now is to show that the off diagonal terms, $k\neq l$, appearing in the definitions of $K_i, G_i$ are subleading and that on the diagonal, one may replace $\bR^{(i)}_{kk}, \bC^{(i)}_{kk}$ with the full resolvent and correlations $\rsp_{kk}, \csp_{kk}$. The main obstacle is that the self-energies, $\bSigma^{(i)}, \bLambda^{(i)}$ appearing in the definition of $\bR^{(i)}, \bC^{(i)}$, are calculated on the full $N$-dimensional network, and thus, $\delta\bJ_{ik}, \delta\bJ_{li}$ are correlated with $\bR^{(i)}, \bC^{(i)}$. This makes the large-$N$ scaling of the sums in $K_i, G_i$ difficult to control.

The next proposition shows that this correlation is weak enough that we may replace $\bR^{(i)}, \bC^{(i)}$ with the ``cavity" variables $\Rc, \Cc,$ obtained by considering the dynamics of the network with the $i$th unit deleted. The resulting cavity variables are fully decoupled from $\delta\bJ_{ik}, \delta\bJ_{li}.$

\begin{prop}
    Let $\Sc, \Lc, \Rc, \Cc$ denote the self-energies and correlation functions of the cavity network obtained by deleting the $i$th unit from the network, that is, by removing every interaction term $i\hat \bx_i \delta\bJ_{ik} \bpsi_k$ and $i\hat \bx_k \delta\bJ_{ki} \bpsi_i$ from $S_{int}$ while keeping the background $S_0$, and in particular the saddle-point values of $\bkappa,\hat\bkappa$ in Eq.\ \eqref{eq:kapsaddle}, fixed at their full-network values. The self-energies of the full network satisfy
    \begin{gather}
        \bSigma^{s}_{kk} = \Sc_{kk} + O(N^{-1/2}\omega^{-p_0}), \nonumber
        \\ \bLambda^{s}_{kk} = \Lc_{kk} + O(N^{-1/2}),
        \label{eq:cavse}
    \end{gather}
    where $k\neq i$. Moreover for $k,l\neq i$ one has
    \begin{gather}
        \rsp_{kl} = \Rc_{kl} + O([N^{-1/2}\delta_{kl} + N^{-1}]\omega^{-p_0}), 
        \\
        \csp_{kl} = \Cc_{kl} + O(N^{-1/2}\delta_{kl} + N^{-1}).
    \end{gather}
    Consequently the self-energy kernels for the $i$th unit satisfy
    \begin{gather}
            K_i = \sum_{k\neq i } \delta\bJ_{ik} \Rc_{kk} \delta\bJ_{ki} + O(N^{-1/2}\omega^{-p_0}),
            \label{eq:cavitysekernels}
            \\ G_i = \sum_{k\neq i} |\delta\bJ_{ik}|^2\Cc_{kk} + O(N^{-1/2}).
        \nonumber
    \end{gather}
\end{prop}
\begin{proof}
    We begin by expanding both sides of Eq.\ \eqref{eq:cavse} in diagrams and considering the difference $\bSigma^{s}_{kk} - \Sc$ and the same for $\bLambda^{s}, \rsp, \csp$. The result is a sum over diagrams in which at least one non-rooted vertex is pinned to the deleted unit $i$. Applying Prop.\ \ref{prop:restrictse} gives the stated scaling for $\bSigma^{s},\bLambda^{s}, \rsp,$ and $\csp$.
    
    The scaling for the kernels follows by a similar construction. The first step is to show that we may replace $\bR^{(i)}, \bC^{(i)}$ by their cavity values  in Eq.\ \eqref{eq:sekernels}. To do this, note that the sum $\sum_{k,l\neq i}\delta\bJ_{ik} (\bR^{(i)} -\Rc)_{kl} \delta\bJ_{li}$ may be represented as a sum over the pinned diagrams of Prop.\ \ref{prop:restrictse} with the provision that we multiply by factors of $\delta\bJ_{ik}, \delta\bJ_{li}$ and sum over the indices $k,l$. To the various terms in the resulting series, we can once again associate multigraphs $M$ to the mean-square of each term based on the pattern of repeated indices. The only difference in the power counting is that the indices $k,l$ are no longer held fixed.
    
    The resulting multigraphs $M$ have a single rooted index, $i$. For $k\neq l,$ the path $i \to k \to \cdots \to l \to i$ must contain a cycle. For $k=l$, we may use the fact that the $i$ index occurs ``internally" as $i\to k \to \cdots \to i \to \cdots \to l \to i$. This implies that there is either a cycle or an edge $m_E$ with multiplicity greater than two. The end result is that these terms scale as $O(N^{-1/2})$ by Prop.\ \ref{prop:count}. The same reasoning generalizes to the $G_i$ kernel. 
    
    Replacing $\bR^{(i)}, \bC^{(i)}$ with the cavity variables, $\Rc, \Cc$ in the kernel definitions,  we may now neglect the off-diagonal terms by estimating the sum $\langle |\sum_{k\neq l} \delta\bJ_{ik} \Rc_{kl} \delta\bJ_{li}|^2\rangle_{\delta\bm J}$ and using the fact that in the cavity construction, $\delta\bJ_{ik}, \delta\bJ_{li}$ are uncorrelated with $\Rc_{kl}$ for $k,l\neq i$ to factorize the average.
    
\end{proof}

Inverting the formula Eq.\ \eqref{eq:diagresponse} and applying the above estimates, we find that we may replace $\bR_{kk}^{(i)}, \bC_{kk}^{(i)}$ in the definition of the kernels in Eq.\ \eqref{eq:sekernels} with the full saddle point values.
\begin{prop}
    Let $\bar \bR_{kk}, \bar \bC_{kk}$ satisfy, for every site $k$,
    \begin{gather}
        \bar \bR_{kk} = \rsp_{kk} + O(N^{-1/2}\omega^{-p_0}), \nonumber
        \\
        \bar \bC_{kk} = \csp_{kk} +O(N^{-1/2}).
    \end{gather}
    The self-energies satisfy the self-consistent equations
    \begin{gather}
        \bSigma^{s}_{ii} = \frac{ \bar \bR_{ii}}{1+  \bar \bR_{ii} \bar K_i} + O(N^{-1/2}\omega^{-p_0})
        \\ \bLambda^{s}_{ii} = \frac{ \bar \bC_{ii} - | \bar \bR_{ii} |^2 \bar G_i}{|1 +  \bar K_i  \bar \bR_{ii}|^2} + O(N^{-1/2}),
    \end{gather}
    where we have replaced the original kernels in Eq.\ \eqref{eq:sekernels} with their values evaluated on $\bar \bR_{ii}, \bar \bC_{ii}$: 
    \begin{gather}
        \bar K_i = \sum_{k} \delta\bJ_{ik} \bar \bR_{kk} \delta\bJ_{ki}, \quad \bar G_i = \sum_k |\delta\bJ_{ik}|^2 \bar \bC_{kk}.
        \label{eq:kersadp}
    \end{gather}
    \label{prop:sadpse}
\end{prop}
\begin{proof}
    We start by inverting Eq.\ \eqref{eq:diagresponse} to obtain the equations
    \begin{gather}
        \bSigma^{s}_{ii} = \frac{ \rsp_{ii}}{1+  \rsp_{ii}  K_i},
        \quad \bLambda^{s}_{ii} = \frac{ \csp_{ii} - | \rsp_{ii} |^2  G_i}{|1 +  K_i  \rsp_{ii}|^2},
    \end{gather}
    with $G_i, K_i$ as in Eq.\ \eqref{eq:sekernels}. Using the previous proposition to replace the kernels $G_i, K_i$ with their cavity sums, Eq.\ \eqref{eq:cavitysekernels} and then replacing $\Rc_{kk}, \Cc_{kk}$ with $\bar \bR_{kk}, \bar \bC_{kk}$ for every $k\neq i$ incurs an error which is $O(N^{-1/2}\omega^{-p_0})$ and $O(N^{-1/2})$ respectively, since $\sum_k |\delta\bJ_{ik}\delta\bJ_{ki}|$ and $\sum_k |\delta\bJ_{ik}|^2$ are $O(1)$. Using the fact that $\rsp_{ii} =O(\omega^{-p_0})$ and $\csp_{ii} = O(1)$ and Taylor expanding then gives the stated error bounds.
\end{proof}

We may therefore plug the diagonals $\rsp_{ii}, \csp_{ii}$ into the expression for the kernels $\bar K_i, \bar G_i$:
\begin{gather}
    \bSigma^{s}_{ii} = \frac{ \rsp_{ii}}{1+  \rsp_{ii} \bar K_i}{}+O(N^{-1/2}\omega^{-p_0}), \nonumber
        \\ \bLambda^{s}_{ii} = \frac{ \csp_{ii} - | \rsp_{ii} |^2  \bar G_i}{|1 +  \bar K_i  \rsp_{ii}|^2}{}+O(N^{-1/2}).
    \label{eq:sadpse}
\end{gather}
In the case in which $\bM=0$ so that $\bkappa =0$ exactly, we can see that the above formulae together with the Dyson equations (Eq.\ \eqref{eq:dysonresum}) and $\bSigma^{s}_{ii}\delta\bJ_{ij} = O(N^{-1/2}\omega^{-p})$  gives the stated result, Eq.\ \eqref{eq:twopoint}. Incorporating the effects of a non-zero $\bM$ is the subject of the next two sections.

\subsection{Higher cumulants}
\label{sec:highercumulants}

The power-counting arguments developed in Sec.\ \ref{sec:largeN} allow us to estimate the scaling of arbitrary cumulants. Here we prove a few lemmas that will be important in the Hessian analysis to follow.

\begin{prop}
    Consider a cumulant of arbitrary intensive order $b\ll N$, evaluated at the $\bkappa$ saddle point: 
    \begin{gather}
        F_{\bm i} = \langle \bzeta_{i_1}(\omega_1) \cdots \bzeta_{i_b}(\omega_b)\rangle_c,
    \end{gather}
    with $\bzeta_i$ equal to either $i\hat \bx_i$ or $\bpsi_i$ and with $1\leq r\leq b$ unique indices among the $i_1, \cdots, i_b$. Then
    \begin{gather}
        F_{\bm i} = O(N^{(1-r)/2}\gamma),
        \quad
        \langle F_{\bm i}\rangle_{\delta\bJ}= O(N^{1-r}\gamma).
        \label{eq:conncorrnosumbase}
    \end{gather}
    where the fixed-$\delta\bJ$ estimate should, as before, be understood to mean $\langle |F_{\bm i}|^2\rangle = O(N^{1-r}\gamma^2),$ and $\gamma(\omega_1,\cdots,\omega_b)$ describes the frequency scaling. Defining $\mathcal I \subset \{i_1,\cdots,i_b\}$ as the set of unique indices, and $\bar\omega_i = \sum_{s:i_s =i}\omega_s$ as the sum over frequencies at a site $i$, we have
    \begin{gather}
        \gamma(\omega_1,\cdots,\omega_b) = \delta\bigg(\sum_s \omega_s\bigg)(1 + \max_{i \in \mathcal I} |\bar\omega_i|)^{-p} \nonumber
        \\ \times \prod_{s:\, \bzeta_{i_s} = i\hat \bx_{i_s}}  (1+|\omega_{s}|)^{-p_0}  .
    \end{gather}
    In other words, $\gamma$ factorizes into a product that counts the number of $i\hat \bx$ legs attached to each index $i\in \mathcal I$, together with an additional factor of $\bar\omega^{-p}$ with $\bar\omega$ the largest net frequency attached to a site.

    Furthermore, writing $\delta F_{\bm i} = F_{\bm i} - \langle F_{\bm i}\rangle_{\delta\bJ}  $, one has for two distinct connected cumulants with a total of $r_\cup$-many fixed indices:
    \begin{gather}
        \langle \delta F_{\bm i} \delta F_{\bm j}^*\rangle _{\delta\bJ} = O(N^{1-r_\cup}\gamma_{\bm i}\gamma_{\bm j}),
    \end{gather}
    with $\gamma_{\bm i}, \gamma_{\bm j}$ the frequency coefficients corresponding to each factor of $\delta F.$
    \label{prop:conncorrnosum}
\end{prop}
\begin{proof}
    The first part is an immediate consequence of Prop. \ref{prop:count}: expanding the connected correlation function in diagrams, one considers a sum over diagrams with $r$ rooted vertices. We then square and $\delta\bJ$ average to estimate the magnitude, $F \sim \langle |F|^2\rangle_{\delta\bJ}^{1/2}$. Associating multigraphs to each partition of the resulting index sums and using their connectedness, the same argument as in Prop. \ref{prop:count} shows that the resulting mean-square is $O(N^{1-r}),$ so that $F_{\bm i} = O(N^{(1-r)/2})$ in root mean square. The estimate on $\langle F_{\bm i}\rangle_{\delta\bJ} $ follows the same line of reasoning.

    The frequency scaling follows from the fact that each fixed $i\hat \bx(\omega)$ leg brings a factor of $(1+|\omega|)^{-p_0}$ and each non-rooted one $(1+|\omega|)^{-p}$ (Sec.\ \ref{sec:largeN}), while each index $i\in\mathcal I$ creates an additional $i\hat \bx$ whose frequency scales as  $\bar\omega_i=\sum_{s:i_s=i} \omega_s$. This additional large $\bar\omega_i$ leg carrying $i\hat \bx$ can either insert into the rooted vertex or elsewhere in the graph. Either way, there is an additional factor of $(1+\max_{i\in\mathcal I}|\bar\omega_i|)^{-p}$ generated by the index with the largest $\bar\omega_i$ (again the decay survives any internal frequency integration since $\int da (1+|a|)^{-p}(1+|\bar\omega-a|)^{-p} \leq C_p(1+|\bar\omega|)^{-p}$ for $p>1$, with constants that depend on the diagram). Finally, the $\delta(\sum_i \omega_i)$ factor follows by stationarity. 

    Let us then consider $\langle \delta F_{\bm i} \delta F_{\bm j}\rangle _{\delta\bJ} $. Expand the product in diagrams and associate again a multigraph to the various index contractions based on the pattern of repeated indices. Since the graphs corresponding to the $F_{\bm i}, F_{\bm j}$ components are individually connected, the only way for the resulting multigraphs to be disconnected is if there are no common indices between the $\delta F_{\bm i}, \delta F_{\bm j}$. In this case the $\delta\bJ$ average factorizes, and gives a factor of zero. If, on the other hand the graph is connected, the same estimates as above give a scaling of $O(N^{1-r_\cup}).$ The frequency contributions follow by the same logic as above.

\end{proof}

We now use the above to state the basic proposition that will allow us to control the fluctuations around the Hessian later on: 
\begin{prop}
    Let $\bw^{(1)}, \cdots, \bw^{(\ell)} \in \mathbb C^N$ be arbitrary, deterministic vectors (possibly frequency dependent and independent of $\delta\bJ$). Then
    \begin{gather}
        \sum_{i_1 \cdots i_{\ell}} \bw^{(1)}_{i_1} \cdots \bw_{i_{\ell}}^{(\ell)} F_{i_1 \cdots i_{\ell} \cdots i_{b}} \nonumber
        \\ = O\bigg ( N^{(1-r_0)/2}\bar\gamma\prod_{i} w_{max}^{(i)} \bigg) ,
    \end{gather}
    where $r_0 \geq 1$ denotes the number of unique indices in $F$ that are held fixed after contracting with the weight vectors, and $\bar\gamma$ is the worst-case among the $\gamma$ associated to each $F_{i_1 \cdots i_{\ell} \cdots i_{b}}$ (Prop. \ref{prop:conncorrnosum}). In the special case in which $r_0 = 0$, one has the looser bound
    \begin{gather}
        \sum_{i_1 \cdots i_{\ell}}   \bw^{(1)}_{i_1} \cdots \bw_{i_{\ell}}^{(\ell)} F_{i_1 \cdots i_{b}} \nonumber
        \\ = O\bigg ( N^{1} \bar\gamma\prod_{i} w_{max}^{(i)} \bigg) ,
    \end{gather}
    \label{prop:conncorr}
\end{prop}
\begin{proof}
    We estimate the mean and variance separately. The scaling for a given choice of $r_0, \ell$ is the largest of the two terms. From the above proposition the mean term brings a factor $O(N^{1-r_0}\prod_i w_{max}^{(i)})$. We will see that this is the leading term when $r_0=0.$ The frequency scaling follows from the previous proposition. 
    
    The variance term is given by 
    \begin{gather}
        \sum_{i_1, \cdots, i_{\ell}} \sum_{i'_1, \cdots, i'_{\ell}} \bw^{(1)}_{i_1} \cdots \bw^{(\ell)}_{i_{\ell}} \nonumber
        \\ \times [\bw^{(1)}_{i'_1} \cdots \bw^{(\ell)}_{i'_{\ell}}]^*  \langle \delta F_{\bm i} \delta F_{\bm i'}^*\rangle.
    \end{gather}
We group the various terms in this sum based on the pattern of repeated indices. There are $O(N^{f})$ terms in which the $2\ell$ summed indices take $f$ distinct values not equal to the fixed indices, $0 \leq f \leq 2\ell$. The number of unique indices for each such term is $f + r_0$ (the factor $r_0$ comes from the fact that the fixed indices are shared between the two copies $\delta F_{\bm i}, \delta F_{\bm i'}$). By the above proposition, each of these terms scales as $O(N^{1-(f+r_0)}[\prod_i w^{(i)}_{max}]^2)$. Taking the product and square-rooting shows that the fluctuating part is at most $O(N^{(1-r_0)/2}\prod_iw^{(i)}_{max})$, which dominates the mean scaling for $r_0 \geq 2.$
\end{proof}

Neglecting the Hessian fluctuations, the above arguments show that higher moments with all indices distinct factorize to leading order into Wick pairings.

\subsection{Hessian terms}
\label{sec:hess}

In the previous sections, we have set $\bkappa, \hat\bkappa$ to their saddle point values after averaging over the $\delta\bJ$ (see Eq.\ \eqref{eq:kapsaddle}). In this section, we show that incorporating Hessian fluctuations generates the remaining terms in the correlation functions that depend on $\bM$ (Eq.\ \eqref{eq:twopoint}). We go on to show that the higher order fluctuations in $\bkappa$ are subleading. Throughout, we will factorize $\tilde \bM=\bU\bY^\top$. To start, let us write the average over an arbitrary observable $\obs$ as
\begin{widetext}
\begin{gather}
    \langle \obs \rangle = \int D(\bkappa, \hat \bkappa)e^{-i \hat \bkappa \bkappa } \int D(\bx,\hat \bx, \bpsi, \hat \bpsi) e^{\sum_i S^0_i + \sum_{ij} i\hat \bx_i \delta\bJ_{ij} \bpsi_j + \sum_{i\rho}  i\hat \bkappa^\rho \by^\rho_i \bpsi_i + \sum_{i\rho} i\hat \bx_i \bu^\rho_i \bkappa^\rho }\obs .
\end{gather}
Multiplying and dividing by the appropriate normalization $Z[\bkappa, \hat \bkappa]$, we may write this as an integral over the average over $\obs $ at a fixed value of $\bkappa, \hat \bkappa$, denoted by $\langle \obs \rangle_{\bkappa, \hat \bkappa}$:
\begin{gather}
    \int D(\bkappa,\hat\bkappa)e^{-i\hat \bkappa \bkappa  + \ln Z[\bkappa,\hat\bkappa]} \langle \obs\rangle_{\bkappa, \hat \bkappa} , 
    \label{eq:Oavg}
\end{gather} 
where 
\begin{gather}
    Z[\bkappa, \hat\bkappa] = \int D(\bx,\hat \bx, \bpsi, \hat \bpsi) e^{\sum_i S^0_i + \sum_{ij} i\hat \bx_i \delta\bJ_{ij} \bpsi_j + \sum_{i\rho}  i\hat \bkappa^\rho \by^\rho_i \bpsi_i + \sum_{i\rho} i\hat \bx_i \bu^\rho_i \bkappa^\rho },
    \\
    \langle \obs\rangle_{\bkappa, \hat \bkappa} = \frac 1 {Z[\bkappa, \hat\bkappa]}\int D(\bx,\hat \bx, \bpsi, \hat \bpsi) e^{\sum_i S^0_i + \sum_{ij} i\hat \bx_i \delta\bJ_{ij} \bpsi_j + \sum_{i\rho}  i\hat \bkappa^\rho \by^\rho_i \bpsi_i + \sum_{i\rho} i\hat \bx_i \bu^\rho_i \bkappa^\rho } \obs.
\end{gather}
\end{widetext}
We now set $\bkappa = \barkap + \delta \bkappa, \hat \bkappa = \delta \hat \bkappa $ and expand both the exponential and the $O$-average to quadratic order around the $\delta\bJ$-averaged values, $\barkap, 0$. Neglecting cubic and higher terms in the expansion of $\langle \obs \rangle_{\bkappa, \hat\bkappa} $, we have
\begin{widetext}
\begin{gather}
    \langle \obs\rangle
    = \big\langle \obs\big\rangle_{\sadp}
    + \int d\nu \bigg[ \langle \delta\bkappa_\rho(\nu) \rangle\frac{\delta \langle \obs\rangle_{\bkappa,\hat\bkappa}}{\delta \bkappa_\rho(\nu)}\bigg|_{\sadp} + \langle \hat \bkappa_\rho(\nu)\rangle \frac{ \delta\langle \obs \rangle_{\bkappa,\hat\bkappa}}{\delta \hat\bkappa_\rho(\nu)}\bigg|_{\sadp}
    \label{eq:Oavgexpansion}
    \\ + \frac12 \big\langle \dk_\rho(\nu)\dk^*_\sigma(\nu)\big\rangle\,
        \frac{\delta^2\langle \obs\rangle_{\bkappa,\hat\bkappa}}{\delta\bkappa_\rho(\nu)\delta\bkappa_\sigma^*(\nu)}\bigg|_{\sadp}
    +\big\langle \dk_\rho(\nu)\dkh^*_\sigma(\nu)\big\rangle\,
        \frac{\delta^2\langle \obs\rangle_{\bkappa,\hat\bkappa}}{\delta\bkappa_\rho(\nu)\delta\hat\bkappa_\sigma^*(\nu)}\bigg|_{\sadp}\bigg], 
    \nonumber
\end{gather}
\end{widetext}
where $\langle \obs \rangle_{\sadp}$ is the value of $\obs$ at $\bkappa_\rho = \bar\bkappa_\rho= \sum_j \by^\rho_j\langle \bmm_j\rangle_{\delta\bJ}, \hat\bkappa_\rho =0$, and we use $|_{\sadp}$ to denote evaluation at these $\delta\bJ$-averaged saddle point values. Note that the parameter $\nu$ is a dummy frequency, and all $\bkappa, \hat\bkappa$ are evaluated in the Fourier domain. We now expand the $\ln Z$ factor in Eq.\ \eqref{eq:Oavg} to second order, allowing us to evaluate the $\delta \bkappa$ moments above.

The first order variations in $Z[\bkappa,\hat\bkappa]$ bring 
\begin{gather}
    \frac{\delta \ln Z}{\delta \bkappa_\rho(\nu)} =  \sum_i \bu^\rho_i \langle i \hat \bx_i(\nu)\rangle_s = 0, \nonumber
    \\ \frac{\delta \ln Z}{\delta \hat \bkappa_\rho(\nu)} =  i\sum_i \by^\rho_i \ms_{i}(\nu),
\end{gather}
where $\bm{m}^s_i$ denotes the average $\langle \bpsi_i\rangle_s$ calculated at the $\bkappa$ saddle. The quadratic factors in the expansion of $Z[\bkappa,\hat\bkappa]$ bring
\begin{align}
    \frac{\delta^2 \ln Z}{\delta\bkappa_\rho(\nu)\delta\bkappa^*_\sigma(\nu)}
    &= \sum_{ij}\bu^\rho_i [\bu^\sigma_j]^*\big\langle \ih_i(\nu)\,\ih^*_j(\nu )\big\rangle_s = 0,
    \nonumber\\
    \frac{\delta^2 \ln Z}{\delta\bkappa_\rho(\nu)\delta\hat\bkappa^*_\sigma(\nu)}
    &= i\sum_{ij} \by^\sigma_i\, \rsp_{ij}(\nu)\, \bu^\rho_j = i\bOmega_{\sigma\rho}(\nu),
    \nonumber\\
    \frac{\delta^2 \ln Z}{\delta\hat\bkappa_\rho(\nu)\delta\hat\bkappa_\sigma^*(\nu)}
    &= -\sum_{ij} \by^{\sigma}_i [\by^{\rho}_j]^*\csp_{ij}
     = -\ckap_{\sigma\rho}(\nu),
    \label{eq:hessianblocks}
\end{align}
where  we have defined $\bOmega_{\rho\sigma} =(\by^\rho)^\top \rsp \bu^\sigma$, i.e. $\bOmega = \bY^\top \rsp \bU$, and $\ckap_{\rho\sigma} = (\by^\rho)^\top \csp [\by^\sigma]^*$, i.e. $\ckap = \bY^\top \csp \bY^*$. The factor of $-i\hat \bkappa \bkappa + \ln Z[\bkappa,\hat\bkappa]$ in Eq.\ \eqref{eq:Oavg} reads after expanding to second order:
\begin{gather}
    W[\bkappa,\hat\bkappa ] = -i\delta\hat\bkappa_\rho (\bI - \bOmega)_{\rho\sigma} (\delta\bkappa_\sigma - \beps_\sigma) - \frac 1 2 \delta\hat\bkappa_\rho \ckap_{\rho\sigma} \delta\hat\bkappa_\sigma,
    \label{eq:Wbackground}
\end{gather}
where we have defined the deviation of $\bkappa$ from its $\delta\bJ$-averaged value, $\beps_\rho = (\bI - \bOmega)^{-1}_{\rho\sigma} (\bkappa^{\sadp} - \barkap)_\sigma$, where $\bkappa^s_\rho = \sum_i \by^\rho_i \ms_i$ is the mean of the collective mode evaluated at the $\delta\bJ$-averaged saddle $\barkap$. Note that in the above and what follows we again suppress writing the explicit frequency arguments.

It follows that the Fourier transformed means and covariances are given by: 
\begin{gather}
    \langle \delta \bkappa\rangle = (\bI- \bOmega)^{-1} (\bkappa^s - \bar\bkappa) , \quad \langle \delta\hat\bkappa\rangle = 0
    \\
    \langle \dk\dk^\dagger\rangle_c
      = (\bI-\bOmega)^{-1}\,\ckap\,(\bI-\bOmega)^{-\dagger}, \nonumber
    \\
    \langle \dk i\dkh^\dagger\rangle_c = (\bI-\bOmega)^{-1},
    \qquad
    \langle \dkh\dkh^\dagger\rangle_c = 0 .
    \label{eq:collprops}
\end{gather}
The physical meaning of the $\langle \bkappa \hat\bkappa^\dagger\rangle$ correlator follows by rewriting the $\hat\bkappa$ as a derivative with respect to $\bkappa$ and integrating by parts: 
\begin{gather}
    \langle \bkappa_\rho i\hat\bkappa^*_\sigma \rangle = 
    \int D(\bkappa,\hat\bkappa) e^{-i\hat\bkappa\bkappa} \nonumber
    \\ \times \frac \delta{\delta \bkappa} \bigg[ 
    \int D(\bx,\hat \bx,\bpsi,\hat\bpsi)e^{S} \bkappa 
    \bigg].
    \label{eq:interpretkappahatkappa}
\end{gather}
The derivative against the $e^S$ term brings the $\rkap$ response of Eq.\ \eqref{eq:rkap} and the other term brings an identity factor, so that $\langle \bkappa i\hat\bkappa^\dagger\rangle = (\bI + \rkap)$. The $\bkappa,\hat\bkappa$ correlator is precisely the factor appearing in the balance assumption, Eq.\ \eqref{eq:balance}. The key point is that under this assumption, the scaling of the Hessian contributions can be controlled. 

\begin{prop}
    The means and covariances of the $\delta \bkappa, \delta \hat\bkappa$ are suppressed in powers of $N$. In particular: 
    \begin{align}
        \langle \delta\bkappa_\rho(\nu)\rangle &= \delta(\nu)\,O(\tilde u_{\max}^{-1}N^{-1/2}),
        \\
        \langle \delta \bkappa_\rho(\nu) \delta\hat\bkappa^*_\sigma(\nu)\rangle &= O(N^{-1}\tilde u^{-1}_{\max} \tilde y_{\max}^{-1}\bet(\nu)), 
        \\
        \langle \delta \bkappa_\rho(\nu) \delta \bkappa^*_\sigma(\nu) \rangle &= O(N^{-1}\tilde u_{\max}^{-2}\bet(\nu)^2),
    \end{align}
    where $\bet(\nu)$ is a model-specific parameter as in Eq.\ \eqref{eq:balance}. 
    \label{prop:kappabounds}
\end{prop}
\begin{proof}
The mean follows by a simple summation, 
\begin{gather}
 \langle \delta \bkappa_\rho\rangle = \sum_{\sigma} (\bI-\bOmega)_{\rho\sigma}^{-1} \sum_i \by^{\sigma}_i (\ms_i- \langle \ms_i\rangle_{\delta\bJ}) \nonumber
 \\ = O(||(\bI-\bOmega)^{-1}||_{op}\, \tilde y_{\max} N^{1/2}),
\end{gather}
together with our assumption on the $\bkappa$ response function, Eq.\ \eqref{eq:balance}. The factor of $N^{1/2}$ follows because the summand is centered: by Prop.\ \ref{prop:conncorrnosum} the sum has variance $\sum_{ii'}\by^\sigma_i[\by^\sigma_{i'}]^*\langle\delta\ms_i\,\delta\ms_{i'}\rangle_{\delta\bJ} = O(N\tilde y_{\max}^2)$.  In the above, we have used $(\bI-\bOmega)^{-1} = \bI+\rkap$ (we verify that the saddle-point and exact $\bOmega$ matrices can be interchanged below). The estimate on $\langle \delta \bkappa \delta \hat \bkappa \rangle$ follows from the same $\bkappa$-response assumption. 

For the $\delta\bkappa$ covariance, note that by Prop. \ref{prop:conncorr} one has the following bound on $\ckap$:
 \begin{gather}
     \ckap_{\rho\sigma} =  \sum_{ij }\by^\rho_i [\by^\sigma_j]^* \csp_{ij}= O(Ny_{\max}^2).
 \end{gather}
Applying once again the assumption Eq. \eqref{eq:balance} and using $||\prod_i \bA_i||_{op} \leq \prod_i  || \bA_i||_{op}$ then establishes the bound on $\delta \bkappa_\rho \delta\bkappa_\sigma^*.$
\end{proof}

We will now use these bounds to estimate the Hessian corrections and higher order terms. The key point is that each time a factor of $\delta\bkappa$ appears, it comes from differentiating either $\ln Z$ or $\langle \obs \rangle_{\bkappa,\hat\bkappa}$. As such, it comes with a factor of $\sum \bu_i^\rho i\hat \bx_i$. Similarly $\delta\hat\bkappa$ always brings with it a factor of $\sum_i \by^\rho_i \bpsi_i$. The bounds above will ensure that when the coefficients are included, the Hessian fluctuations will at most match the order of $\langle \obs \rangle_s$, and higher corrections will be subleading.

We now move on to estimating the contribution of the Hessian terms to the one- and two-point functions. The derivatives for the mean are:
\begin{align}
    \frac{\delta \langle \bpsi_k\rangle_s}{\delta \bkappa_\rho(\nu)} &= \sum_j \bu^\rho_j \langle \bpsi_k i\hat \bx_j \rangle_s = O(\tilde u_{\max}^\rho)
    \\
    \frac{\delta^2\langle\bpsi_k\rangle_s}{\delta\bkappa_\rho(\nu) \delta\bkappa_\sigma^*(\nu)}
    &= \sum_{ij}\bu^\rho_i\bu^\sigma_j\,\big\langle\bpsi_k\ih_i\ih_j\big\rangle_s = O(\tilde u^\rho_{\max} \tilde u^\sigma_{\max}),
    \\ 
    \frac{\delta^2\langle\bpsi_k\rangle_s}{\delta\bkappa_\rho(\nu)\delta\hat\bkappa_\sigma^*(\nu)}
    &= i\sum_{ij}\bu^\rho_i\by^\sigma_j\langle\bpsi_k\ih_i\delta\bpsi_j\rangle_s \nonumber
    \\ &= O(\tilde u^\rho_{\max} \tilde y^\sigma_{\max}).
\end{align}
Note that the $\delta\bpsi_j$ variables are centered from the derivative hitting the $Z[\bkappa, \hat\bkappa]$ in the denominator of $\langle O\rangle_{\bkappa,\hat\bkappa},$ and all derivatives are evaluated at the saddlepoint, $\delta\bkappa=\delta\hat\bkappa=0.$ In addition, we have dropped all frequency dependence from the error estimates, as the mean is by supposition constant in time. Applying Prop. \ref{prop:conncorr} then gives the stated error bounds. Inserting the above into Eq.\ \eqref{eq:Oavgexpansion}, it follows that the means $\langle \bpsi_k\rangle$ are unaffected by the fluctuations around the saddle point up to $O(N^{-1/2})$ errors.

We now turn to the two-point function $\langle \bpsi_i \bpsi_j\rangle_c.$ It is convenient at this point to break up the connected correlation function into two terms, $T_{1}$ and $T_{2}$:
\begin{gather}
    \langle \bpsi \bpsi^\dagger\rangle_{(\bkappa, \hat \bkappa)} - \langle \bpsi\rangle \langle \bpsi^\dagger\rangle = \nonumber
    \\ \underbrace{\langle \bpsi \bpsi^\dagger \rangle_{\bkappa, \hat \bkappa} - \langle \bpsi\rangle_{\bkappa, \hat \bkappa} \langle \bpsi^\dagger\rangle_{\bkappa, \hat \bkappa}}_{T_{1}} \nonumber
    \\ + \underbrace{\langle \bpsi\rangle_{\bkappa, \hat \bkappa} \langle \bpsi^\dagger\rangle_{\bkappa, \hat \bkappa} - \langle \bpsi \rangle \langle \bpsi ^\dagger\rangle}_{T_{2}}
    \label{eq:T1T2-decomp}
\end{gather}
At the saddle point value, $T_{1}$ merely reproduces $\csp$ (i.e. Eq. \eqref{eq:twopoint} with $\bM=0$) while $T_{2}$ vanishes.  Let us then consider the fluctuations in each term.

Taking the $\bkappa, \hat\bkappa $ derivatives of $T_{1}$ leaves a sum over connected three and four point functions:
\begin{align}
    \frac{\delta}{\delta \bkappa_\rho(\nu)} T_{1} &= \sum_k \bu^\rho_k \langle \bpsi_i(\omega) \bpsi_j(-\omega) i\hat \bx_k(-\nu) \rangle_{c,s} \\
    &= O(\tilde u_{\max}\bar\gamma [\delta_{ij} + N^{-1/2}]) \nonumber
    \\ 
    \frac{\delta^2}{\delta \bkappa_\rho (\nu) \delta \bkappa^*_\sigma(\nu)} T_{1}
    &= \sum_{kl}  \bu_k^\rho \bu_l^\sigma \nonumber
    \\ &\times \langle \bpsi_i(\omega) \bpsi_j(-\omega) i \hat \bx_k(-\nu) i\hat \bx_l(\nu) \rangle_{c,s } \nonumber
    \\ &= O([\delta_{ij} + N^{-1/2}] \tilde u^\rho_{\max} \tilde u^\sigma_{\max}\bar\gamma )
    , \nonumber 
    \\
    \frac{\delta^2}{\delta \bkappa_\rho(\nu) \delta\hat  \bkappa^*_\sigma(\nu)} T_{1}
    &=  i\sum_{kl} \bu^\rho_k \by^\sigma_l \nonumber
    \\ &\times \langle \bpsi_i(\omega) \bpsi_j(-\omega) i \hat \bx_k(-\nu) \bpsi_l(\nu) \rangle_{c,s } \nonumber 
    \\ & = O([\delta_{ij}  + N^{-1/2}]\tilde u^\rho_{\max} \tilde y^\sigma_{\max} \bar\gamma), \nonumber
\end{align}
where we have used Prop. \ref{prop:conncorr} with 1 fixed index $r_0=1$ for $i=j$ and $r_0=2$ otherwise, and $\bar\gamma$ is the worst-case frequency scaling across all choices of summed indices $k,l$. 

Multiplying by the corresponding moments and integrating over $\nu$, we find that the contribution of these terms is subleading in $N$: 
\begin{gather}
    \langle \delta\bkappa_\rho\rangle \frac{\delta}{\delta \bkappa_\rho} T_{1}\bigg|_{\sadp} + \frac 1 2 \langle \delta\bkappa_\rho \delta\bkappa_\sigma^*\rangle \frac{\delta^2}{\delta \bkappa_\rho \delta \bkappa_\sigma^*} T_{1} \bigg|_{\sadp} \nonumber
    \\ + \langle \delta\bkappa_\rho \delta\hat \bkappa^*_\sigma\rangle \frac{\delta^2}{\delta \bkappa_\rho \delta \hat \bkappa^*_\sigma} T_{1} \bigg|_{\sadp} \nonumber
    \\ = O(N^{-1/2}[ \delta_{ij}  + N^{-1/2}\omega^{-p}]),  
\end{gather}
where we use $\beta(\nu) \leq \min\{\sqrt{N}, (1+|\nu|)^p\}$ (Eq.\ \eqref{eq:balance}) and $p>1$, together with Prop.\ \ref{prop:conncorr}, to estimate the contribution from the $\nu$ integrals. This allows us to estimate $\int d\nu \beta(\nu)^{n}/[(1+|\nu|)^{np}(1+|\nu+\omega|)^{p}] = O(\sqrt{N}\omega^{-p})$ for positive integer $n$. The Hessian corrections to $T_{1}$ may therefore be treated as error terms. Repeating the analysis for the response function, one finds errors scaling as $O(N^{-1/2}[\delta_{ij} \omega^{-p_0} + N^{-1/2}\omega^{-p-p_0}])$---i.e. with errors multiplied by an additional factor of $\omega^{-p_0}$, the decay of the rooted response leg.

On the other hand, the $T_{2}$ terms generate the $\bM$ corrections in Eq. \eqref{eq:twopoint}.  To see this, we first note that
\begin{gather}
    \frac{\delta}{\delta \bkappa_\rho} \langle \bpsi_i \rangle_{\bkappa, \hat\bkappa} = \sum_j \rsp_{ij} \bu^\rho_j, \nonumber
    \\ \frac{\delta}{\delta \hat \bkappa_\rho} \langle \bpsi_i \rangle_{\bkappa, \hat\bkappa} = i\sum_j \csp_{ij} \by^\rho_j ,
\end{gather}
Let us write the contribution of $T_{2}$ as
\begin{gather}
    T_{2} \bigg|_{\sadp} + \langle \delta \bkappa_\rho \rangle \frac {\delta}{\delta \bkappa_\rho} T_{2}\bigg |_{\sadp} \nonumber
    \\ + \frac 12 \langle \delta \bkappa_\rho \delta\bkappa_\sigma^*\rangle \frac{\delta^2}{\delta \bkappa_\rho \delta \bkappa_\sigma^*} T_{2} \bigg |_{\sadp} \nonumber
    \\ + \langle \delta \bkappa_\rho \delta\hat\bkappa_\sigma^*\rangle  
\frac{\delta^2}{\delta \bkappa_\rho \delta \hat\bkappa_\sigma^*} 
T_{2}\bigg |_{\sadp}, \label{eq:T2exp}
\end{gather}
where again the frequency argument $\nu$ of the $\bkappa,\hat\bkappa$ are integrated out. Expanding out the $\delta\bJ$-averaged product of means $-\langle \bpsi_i \rangle \langle \bpsi_j\rangle $ appearing in Eq.\ \eqref{eq:T1T2-decomp} in powers of $\delta\bkappa, \delta\hat\bkappa$ cancels the term $\langle \bpsi_i \rangle_{\sadp} \langle \bpsi_j\rangle_{\sadp} $ in $T_2 |_s$. The remaining powers in this expansion of $-\langle \bpsi_i \rangle\langle\bpsi_j\rangle$ cancel all of the remaining terms in the expansion of $T_2$ in which the differential operators $\delta /\delta\bkappa ,\delta/\delta \hat \bkappa$ apply only to either $\langle \bpsi_i\rangle_{\bkappa,\hat\bkappa}$ or $\langle \bpsi_j\rangle_{\bkappa,\hat\bkappa}$. This immediately eliminates the first order term $\langle \delta \bkappa\rangle \delta T_2/\delta\bkappa$ in the expansion of $T_{2}$. The disconnected (mean) terms in the second-order $\bkappa,\hat\bkappa$ moments are similarly eliminated.

Neglecting canceled terms, the remaining factors in the second derivatives are: 
\begin{gather}
    \frac{\delta^2}{\delta \bkappa_\rho(\nu) \delta \bkappa_\sigma^*(\nu)}\langle \bpsi_i\rangle_{\bkappa, \hat \bkappa} \langle \bpsi_j\rangle_{\bkappa, \hat \bkappa} \nonumber
    \\ = 
    \sum_{kl} \bu^\rho_k \rsp_{ik} [\rsp_{jl}]^* [\bu^\sigma_l]^*\delta(\nu-\omega) + (i \leftrightarrow j,\ \omega\to-\omega)  ,
    \\
    \frac{\delta^2}{\delta \bkappa_\rho(\nu) \delta \hat \bkappa_\sigma^*(\nu)}\langle \bpsi_i\rangle_{\bkappa, \hat \bkappa} \langle \bpsi_j\rangle_{\bkappa, \hat \bkappa} \nonumber
    \\ = 
    i\sum_{kl} \bu^\rho_k \rsp_{ik} \csp_{lj} \by^\sigma_l\delta(\nu-\omega)  + (i\leftrightarrow j,\ \omega\to-\omega).
\end{gather}
Note the crucial difference with the $T_1$ derivatives: here, the differential operators create two-point functions with one $\omega$ and one $\nu$ argument: $\langle \bpsi_i(\omega) i\hat \bx_i (\nu)\rangle, \langle \bpsi_i(\omega)\bpsi_k (\nu)\rangle,$ and so on. The result is the delta function $\delta(\nu\mp\omega)$ clipping $\nu$ to $\pm\omega$. In this way, the $T_2$ Hessian terms  have a distinct frequency-scaling.

Combining these formulae with the Hessian contributions gives after contracting the $\bkappa,\hat\bkappa$ indices and setting $\nu = \omega$: 
\begin{widetext}
\begin{gather}
    \bC = \csp + \rsp \bU(\bI -\bOmega)^{-1} \bY^\top  \csp \bY^* (\bI-\bOmega)^{-\dagger} \bU^{\dagger} (\rsp)^\dagger  
    \nonumber
    \\ + \rsp \bU (\bI-\bOmega)^{-1} \bY^\top \csp + \csp \bY^* (\bI-\bOmega)^{-\dagger} \bU^{\dagger} (\rsp)^\dagger + O(N^{-1/2}[\delta_{ij} +N^{-1/2}\omega^{-p}]).
    \label{eq:fullhess}
\end{gather}
The same line of reasoning gives for the response
\begin{gather}
    \bR = \rsp + \rsp \bU(\bI - \bOmega)^{-1}\bY^\top \rsp  + O(N^{-1/2}(\delta_{ij} +N^{-1/2}\omega^{-p})\omega^{-p_0})
    \label{eq:fullhessR}
\end{gather}
\end{widetext}
Let us consider the scaling of the Hessian corrections from $T_2$ (i.e. the second terms in the sum above). Applying Prop.\ \ref{prop:conncorr} to the contractions together with the balance assumption, Eq.\ \eqref{eq:balance}, shows that the Hessian corrections scale as 
\begin{align}
    [\bC- \csp]_{ij} &=  O\big( N^{-1}(\omega^{-p}\beta(\omega) + \omega^{-2p} \beta^2)\big) \nonumber
    \\
    [\bR-\rsp]_{ij} &= O\big( N^{-1}\omega^{-p-p_0}\beta\big)  
    \label{eq:hessscale}
\end{align}
Thus for $\omega = O(1),$ the corrections are comparable to the existing error terms. In other words, $\bC=\csp$ and $\bR=\rsp$ within the existing error budget. However, if $\beta(\omega)$ grows with $\omega$, it is possible that the off-diagonal Hessian terms rise above the errors at large $N,\omega$. We show that tightly balanced E/I networks provide one such example in Sec.\ \ref{sec:Mex}.

Applying the Woodbury identity to sum these contributions gives
\begin{align}
    \bC(\omega) &= (\bI - \bSigma^{s}[\delta\bJ+\tilde \bM])^{-1}\bLambda^{s} ( \bI -\bSigma^{s} [\delta\bJ+\tilde \bM])^{-\dagger } \nonumber
    \\ &+ O(N^{-1/2}\delta_{ij} + N^{-1}\omega^{-p}),
    \nonumber
    \\
    \bR(\omega) &= ( \bI - \bSigma^{s} [\delta\bJ+\tilde \bM])^{-1}\bSigma^{s} \nonumber
    \\ &+ O(N^{-1/2}\omega^{-p_0}\delta_{ij} + N^{-1}\omega^{-p-p_0}).
    \label{eq:fullhess2}
\end{align}
At this point, the only difference between the above and Eq.\ \eqref{eq:twopoint} is that the self energies use the diagonal correlation and responses evaluated at the $\bkappa$ saddle, $\csp_{ii}, \rsp_{ii}$, rather than $\bC_{ii}, \bR_{ii}$. The final step is to replace the self-energies $\bLambda^{s},\bSigma^{s}$ with their full self-consistent values, $\bLambda,\bSigma$ of Eq.\ \eqref{eq:fullse}. In other words, we replace $\csp, \rsp$ with $\bC,\bR$ in the self-energy definitions, Eq.\ \eqref{eq:sadpse}. This is justified for the saddle-point terms $\csp,\rsp$ (i.e. the first terms in the sums of Eqs.\ \eqref{eq:fullhess}  \& \eqref{eq:fullhessR}) by the same arguments we previously used to obtain the self-energy equations (Prop.\ \ref{prop:sadpse}).

For the Hessian terms, we define $\bar\bOmega = \bY^\top (\bI-\bSigma\delta\bJ)^{-1}\bSigma \bU$ with $\bSigma$ the self-energy defined on the full $\bR_{ii}$ as in Eq.\ \eqref{eq:fullse} and expand the resolvent $(\bI - \bar\bOmega + \delta\bOmega)^{-1}$ in powers of $\delta\bOmega = \bar\bOmega - \bOmega$. Since $\bSigma_{ii} - \bSigma^{s}_{ii} = O(N^{-1/2}\omega^{-p_0})$ by Prop.\ \ref{prop:sadpse}, one has $\delta\bOmega_{\rho\sigma} = \sum_i \by^\rho_i (\bSigma_{ii} - \bSigma^{s}_{ii}) \bu^\sigma_i + \cdots = O(\omega^{-p})$, and the balance condition gives $||(\bI-\bOmega)^{-1}\delta\bOmega||_{op} = O(N^{-1/2})$, which verifies the assumption made in the proof of Prop.\ \ref{prop:kappabounds}. Applying the balance condition and $\beta \leq \min\{\omega^p, \sqrt{N}\}$, together with Prop.\ \ref{prop:conncorr} to deal with the flanking $\rsp,\csp$ terms, it follows that the corrections are at most $O(\omega^{-p-p_0}N^{-1})$ for $\bR$ and $O(\omega^{-p}N^{-1})$ for $\bC$, matching the existing errors. Assuming the neglected terms in the Hessian expansion are indeed subleading (see below), we have derived Eq.\ \eqref{eq:twopoint}.

We now obtain the high-frequency expansion in Eq.\ \eqref{eq:highfreq}. We begin by noting that by Prop.\ \ref{prop:conncorr} and using $\bu^\rho_i \by^\sigma_j = O(N^{-1/2})$, it follows that $||\bOmega||_{op} = O(\sqrt{N} \omega^{-p})$. Hence, if we take $N, \omega \gg 1$ with $\omega^p \gg \sqrt{N}$, one may expand the resolvent $(\bI-\bOmega)^{-1}$ appearing in Eqs.\ \eqref{eq:fullhess} \& \eqref{eq:fullhessR} to arrive at 
\begin{align}
    \bC &= \bLambda + \bV\bLambda +(\bV\bLambda)^\dagger \nonumber
    \\ &+ O(\omega^{-2p} + N^{-1/2} \bI +  \omega^{-p} N^{-1}\bm 1), 
    \\
    \bR &= \bSigma + \bV\bSigma \nonumber
    \\ &+ O(\omega^{-2p-p_0} + [N^{-1/2} \bI +  \omega^{-p} N^{-1}\bm 1] \omega^{-p_0}).
    \label{eq:highfreqhess}
\end{align}
Note that when the Hessian terms from $T_2$ rise above the error terms, we must have $\omega^p \gg \sqrt{N}$ to ensure that the higher order terms $\bU[\bOmega + \bOmega^2 + \cdots]\bY^\top$ are subleading.

We conclude the calculation by showing that we are justified in neglecting the higher order terms in the expansion around the $\bkappa$ saddle and that higher cumulants remained suppressed in powers of $N$ as in Prop.\ \ref{prop:conncorrnosum}. The dropped terms have the  form 
\begin{gather}
    \sum_{z\geq 0, w\geq 1} \int d^{n_\nu}\nu \bigg(  \prod_{a=1}^z \sum_{k_a l_a}  \frac{\delta^{k_a+l_a}}{\delta \bkappa^{k_a} \delta \hat\bkappa^{l_a}} \ln Z \bigg)  \nonumber
   \\ \times  \bigg( \sum_{m+n=w} \frac{\delta^{m+n}}{\delta\bkappa^m \delta\hat\bkappa^n} \langle \obs \rangle_{\bkappa,\hat\bkappa}\bigg) \nonumber
    \\ \times \langle   \delta\bkappa^{m+\sum_a k_a} \delta\hat\bkappa^{n + \sum_a l_a}\rangle ,
    \label{eq:errorseries}
\end{gather}
where we drop factorial factors that do not affect the scaling, and the terms with $(z=0, w=1)$ and $(z=0,w=2)$ are not included in the sum. Note too that we suppress the contractions over $\sigma$ and $\rho$, and each power of $\bkappa, \hat\bkappa$ is assigned its own frequency $\nu$ (i.e. $n_\nu=\sum_a(k_a+l_a)+m+n$ above). Finally, we note that the $w=0$ terms do not contribute, owing to the normalization of the MSRDJ functional, $\int D(\bkappa,\hat\bkappa)e^{-i\bkappa\hat\bkappa + \ln Z} = 1$ (and of its quadratic part, $\int D(\delta\bkappa,\delta\hat\bkappa) e^{W} = |\det(\bI-\bOmega)|^{-1} = 1$, since $\bOmega$ is strictly causal and the collective modes are stable). The same mechanism removes all Wick pairings in which any $\delta\bkappa,\delta\hat\bkappa$ variables arising from the $\ln Z$ derivatives are not connected to the $\obs$ factors through a chain of pairings (i.e. any term containing a vacuum graph; see Ref.\ \cite{hertz2017path}).

The $\ln Z$ factors contribute 
\begin{widetext}
\begin{align}
    \frac{\delta^{k_a+l_a}}{\delta \bkappa^{k_a} \delta \hat\bkappa^{l_a}} \ln Z &= \bigg\langle 
    \prod_{s=1}^{k_a} \bigg(\sum_i i\hat \bx_i(\nu^a_s) \bu^\rho_i\bigg) \prod_{s=1+k_a}^{k_a+l_a} \bigg(\sum_i \bpsi_i(\nu^a_s) \by^\rho_i \bigg) \bigg\rangle_c  
    \\
    &= \delta\bigg(\sum_s \nu^a_s\bigg)\,O\bigg(N \tilde u_{\max}^{k_a} \tilde y_{\max}^{l_a} \prod_{s=1}^{k_a}  (1+|\nu^a_s|)^{-p}  \bigg) ,
\end{align}
\end{widetext}
where we used Prop.\ \ref{prop:conncorr}. We can see that all non-zero $\ln Z$ terms have $l_a \geq 1$ by causality, and $k_a + l_a\geq 3$, since we previously kept quadratic terms. We set $\obs$ to be a connected correlation function with $r_0$ distinct indices whose frequencies sum to zero and individually scale with a parameter $\omega$. The $\obs$ derivatives contribute at order $m,n$
\begin{gather}
 \delta\bigg(\sum_t \nu_t\bigg)\,O\bigg(N^{(1-r_0)/2} \tilde u_{\max}^{m} \tilde y_{\max}^n   \bar\gamma\bigg)
\end{gather}
where $\bar\gamma$ is the worst-case frequency scaling of the individual terms as in Prop.\ \ref{prop:conncorr}.

Finally, we estimate the contribution from the $\delta \bkappa,\delta\hat\bkappa$ and the $\nu$ integrals. By Prop.\ \ref{prop:kappabounds}, each $\delta \bkappa$ effectively brings a factor of $N^{-1/2} \tilde u_{\max}^{-1} \beta(\nu)$, while each $\delta \hat\bkappa$ brings a factor of $N^{-1/2} \tilde y_{\max}^{-1}.$ Moreover, the average $\langle \delta\bkappa^{m +\sum_ak_a} \delta\hat\bkappa^{n+\sum_al_a}\rangle$ factorizes into a sum over Wick pairings, with arbitrary replacements of $\delta \bkappa$ by its mean. We now estimate the contributions from these pairings.

Let us consider the contribution of a term with $P$ Wick pairs and $Q$ mean replacements using Feynman diagrams. We draw a vertex for each $\ln Z$ factor with $k_a + l_a$ legs, a single vertex with $m+n$ legs for the $\obs$ derivative, and we remove $\delta\bkappa$ legs replaced by the mean. Connecting paired legs, the stationarity of $W[\delta\bkappa,\delta\hat\bkappa]$ (Eq.\ \eqref{eq:Wbackground}) forces the associated $\nu$ variables to sum to zero. The vanishing of vacuum graphs under the MSRDJ action \cite{hertz2017path} implies that the resulting diagram is connected. 

Following standard arguments for connected graphs (see e.g. Ref.\ \cite{coleman1988aspects} Ch.\ 5), it follows that the $\delta$ functions sitting at the $\ln Z$ vertices reduce the number of independent frequency integrals from $P$ down to $L=P-z$. We may bound the $z$ redundant frequency integrals via $\beta(\bar\nu)^n (1+|\bar\nu|)^{-np}\leq 1$, where $\bar\nu = \sum_i \nu_i$ is a sum over frequency integrals created by a $\delta$ function and $n\geq 1$ a positive integer. For the remaining terms, the $\ln Z$ factors produce frequency integrals of the form $\int d\nu \beta(\nu)^n/(1+|\nu|)^{np}$ for some integer $n\geq1.$ Splitting the integrals at $\nu=\sqrt{N},$ we can see that each of these contributes a factor that is $O(N^{1/2})$ for $p> 1$. 

The $\nu$ integrals from the $\obs$ factors are similar but contain additional terms of the form $(1 + |\omega+S|)^{-p}$, with $S$ some sum over dummy frequencies $\nu$ and $\omega$ one of the fixed frequency variables (Prop.\ \ref{prop:conncorr}). Using $\beta(\nu)^n/(1+|\nu|)^{np} \leq \sqrt N/(1+|\nu|)^{p}$ then allows us to estimate $\int d\nu \beta(\nu)^{n}/[\allowbreak(1+|\nu|)^{np}\allowbreak(1+|\omega+\nu|)^{p}]= O(N^{1/2}\omega^{-p})$. We can see that these terms generate an additional factor of $\omega^{-\alpha}$, where $\omega^{-\alpha}$ corresponds to the frequency scaling of the $\obs$ cumulant evaluated at the $\bkappa$ saddle point (Prop.\ \ref{prop:conncorrnosum}). Meanwhile, the $Q$ mean insertions have no frequency contribution and contribute only the $\langle\delta\bkappa\rangle$ given in Prop.\ \ref{prop:kappabounds}.

Putting this all together, we find that each term scales at most as $N^\chi \omega^{-\alpha}$ where
\begin{gather}
    \chi = z + \frac 1 2(P-z) - \frac{\sum_a k_a + l_a + m +n} 2 + \frac {1-r_0}{2 }.
\end{gather}
Note that $\sum_a k_a + l_a + m +n = 2P + Q$. To finish the argument, note that each $z$ order in Eq.\ \eqref{eq:errorseries} brings at least $3$ instances of $(\bkappa, \delta\bkappa)$. Using this fact together with the connectivity of the graph ($P\geq z$), it follows that $Q+P \geq z+1$, and thus, $\chi \leq -\frac1 2 + \frac 1 2 (1-r_0)$, matching the existing error budgets for the two-point functions, and the connected correlation functions for higher cumulants given in Prop.\ \ref{prop:conncorrnosum}.\footnote{The bound $O(N^{(1-r_0)/2})$ on the derivatives of $\langle\obs\rangle_{\bkappa,\hat\bkappa}$
requires $\obs$ to be connected at fixed $\bkappa,\hat\bkappa$, which holds for $T_{1}$ but not for $T_{2}$,
whose derivatives contain products of one-point cumulants such as
$\partial_{\hat\bkappa}\langle\bpsi_i\rangle\,\partial_{\hat\bkappa}\langle\bpsi_j\rangle = O(N^0)$.
The same estimate nevertheless applies to $T_{2}$ because each one-point factor carries its own frequency
delta function $\delta(\pm\omega+\sum\nu)$ and only diagrams linking the two factors survive (the
others vanish for $\omega\neq0$ or cancel against $-\langle\bpsi_i\rangle\langle\bpsi_j\rangle$), so
there is one loop fewer and at least one more pair, $L=P-z-1$ and $P\stdge z+1$, which reproduces
$\chi=(z-P-Q-1)/2$ with $\chi\stdle-3/2$ for every term beyond the Gaussian one. The frequency count is likewise unchanged: each $\bkappa$-derivative of a one-point factor brings $\omega^{-p}$, and every leg clipped to $\pm\omega$ attaches to such a factor or to a $\bkappa$ leg of a $\ln Z$ vertex, so its $\beta(\omega)$ comes with a $(1+|\omega|)^{-p}$ and contributes at most $\min\{1,\sqrt N\omega^{-p}\}$; every term beyond the Gaussian one is therefore $O(\min\{N^{-3/2},N^{-1}\omega^{-p}\})$, within the error of Eq.~\eqref{eq:fullhess} at all frequencies.}

We have shown that the higher terms in the Hessian analysis can indeed be dropped, and we have shown that the cumulant bounds given in Prop.\ \ref{prop:conncorrnosum} and the main text hold after incorporating fluctuations in $\bkappa,\hat\bkappa.$

\subsection{Cavity construction and DMFT Closure}
\label{sec:dmft}

In this section we construct a DMFT which provides the leading order behavior of the diagonal two-point functions $\bC_{ii}, \bR_{ii}$ appearing in the definitions Eq.\ \eqref{eq:fullse} as well as the one-point functions $\bmm_i = \langle \bpsi_i \rangle.$

Let us consider the average over a single-site observable $\langle \obs \rangle$ with $\obs = \delta \bpsi_i \delta \bpsi_i^*, \bpsi_i\, i\hat \bx_i^*$ or $\bpsi_i$. Integrating out all neurons except for the $i$th gives the representation
\begin{widetext}
\begin{gather}
    \langle \obs\rangle = \int D(\bx_i,\hat \bx_i, \bpsi_i,\hat\bpsi_i) e^{S^0_i +\sum_\rho i\hat \bx_i \bu^\rho_i \bkappa_\rho +  W[i\hat \bx_i \delta\bJ_{ij}, \delta\bJ_{ji}\bpsi_i]} \obs ,
\end{gather}
where $\bkappa^\rho$ is evaluated at the saddle point, and $W[ \bh,\hat \bh]$ is the cumulant generating function of the cavity system obtained by deleting the $i$th unit: 
\begin{gather}
    W[\bh, \hat \bh] = \ln \int \prod_{j\neq i}D(\bx_j, \hat \bx_j, \bpsi_j, \hat \bpsi_j) e^{\sum_{j\neq i} S^0_j + \sum_{jk\neq i} i\hat \bx_j \delta\bJ_{jk} \bpsi_k + \sum_{\rho j} i\hat \bx_j \bu^\rho_j \bkappa^\rho +  \sum_j i\hat \bx_j \hat \bh_j + \bpsi_j\bh_j }.
\end{gather}
Expanding $W$ in powers of $\hat \bh, \bh$, an arbitrary term in the expansion of order $\alpha+\beta$ has the form 
\begin{gather}
    T = \sum_{j_1, \cdots, j_\alpha, k_1, \cdots, k_\beta \neq i} \langle i\hat \bx_i \delta\bJ_{ij_1} \bpsi_{j_1} \cdots i\hat \bx_i \delta\bJ_{ij_\alpha} \bpsi_{j_\alpha} i\hat \bx_{k_1} \delta\bJ_{k_1 i} \bpsi_i \cdots i\hat \bx_{k_\beta} \delta\bJ_{k_\beta i} \bpsi_i \rangle_{\mathrm{cav}, c},
    \label{eq:Texm}
\end{gather}
\end{widetext}
where we use $\langle \cdot \rangle_{\mathrm{cav},c}$ to denote the cumulant with respect to the cavity system. We will show that only the linear and quadratic terms, $\alpha +\beta \leq 2$, survive at large $N$. 

Let us calculate the scaling of an arbitrary term $T$ as above via $\langle |T|^2\rangle_{\delta\bm J}$. Splitting the sum in \eqref{eq:Texm} according to the pattern of repeated indices and associating multigraphs as in Sec.\ \ref{sec:largeN}, we can see that only the terms in which each $\delta\bJ_{ij}$ and $\delta\bJ_{ki}$ is paired with exactly one partner survive at leading order at large $N$ ($m_E =2$ for all edges in the multigraph; Prop.\ \ref{prop:count}). The partner may come from the same copy of $T$ or the conjugated $T^*.$ The result is that the scaling of all $\delta\bJ_{ij}, \delta\bJ_{ji}$ contractions contribute a factor that is at most $O(1).$

Let us then consider terms with $\alpha+\beta >2$. Prop.\ \ref{prop:conncorrnosum} implies that the cumulant in the definition of $T$ scales as $O(N^{(1-r)/2})$, where in this case $r$ counts the number of unique indices among the $j_1,\cdots, j_\alpha, k_1,\cdots,k_\beta.$ Since each $\delta\bJ_{ij}, \delta\bJ_{ji}$ has exactly one partner in a leading graph, one has $r\geq\lceil(\alpha+\beta)/2\rceil.$ Since the contractions with the $\delta\bJ_{ij}, \delta\bJ_{ki}$ contribute $O(1)$, it follows that the higher cumulant terms $\alpha+\beta>2$ are subleading.  The same bound at $\alpha+\beta=2$ with $r=2$ shows that the off-diagonal terms $\delta\bJ_{ij}\Rc_{jk} \delta\bJ_{ki}$ and $\delta\bJ_{ij}\Cc_{jk}\delta\bJ_{ik}^*$ with $j\neq k$ are $O(N^{-1/2})$.

Keeping only the leading order terms, we arrive at the representation
\begin{gather} 
    \langle \obs \rangle = \int D(\bx_i,\hat \bx_i, \bpsi_i, \hat\bpsi_i ) e^{S^i_{(\mathrm{cav})} + O(N^{-1/2})} \obs ,
    \\
        S^i_{(\mathrm{cav})} = S^0_i[\bx_i,\hat \bx_i, \bpsi_i, \hat\bpsi_i] \nonumber
        \\ + \sum_{j} i\hat \bx_i \delta\bJ_{ij} \mc_j + \sum_{\rho } i\hat \bx_i \bu^\rho_i \bkappa^\rho  \nonumber
        \\
        + \sum_{j} i\hat \bx_i \delta\bJ_{ij} \Rc_{jj} \delta\bJ_{ji} \bpsi_i - \frac 12 \hat \bx_i \sum_j |\delta\bJ_{ij}|^2 \Cc_{jj} \hat \bx_i.\nonumber
\end{gather} 
In the large $N$ limit, each $\mc_j$ is close to the full $\bmm_j$ for $j\neq i$. Introducing the cavity variable $\bpsi_i$ into the network dynamics and expanding gives after averaging: $\bmm_j = \mc_j + \Rc_{jj} \delta\bJ_{ji} \bmm_i.$ It follows that up to $O(N^{-1/2})$ errors, one has $\sum_{j} i\hat \bx_i \delta\bJ_{ij} \mc_j = \sum_j i\hat \bx_i \delta\bJ_{ij} \bmm_j - \sum_j i\hat \bx_i \delta\bJ_{ij} \Rc_{jj} \delta\bJ_{ji} \bmm_i + O(N^{-1/2})$ above.

Turning to the two-point functions, the same line of argument shows that we may replace $\Rc_{jj}, \Cc_{jj}$ with their full, limiting values $ \Rd_{jj}, \Cd_{jj}$ up to errors that must vanish at some rate with $N$ (Prop.\ \ref{prop:restrictse}). The key difference is that the corrections from reintroducing the cavity variable enter only at second order. For convenience, we will make the assumption following standard DMFT calculations that the resulting errors between $\bmm_j, \bC_{jj}$ and their large-$N$ limiting values are at most $O(N^{-1/2})$, and that the error between $\bR_{jj}$ and $\Rd_{jj}$ is at most $O(N^{-1/2}\omega^{-p_0})$, uniformly in $\omega$, so that the finite-$N$ corrections to the response decay with frequency as the response itself does (Eq.\ \eqref{eq:omassmptn1}). 

The result is a theory in which $\bpsi_i$ obeys its standard dynamics and receives an additional gaussian drive with mean $\sum_j \delta\bJ_{ij} \bmm_j + \sum_\rho \bu^\rho_i \bkappa^\rho $ and autocorrelation $\sum_j |\delta\bJ_{ij}|^2 \bC_{jj}$, as well as a self-coupling in which $\bx_i$ is driven by a term $\int d\tau\, \bar K_i(t-\tau) (\bpsi_i(\tau)-\bmm_i)$, with $\bar K_i(\omega) = \sum_j \delta\bJ_{ij}(\omega) \bR_{jj}(\omega) \delta\bJ_{ji}(\omega)$ the kernel of Eq.\ \eqref{eq:kersadp} :
\begin{gather}
    S^i_{(\mathrm{DMFT})} = S^0_i + \sum_j i\hat \bx_i \delta\bJ_{ij} \bmm_j + \sum_\rho i\hat \bx_i \bu^\rho_i \bkappa^\rho
    \\ 
    +\sum_j i\hat \bx_i \delta\bJ_{ij} \bR_{jj} \delta\bJ_{ji} (\bpsi_i - \bmm_i)
     -\frac 12 \hat \bx_i \sum_j |\delta\bJ_{ij}|^2 \bC_{jj} \hat \bx_i \nonumber
\end{gather}
The limiting DMFT values obey the self-consistent equations 
\begin{gather}
    \md_i = \langle \bpsi_i\rangle_D, \ \ \bkappa^\rho = \sum_j \by^\rho_j \md_j,
    \\
    \Cd_{ii} = \langle\delta \bpsi_i \delta \bpsi_i^* \rangle_D, \quad \Rd_{ii} = \langle \bpsi_i i\hat \bx_i\rangle_D.  \nonumber 
\end{gather}
where $\langle \cdot \rangle_D$ denotes the average with respect to the corresponding DMFT action $S^i_{(\mathrm{DMFT})},$ and we have repeated the condition on $\bkappa$ for completeness. 

Under this assumption, together with the diagonal bounds $\bR_{ii}-\rsp_{ii}=O(N^{-1/2}\omega^{-p_0})$ and $\bC_{ii}-\csp_{ii}=O(N^{-1/2})$ of Eqs.\ \eqref{eq:fullhessR} and \eqref{eq:fullhess}, we can use Prop.\ \ref{prop:sadpse} to swap the DMFT statistics into the self-energy equations (Eq.\ \eqref{eq:fullse}). In this way, we may obtain the full two-point correlation and response functions without ever observing the system: one first solves the above DMFT for $\Rd_{ii}, \Cd_{ii}, \md_i$ and then plugs them into Eq.\ \eqref{eq:fullse} \& \eqref{eq:twopoint}.

\subsection{Example $\bM$ matrices}
\label{sec:Mex}
We now give a few examples of models that satisfy the hypothesis Eq.\ \eqref{eq:balance}. As before, we will assume that we are in a parameter regime in which the resulting model dynamics and DMFT are stable.

\subsubsection{Balanced EI networks}

We first consider the balanced network connectivity considered in Refs.\ \cite{van1996chaos, KadmonSompolinsky2015}. To summarize, one divides the network into $r$ populations of neurons, each with an extensive number of neurons, and the statistics of the connections are chosen based on the population of the pre- and post-synaptic cell. Formally, if we let $b(\cdot)$ denote the function mapping neuronal indices to the $1, \dots, r$ populations, then 
\begin{gather}
    \bM_{ij} = N^{-1/2} \Mpop_{b(i)b(j)} ,\quad \langle \delta\bW_{ij}^2 \rangle =  N^{-1} \tilde \bsigma^2_{b(i) b(j)}, 
\end{gather}
for some $r\times r$ matrices $\Mpop, \tilde \bsigma$ with $O(1)$ entries. The external input into each neuron varies by population and is given by an $r$ dimensional vector, $\sqrt{N} \bQ$, where the entries of $\bQ$ are $O(1)$.  We may therefore factorize the mean matrix as $\tilde \bM =\Gamma(\omega)N^{-1/2} \bF\Mpop \bF^\top = \bU\bY^\top$ for a common synaptic filter $|\Gamma(\omega)| \leq K(1+|\omega|)^{-g}$ ($\bGamma_{ij}(\omega)=\Gamma(\omega)$ for all $i,j$) with $\bF_{i\rho} = \delta_{b(i) \rho}$ the population indicator function, and $\bU=\sqrt{N}\, \Gamma(\omega)\bF\Mpop, \bY= N^{-1}\bF$.

The scalings on $\bU,\bY$ are chosen so that the resulting collective fields are $O(1).$\footnote{Note that we can additionally enforce $\bM_{ii}=0$ (no autapses) by absorbing an $O(N^{-1/2})$ modification into the definition of $S^i_0$. This generates a negligible $O(N^{-1/2})$ correction to the (diagonal) self-energies.} We will assume that the population-averaged response at the $\bkappa$ saddle point, the matrix $\bchi$ defined below, does not decay faster than the single-unit response: $||\bchi(\omega)^{-1}||_{op}=O(\omega^{p_0})$, and we assume that the mean-field solution is stable against uniform perturbations, as in Ref.\ \cite{KadmonSompolinsky2015}. 

We now estimate $\bI + \rkap$ via the identity $\bI + \rkap = (\bI-\bOmega)^{-1}$ (Eq.\ \eqref{eq:interpretkappahatkappa}). Let us consider $\bOmega_{\rho\sigma} = [\bY^\top (\bI -\bSigma^{s}\delta\bJ)^{-1}\bSigma^{s}\bU]_{\rho\sigma}$. We write this as $\sqrt N \Gamma(\omega)\bchi\Mpop$ where we define the $r\times r$ frequency-dependent matrix $\bchi_{\rho\sigma} =N^{-1} \sum_{i: b(i) =\rho} \sum_{j: b(j)=\sigma} [(\bI - \bSigma^{s}\delta\bJ)^{-1} \bSigma^{s}]_{ij}$; that is, $\bchi_{\rho\sigma}$ is the response of the population average of population $\rho$ to a unit input into every neuron of population $\sigma$, with the collective feedback removed. Let us further separate $\bchi$ into diagonal and off-diagonal contributions
\begin{gather}
    \bchi_{\rho\sigma} = \frac {\delta_{\rho\sigma} } N\sum_{i: b(i)=\rho}  \bR^{s}_{ii} \nonumber
    \\ + \frac 1 N \sum_{i : b(i)=\rho, j: b(j)=\sigma, i\neq j } \bR^{s}_{ij}.
\end{gather}
Applying Prop.\ \ref{prop:conncorr} to estimate the scalings of these two sets of terms gives $\bchi_{\rho\sigma} = O(\delta_{\rho\sigma} \omega^{-p_0} + \omega^{-p-p_0} )$.  Using $||(\Gamma\bchi)^{-1}(\omega)||_{op} = O(\omega^p)$ with $p=p_0+g$ and assuming invertibility of $ \Mpop,$ one therefore has $(\bI - \bOmega)^{-1} = (\bI - \sqrt{N} \Gamma\bchi\Mpop)^{-1} = O(\min\{1, N^{-1/2}\omega^p\})$ with $p=p_0+g>1$ by assumption (for an LIF network driven by white noise, with exponential synapses, $p_0=1/2$, $g=1$, so that $p=3/2$). It follows that the balance assumption, Eq.\ \eqref{eq:balance}, holds with $\beta(N,\omega) =  \min\{\sqrt{N} , 1+\omega^p \}$. Thus, within a band whose upper limit grows with $N$, one has $\beta \sim \omega^p$. The result is that the high frequency expansion of Eq.\ \eqref{eq:highfreq} is valid in the limit $N ,\omega \gg 1$ so long as one imposes $\omega^{p} \gg \sqrt N.$

\subsubsection{Low-rank networks}

We now consider the low-rank networks of Ref.~\cite{MastrogiuseppeOstojic2018}. We take $\tilde \bM=\bU\bY^\top$ with $\bu^\rho_i=O(1)$ and
$\by^\rho_i=O(N^{-1})$, so that $\tilde u_{\max}\tilde y_{\max}=O(N^{-1})$, and we assume collective stability:
$\|(\bI-\bOmega(\omega))^{-1}\|_{op}=O(1)$ uniformly in $N$ for real $\omega$, which is the
linear-stability condition for the collective modes of such networks \cite{MastrogiuseppeOstojic2018}. 

By
Prop.~\ref{prop:conncorr} with $r_0=0$,
$\|\bOmega(\omega)\|_{op}=\|\bY^\top\rsp \bU\|_{op}=O(N\tilde u_{\max}\tilde y_{\max}\omega^{-p})=O(\omega^{-p})$
uniformly in $N$. For $\omega\gg1$ one therefore has $\|(\bI-\bOmega)^{-1}\|_{op}\le2$ with no
further assumption, while for $\omega=O(1)$ stability gives $\|(\bI-\bOmega)^{-1}\|_{op}=O(1)$. Since
$\bI+\rkap=(\bI-\bOmega)^{-1}$ (Eq.~\eqref{eq:interpretkappahatkappa}), the balance assumption Eq.~\eqref{eq:balance}
holds with $\beta=1$: $\tilde u_{\max}\tilde y_{\max}\|\bI+\rkap\|_{op}=O(N^{-1})$ at every frequency.
Consequently the $\bM$-dependent terms of Eq.~\eqref{eq:fullhess} are
$O(N^{-1}\omega^{-p-p_0})$ for $\bR$ and $O(N^{-1}\omega^{-p})$ for $\bC$ (Eq.~\eqref{eq:hessscale} with $\beta=1$) entrywise at all frequencies and never rise above the off-diagonal error. That is,  for low-rank networks the contributions from the mean connectivity sits inside the error terms in Eq.\ \eqref{eq:twopoint}.

\setcounter{footnote}{\savedfootnotecount}
\endgroup

\section{\label{SM:inverse}Uniqueness of the inverse}

In the main text, we showed that the knowledge of the effective interaction matrix $\bV$ allows one to compute the spike covariance $\bC$ in the large-$N$ limit via the formula
\begin{equation}\label{eq:S1}
\bC(\omega) = [\bI - \bV(\omega)]^{-1} \bLambda(\omega) [\bI - \bV(\omega)]^{-\dag}\, ,
\end{equation}
where $\bLambda$ is diagonal.
In this section, we show that the inverse problem, reconstructing $\bV$ and $\bLambda$ from $\bC$, has a unique functional solution valid for all frequencies, provided the following three physically motivated conditions are met: (i) synaptic causality, (ii) network stability, (iii) source diagonality. We will elaborate on each ingredient in a short while. 

\subsection{Assumptions}

\textbf{Synaptic causality.}
We assume that every synapse transmits signals causally and not instantaneously.
Mathematically, each
entry of $\bV$ is a bounded causal filter with no zero-lag components.
A filter $f(\tau)$ is called \textit{causal} if it vanishes for $\tau<0$, i.e., makes the output depend only on the inputs preceding it. In frequency domain, $f(\omega) = \int_{-\infty}^{\infty} \e^{i \omega \tau} f(\tau)\, \d \tau$ must therefore be analytic in the upper half-plane (UHP), defined as $\text{Im }\omega>0$. The requirement of the absence of zero-lag weights implies $\bV(\omega)$ has to vanish as $\text{Im }\omega \to \infty$ in the UHP. A product or convolution of two causal filters is itself a causal filter. For a review on the connection between causality and analyticity, see \cite{Nussenzveig1972}.

For example, a membrane has the causal filter $f(\tau) = \frac{1}{\tau_i}\e^{-\tau/\tau_i} \theta(\tau)$ which in frequency space is $f(\omega) = 1/(1- i\omega \tau_i)$. For every $\tau_i > 0$, it is analytic everywhere except for the pole at $\omega = -i/\tau_i$ in the lower half-plane, and it vanishes when $|\omega| \to \infty$. As another example, consider a pure conduction delay $f(\tau) = \delta(\tau - \tau_{ij})$ with $\tau_{ij}>0$ which in frequency space gives $f(\omega) = \e^{i \omega \tau_{ij}}$. It is a bounded causal filter, translating to analyticity in the UHP and boundedness $|\e^{i \omega \tau_{ij}}| = \e^{- ( \text{Im}\,\omega)\tau_{ij}} \leq 1$. Note that this $f(\omega)$ does not decay along the real $\omega$ axis.

\textbf{Network stability.}
We assume that every bounded input to the network produces a bounded response: the
resolvent $(\bI - \bV)^{-1}$ is a bounded causal filter. Mathematically, it means that $|\det(\bI - \bV)|$ stays bounded away from zero for every frequency in the UHP and the real axis.

\textbf{Source diagonality.} We assume that $\bLambda(\omega)$ is diagonal and each entry is bounded above and away from zero ($0<m\le\bLambda_{ii}(\omega)\le M<\infty$ with constants $m$ and $M$ for all real $\omega$). In other words, the emission sources are private to each neuron.

\textbf{Degeneracy.}
Notice that the inverse problem has a redundancy. Rescaling $\bI - \bV \to \bD( \bI - \bV)$ and $\bLambda \to \bD \bLambda \bD^{\dag}$ with any invertible diagonal matrix $\bD(\omega)$, yields back the same $\bC(\omega)$ unchanged:
\be
\bC \to [\bD(\bI - \bV)]^{-1} (\bD\bLambda\bD^\dag) [\bD (\bI - \bV)]^{-\dag}
= \bC\, .
\ee
The above transformation is therefore a degeneracy with $N$ degrees of freedom preserving diagonality of $\bLambda(\omega)$. A good way of fixing it is to set all the diagonal entries $\bV_{ii}(\omega) = 0$, which fixes $N$ functions and removes the degeneracy. In other words, any synaptic self-connections can be absorbed into the effective source of fluctuations $\bLambda_{ii}(\omega)$ emitted within a given cell $i$. Hence, moving forward, we assume that the diagonal of $\bV(\omega)$ vanishes for every frequency.

\subsection{Leading Born approximation}

Before proceeding with the exact inversion, it is instructive to understand how it works at the leading orders in the Born expansion. Recall that through monosynaptic order, we have
\be
\bC = \bLambda + \bV \bLambda + \bLambda \bV^\dag + \ldots
\ee
From now on, we truncate to this order and suppress the ellipses.
Since $\bV$ has a zero diagonal, and $\bLambda$ is diagonal, we can identify $\bC_{ii} = \bLambda_{ii}$, which is then bounded above and away from zero. Plugging this result into the off-diagonal entries, we get
\be\label{eq:Cij-mono}
\bC_{ij} = \bV_{ij} \bC_{jj} + \bC_{ii} \bV_{ji}^\ast\, .
\ee
The two terms have a distinct physical interpretation since $\bV_{ij}$ propagates modes $j \to i$ while $\bV_{ji}^\ast$ propagates $i \to j$. In the time domain, measuring the time difference $\tau \equiv t_i - t_j$, the filters $\bV_{ij}$ and $\bV_{ji}^\ast$ are supported on $\tau > 0$ and $\tau < 0$ respectively. Henceforth, we will call them causal and anti-causal.

The thing preventing us from cleanly separating the two terms in \eqref{eq:Cij-mono} and solving for $\bV$ is the fact that the $\bC$ is itself not causal: it has support for all $\tau$. The first step is therefore separating $\bC_{ii}$ into two contributions:
\be\label{eq:A-def}
\bC_{ii}(\omega) = c_{i}(\omega)\, c_i^\ast(\omega)\, ,
\ee
where $c_i$ is causal and $c_i^\ast$ is anti-causal. This separation is not unique, since we could multiply $c_i$ by an $\omega$-dependent phase. Let us continue with the computation and see how the choice of the phase affects the arguments. Plugging \eqref{eq:A-def} into \eqref{eq:Cij-mono} and dividing through $c_i c_j^\ast$, we get
\be\label{eq:CAA}
\frac{\bC_{ij}}{c_i c_j^\ast} = \bV_{ij} \frac{c_j}{c_i} + \left(\bV_{ji} \frac{c_i}{c_j}\right)^{\!\ast} .
\ee
The two terms on the right-hand side are causal and anti-causal respectively and hence the only remaining thing to do is to project both sides of this equation onto purely causal contributions. This can be done by convolving with the Fourier transform of the Heaviside step function $\theta(t)$. Let us call the resulting projector $P_+$:
\be
P_{+} \left[ f(\omega) \right] \equiv -\frac{1}{2\pi i }\int_{-\infty}^{\infty} \frac{f(\omega')\, \d \omega' }{\omega -\omega' + i \varepsilon} \, ,
\ee
where $\varepsilon \to 0^+$.
Applying $P_+$ to both sides of \eqref{eq:CAA} and solving for $\bV_{ij}$ gives
\be\label{eq:Vij-mono}
\bV_{ij} = \frac{c_i}{c_j} P_+ \!\left[ \frac{\bC_{ij}}{c_i c_j^\ast} \right].
\ee
Finally, the same projector can be used to compute $c_i$ and $c_j$. The only difference is that they enter multiplicatively in \eqref{eq:CAA} and the projector is linear, so we first apply the logarithm and then exponentiate, giving
\be
c_i = \e^{P_+[\log \bC_{ii}]}\, .
\ee
This gives the formula \eqref{eq:V-ij}.
The logarithm does not have any singularities since $\bC_{ii}$ is bounded above and away from zero. Notice that $1/c_i$ satisfies the same properties and hence both $c_i^{\pm 1}$ are bounded causal filters.

\subsection{Equivalence with least squares}

Let us compare the above approach with the least-squares method described in App.~\ref{app:Numerics}. For every pair of neurons $i$ and $j$, we use 
\be
\mathcal{L}_{ij} = \int_{\omega_l}^{\omega_h} \frac{\d \omega}{2\pi} \frac{| \bC_{ij} - \bV_{ij} \bC_{jj} - \bC_{ii} \bV_{ji}^\ast |^2}{\bC_{ii} \bC_{jj}}\, ,
\ee
which is a continuous version of the cost function from App.~\ref{app:Numerics} where we now perform an integral instead of a sum over frequencies. The goal is to minimize $\mathcal{L}_{ij}$.
Since $|c_i c_j^\ast|^2 = \bC_{ii} \bC_{jj}$, we can rewrite
\be
\mathcal{L}_{ij}[\bm{g}] = \int_{\omega_l}^{\omega_h} \frac{\d \omega}{2\pi} \left| \frac{\bC_{ij}}{c_i c_j^\ast} - \bm{g} \right|^2
\ee
where
\be
\bm{g} (\omega) = \bV_{ij}\frac{c_j}{c_i} + \left(\bV_{ji}\frac{c_i}{c_j}\right)^\ast\, .
\ee
The optimization problem is therefore finding the optimal function $\bm{g} = \bm{g}_0$ belonging to a given Hilbert space $\mathcal{H}$ of admissible whitened cross-correlations (in the simplest case, $\mathcal{H} = L^2(\mathbb{R})$):
\be
\bm{g}_0 = \arg \min_{\bm{g} \in \mathcal{H}} \mathcal{L}_{ij}[\bm{g}]\, .
\ee
An example basis of $\mathcal{H}$ is the raised-cosine basis used in App.~\ref{app:Numerics}. This finally gives
\be
\bV_{ij} = \frac{c_i}{c_j} P_+ \left[ \bm{g}_0 \right]\, .
\ee
If we do not truncate in frequency at all, i.e., set $(\omega_l, \omega_h) = (-\infty,\infty)$, and take $\mathcal{H} = L^2(\mathbb{R})$, the unique optimal solution is $\bm{g}_0 = \bC_{ij}/(c_i c_j^\ast)$ giving \eqref{eq:Vij-mono}. For a non-trivial frequency filter, we have $\bm{g} = \bC_{ij}/(c_i c_j^\ast) + \bm{h}$, where $\bm{h}$ can be any function in the Hilbert space $\mathcal{H}$ that vanishes on the interval $(\omega_l, \omega_h)$. The solution \eqref{eq:Vij-mono} with $\bm{h}=0$ is still the global minimum, but it is no longer unique. In the fits of App.~\ref{app:Numerics}, the minimizer is selected by the low-dimensional choice of $\mathcal{H}$, which acts as a prior on the kernels.
This is an inherent feature of the band-limited reconstruction forced on us by the monosynaptic truncation. It motivates finding a formula for $\bV_{ij}$ valid across all frequencies. We turn to this problem next.

\subsection{\label{app:inverse-general-case}General case}

\textbf{Setup.} Suppose that two pairs $(\bV^{(a)}, \bLambda^{(a)})$ for $a=1,2$ produce the same $\bC$. By construction, we know that at least one such pair exists. We will show that it is unique, i.e., $(\bV^{(1)}, \bLambda^{(1)}) = (\bV^{(2)}, \bLambda^{(2)})$. For each pair, let us construct the matrix
\be
\bc^{(a)} \equiv (\bI - \bV^{(a)})^{-1} \bD_{\bLambda}^{(a)}
\ee
with
\be
\quad\bD_{\bLambda}^{(a)} \equiv \text{diag}(\lambda_i^{(a)}) \quad\text{and}\quad \lambda_i^{(a)} \equiv \e^{P_+[\log \bLambda_{ii}^{(a)}]} 
\ee
so that $|\lambda_i^{(a)}(\omega)|^2 = \bLambda^{(a)}_{ii}(\omega)$ for real $\omega$. With these definitions, we can write
\be
\bC = \bc^{(a)}\, \bc^{(a)\dag}
\ee
on the real axis. Using synaptic causality on $\bI - \bV^{(a)}$ and network stability for $(\bI - \bV^{(a)})^{-1}$, both $\bc^{(a)}$ and their inverses are bounded causal filters.

\textbf{Spectral factorization.}
The ratio of the two matrices $\bc^{(a)}$ is
\be\label{eq:U-omega}
\bU(\omega) \equiv [\bc^{(1)}(\omega)]^{-1} \bc^{(2)}(\omega)\, .
\ee
Both $\bU$ and $\bU^{-1}$ are bounded causal filters. Notice that for real frequencies $\bU(\omega)$ is unitary:
\be
\bU \bU^\dag = [\bc^{(1)}]^{-1} \underbrace{\bc^{(2)} \bc^{(2)\dag}}_{\bc^{(1)} \bc^{(1)\dag}}[\bc^{(1)\dag}]^{-1} = \bI\, ,
\ee
which means
\be
\bU^{-1}(\omega) = \bU^\dag(\omega)\, .
\ee
The right-hand side is anti-causal, because Hermitian conjugate maps a filter $f(\tau)$ to $f^\ast(-\tau)$ and hence reverses the time lag. A filter that is simultaneously causal and anti-causal can be only supported at zero lag: it must be a finite combination of $\delta(\tau)$ and its derivatives. In frequency space, $\bU(\omega)$ has to be a polynomial in $\omega$, but a bounded polynomial has to be a constant. We conclude that $\bU(\omega) = \bU$ is a constant unitary matrix.

Since $\bU$ is constant, we can evaluate the right-hand side of \eqref{eq:U-omega} anywhere in the UHP.
Let us pick $\text{Im } \omega \to \infty$ where $\bV^{(a)}(\omega)$ both vanish and hence $\bc^{(a)}(\omega) \to \bD_{\bLambda}^{(a)}(\omega)$ which are diagonal matrices bounded above and away from zero. This implies $\bU$ can be written as
\be
\bU = [\bD_{\bLambda}^{(1)}(\omega)]^{-1}  \bD_{\bLambda}^{(2)}(\omega)
\ee
for $\text{Im}\, \omega \to \infty$
and hence it is diagonal with $|\bU_{ii}| = 1$.

\textbf{Uniqueness.}
Massaging \eqref{eq:U-omega}, we find
\be\label{eq:one-minus-V}
\bI - \bV^{(1)}(\omega) =
\bD_{\bLambda}^{(1)}(\omega) \bU [\bD_{\bLambda}^{(2)}(\omega)]^{-1} [\bI - \bV^{(2)}(\omega)]\, .
\ee
Since $\bV^{(a)}$ have zero diagonals, taking the diagonal of \eqref{eq:one-minus-V} gives
\be\label{eq:one-minus-V2}
\bI  =
\bD_{\bLambda}^{(1)}(\omega) \bU [\bD_{\bLambda}^{(2)}(\omega)]^{-1} \, .
\ee
for every $\omega$. 
Combined with the fact that $\bU$ is unitary, this result tells us that the two source matrices must be the same:
\begin{align}
\bLambda_{ii}^{(1)}(\omega) &= |\lambda_i^{(1)}(\omega)|^2 \\
&= |\lambda_i^{(2)}(\omega) \bU_{ii}^{-1}|^2 = |\lambda_i^{(2)}(\omega)|^2 = \bLambda_{ii}^{(2)}(\omega)\, .
\end{align}
Finally, substituting \eqref{eq:one-minus-V2} into \eqref{eq:one-minus-V} gives
\be
\bV^{(1)}(\omega) = \bV^{(2)}(\omega)\, .
\ee
This concludes the proof.

\subsection{Explicit formula}

The above proof can be used to construct an explicit solution for $\bV$. Let us use the notation of Sec.~\ref{app:inverse-general-case}, but drop the superscript ${}^{(a)}$ since the solution is unique. We are going to consider the matrix $\bc(\omega) = [\bI - \bV(\omega)]^{-1} \bD_{\bLambda}(\omega)$. In practice, it cannot be computed from data since it depends on the unknowns $\bLambda$ and $\bV$. 
However, one can compute a spectral factorization
\be
\bC(\omega) = \bb(\omega)\, \bb(\omega)^\dag
\ee
in which $\bb$ and $\bb^{-1}$ are bounded causal filters, for instance with Wilson's algorithm \cite{Wilson1972, Wilson1978, Dhamala2008b}. Such $\bb$ is not unique. Applying the arguments around \eqref{eq:U-omega}, any such $\bb$ is related to $\bc$ by $\bb(\omega) = \bc(\omega)\,\bU$, where $\bU$ is a constant unitary matrix. For two frequencies $\omega$ and $\omega'$, let us define
\begin{align}
\bv(\omega', \omega) &\equiv \bb(\omega')\, \bb(\omega)^{-1} \\
&= [\bI - \bV(\omega')]^{-1} \bD_{\bLambda}(\omega')\, \bD_{\bLambda}(\omega)^{-1} [\bI - \bV(\omega)]\, ,\nn
\end{align}
which depends on both frequencies but not on which $\bb$ the algorithm returned. Note that $\bU$ canceled out. As $\text{Im}\,\omega' \to \infty$, $\bV(\omega')$ vanishes and hence
\be
\lim_{\text{Im}\,\omega' \to \infty} \bv_{ij}(\omega', \omega) = \frac{\lambda_i(\omega')}{\lambda_i(\omega)}\, [\delta_{ij} - \bV_{ij}(\omega)]\, .
\ee
The prefactor is common to the whole row $i$ and $\bV_{ii} = 0$, so for $i \neq j$
\be
\bV_{ij}(\omega) = - \lim_{\text{Im}\,\omega' \to \infty} \frac{\bv_{ij}(\omega', \omega)}{\bv_{ii}(\omega', \omega)}\, .
\ee
Since $\bv$ was obtained solely in terms of the covariance matrix $\bC$, this equation gives an exact inversion formula for $\bV$.

\subsection{Subleading Born terms}
Finally, let us demonstrate how to consistently find subleading Born expansion terms for $\bV_{ij}$. 
Let us first reorganize \eqref{eq:S1} to
\be
(\bI - \bV) \bC = \bLambda (\bI - \bV)^{-\dag}\, ,
\ee
whose off-diagonal entries $i \neq j$ are
\be\label{eq:Cij-full}
\bC_{ij} - \bV_{ij} \bC_{jj} - \sum_{k \neq i,j}\bV_{ik} \bC_{kj} = \bLambda_{ii} [(\bI - \bV)^{-1}]_{ji}^\ast \, .
\ee
We can split the diagonals of $\bC$ and $\bLambda$ into their causal and anti-causal contributions:
\begin{subequations}
\begin{align}
\bC_{jj} = c_j c_j^\ast \qquad &\text{with} \qquad c_j = \e^{P_+[\log \bC_{jj}]}\, ,\\
\bLambda_{ii} = \lambda_i \lambda_i^\ast \qquad &\text{with}\qquad \lambda_i = \e^{P_+[\log \bLambda_{ii}]}\, .
\end{align}
\end{subequations}
Both are bounded above and away from zero.
Substituting it into \eqref{eq:Cij-full} and dividing by $\lambda_i c_j^\ast$, we find
\be
\frac{\bC_{ij} - \sum_{k \neq i,j}\bV_{ik} \bC_{kj}}{\lambda_i c_j^\ast} = \bV_{ij} \frac{c_j}{\lambda_i} + \left( \frac{\lambda_i}{c_j} [(\bI - \bV)^{-1}]_{ji} \right)^{\!\ast} \, .
\ee
The first term on the right-hand side is causal and vanishes as $\text{Im}\, \omega \to \infty$ since the ratio $c_j/\lambda_i$ is bounded and $\bV_{ij}$ decays at infinity. The second term on the right-hand side is anti-causal and vanishes as $\text{Im}\,\omega \to -\infty$ for similar reasons. Applying the causal projector $P_+$ therefore gives
\be
\bV_{ij} = \frac{\lambda_i}{c_j} P_+ \!\left[ \frac{\bC_{ij} - \sum_{k \neq i,j}\bV_{ik} \bC_{kj}}{\lambda_i c_j^\ast} \right] .
\ee
This formula can be solved order by order in powers of $\bV$ (beyond second order, $\lambda_i$ must itself be expanded in $\bV$ through $\bLambda = (\bI - \bV)\bC(\bI - \bV)^\dag$). Let us write
\be
\bV = \bV^{(1)} + \bV^{(2)} + \ldots
\ee
where the superscript denotes the number of hops. Using the notation
\be
\mathcal{P}_{ij}[X] \equiv \frac{c_i}{c_j}\, P_+ \!\left[ \frac{X}{c_i\, c_j^\ast} \right]\, ,
\ee
the first couple of terms can be written as
\begin{subequations}\label{eq:V-born}
\begin{align}
\bV^{(1)}_{ij} &= \mathcal{P}_{ij}\big[ \bC_{ij} \big]\, , \label{eq:V-born-1}\\
\bV^{(2)}_{ij} &= -\mathcal{P}_{ij}\Big[ \sum_{k \neq i,j} \bV^{(1)}_{ik}\, \bC_{kj} \Big]\, . \label{eq:V-born-2}
\end{align}
\end{subequations}
Here we used the fact that, by \eqref{eq:S1}, $\bLambda_{ii} = \bC_{ii}[1 + \mathcal{O}(\bV^2)]$, so $\lambda_i = c_i[1 + \mathcal{O}(\bV^2)]$.
As expected, the first one coincides with the monosynaptic result \eqref{eq:Vij-mono}. The second one is the disynaptic correction.

\end{document}